\documentclass[12pt]{article}
\usepackage{graphicx}

\usepackage{scicite}
\usepackage{times}
\usepackage[usenames, dvipsnames]{color}
\usepackage{amsmath}
\usepackage{graphicx}
\usepackage{subcaption}

\usepackage[colorinlistoftodos]{todonotes}
\usepackage[colorlinks=true, allcolors=blue]{hyperref}
\usepackage{amssymb}
\usepackage{url}
\usepackage{rotating}

\usepackage[utf8]{inputenc}
\usepackage[T1]{fontenc}

\newcommand{\Fermi}{\emph{Fermi}}
\newcommand{\g}{$\gamma$}

\newcommand{\apj}{Astrophys. J.}
\newcommand{\apjl}{Astrophys. J. Lett.}
\newcommand{\apjs}{Astrophys. J. Suppl.}
\newcommand{\procspie}{Proc. SPIE}
\newcommand{\aj}{Astron. J.}
\newcommand{\pasj}{Publ. Astron. Soc. Jpn.}
\newcommand{\pasp}{Publ. Astron. Soc. Pac.}
\newcommand{\aap}{Astron. Astrophys.}
\newcommand{\app}{Astropart. Phys.}
\newcommand{\prl}{Phys. Rev. Lett.}

\newcommand{\mnras}{Mon. Notices Royal Astron. Soc.}
\newcommand{\nimp}{Nucl. Instr. Meth. Phys. Res.}
\newcommand{\jinst}{J. Instrum.}
\newcommand{\nustar}{\textit{NuSTAR}}
\newcommand{\swift}{\textit{Swift}}

\newcommand{\citesm}{\cite{supplementary_material}}

\newenvironment{sciabstract}{%
\begin{quote} \bf}
{\end{quote}}

\title{Multi-messenger observations of a flaring blazar coincident with high-energy neutrino IceCube-170922A}

\author{The IceCube, \Fermi-LAT, MAGIC, {\it AGILE}, ASAS-SN, HAWC, H.E.S.S, \\ 
{\it INTEGRAL}, Kanata, Kiso, Kapteyn, Liverpool telescope, Subaru, {\it Swift}$/$\nustar, \\
 VERITAS, and VLA/17B-403 teams~\footnote{The full lists of participating members for each team and their affiliations are provided in the supplementary material.}~\footnote{Email: analysis@icecube.wisc.edu}}

\date{}

\begin{document} 

\maketitle

\begin{sciabstract}

Individual astrophysical sources previously detected in neutrinos are limited to the Sun and the supernova 1987A, whereas the origins of the diffuse flux of high-energy cosmic neutrinos remain unidentified. On 22 September 2017 we detected a high-energy neutrino, IceCube-170922A, with an energy of $\sim$290~terra--electronvolts.  Its arrival direction was consistent with the location of a known \g-ray blazar TXS~0506+056, observed to be in a flaring state. An extensive multi-wavelength campaign followed, ranging from radio frequencies to \g-rays. These observations characterize the variability and energetics of the blazar  and include the first detection of TXS~0506+056 in very-high-energy \g-rays. This observation of a neutrino in spatial coincidence with a \g-ray emitting blazar during an active phase suggests that blazars may be a source of high-energy neutrinos.

\end{sciabstract}

Since the discovery of a diffuse flux of high-energy astrophysical neutrinos~\cite{Aartsen:2013jdh,Aartsen:2016xlq}, IceCube has searched for its sources. The only non-terrestrial neutrino sources identified previously are the Sun and the supernova 1987A, producing neutrinos with energies millions of times lower than the high-energy diffuse flux, such that the mechanisms and the environments responsible for the high-energy cosmic neutrinos are still to be ascertained~\cite{Aartsen:2016oji,Aartsen:2016lir}. 
Many candidate source types exist, with Active Galactic Nuclei (AGN) among the most prominent~\cite{Stecker1991}, in particular the small fraction of them designated as radio-loud~\cite{MANNHEIM1995295}.
In these AGNs, the central super-massive black hole converts gravitational energy of accreting matter and/or the rotational energy of the black hole into powerful relativistic jets, within which particles can be accelerated to high energies.
If a number of these particles are protons or nuclei, their interactions with the radiation fields and matter close to the source would give rise to a flux of high-energy pions that eventually decay into photons and neutrinos~\cite{Petropoulou:2015upa}. In blazars~\cite{Urry:1995mg} -- AGNs that have one of the jets pointing close to our line of sight --  the observable flux of neutrinos and radiation is expected to be greatly enhanced owing to relativistic Doppler boosting. Blazar electromagnetic (EM) emission is known to be highly variable on time scales from minutes to years~\cite{Ulrich:1997rf}. 

Neutrinos travel largely unhindered by matter and radiation. Even if high-energy photons (TeV and above) 
are unable to escape the source owing to intrinsic absorption, or are absorbed by interactions with the extragalactic background light (EBL) \cite{Hauser2001, Stecker1992}, high-energy neutrinos may escape and travel unimpeded to Earth. An association of observed astrophysical neutrinos with blazars would therefore imply that high-energy protons or nuclei up
to energies of at least tens of PeV are produced in blazar jets, suggesting that they may be the birthplaces of the most energetic particles observed in the Universe, the ultra-high energy cosmic rays~\cite{Agashe:2014kda}. If neutrinos are produced in correlation with photons, the coincident observation of neutrinos with electromagnetic flares would greatly increase the chances of identifying the source(s). Neutrino detections must therefore be combined with the information from broad-band observations across the electromagnetic spectrum (multi-messenger observations). 

To take advantage of multi-messenger opportunities,
the IceCube neutrino observatory~\cite{Aartsen:2016nxy} has established a system of real-time alerts that rapidly
notify the astronomical community of the direction of astrophysical neutrino candidates~\cite{Aartsen:2016lmt}. 
From the start of the program in April 2016 through October 2017, 10 public alerts have been issued for high-energy neutrino candidate events with well-reconstructed directions~\cite{amongcnref}.

We report the detection of a high-energy neutrino by IceCube and the multi-wavelength/multi-instrument observations of a flaring \g-ray blazar, TXS~0506+056, which was found to be positionally coincident with the neutrino direction \cite{2017Atel_Fermi}. Chance coincidence of the IceCube-170922A event with the flare of TXS~0506+056 is statistically disfavored at the level of 3$\sigma$ in models evaluated here associating neutrino and \g-ray production.

\section*{The neutrino alert}
\label{sec:alert}

\begin{figure*}
\begin{center}
\includegraphics[width=0.90\textwidth]{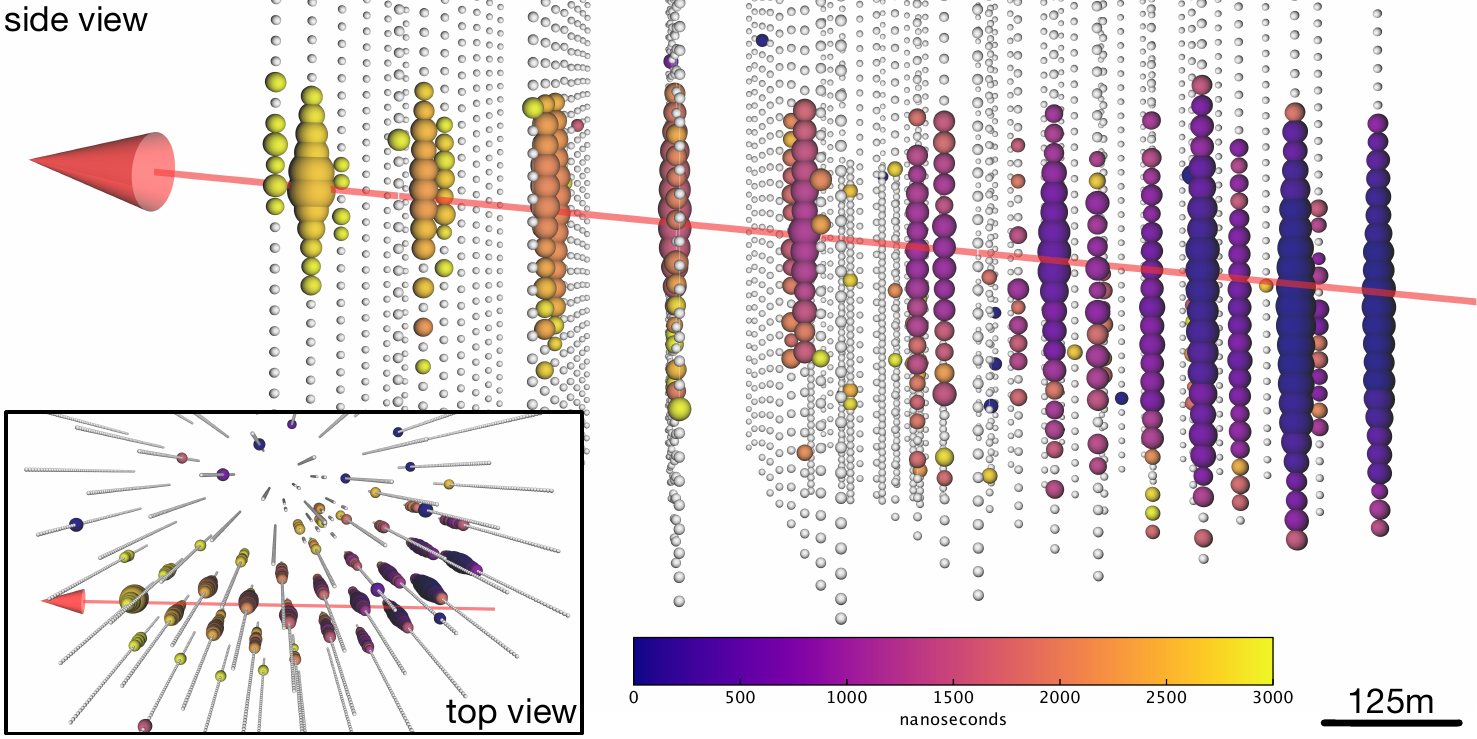} 
  \end{center}
\caption{{\bf Event display for neutrino event IceCube-170922A.}  The time at which a DOM observed a signal
  is reflected in the color of the hit, with dark blues for earliest hits and yellow for latest.  Time shown are relative
  to the first DOM hit according to the track reconstruction, and earlier and later times are shown with the same
  colors as the first and last times, respectively.  The total time the event
  took to cross the detector is $\sim$3000~ns.  The size of a colored sphere is proportional to the logarithm of the amount of light observed at the DOM, with larger spheres corresponding to 
larger signals.  The total charge recorded is $\sim$5800 photoelectrons. Inset is an overhead perspective view of the event. 
The best-fitting track direction is shown as an arrow, consistent with a zenith angle $5.7^{+0.50}_{-0.30}$\,degrees
below the horizon.}
\label{fig:eview}
\end{figure*}

IceCube is a neutrino observatory with more than 5000 optical sensors embedded in 1~km$^{3}$ of the Antarctic ice-sheet
close to the Amundsen-Scott South Pole Station. The detector consists of 86 vertical strings frozen into the ice 125~m apart, each equipped with 60 digital
optical modules (DOMs) at depths between 1450 and 2450~m. When a high-energy muon-neutrino interacts with an atomic nucleus in or close to the detector array, a muon is produced moving through the ice at superluminal speed and creating Cherenkov radiation detected by the DOMs. On 22 September 2017 at 20:54:30.43 Coordinated Universal Time (UTC), a high-energy neutrino-induced muon track event was detected in an automated analysis that is part of IceCube's real-time alert system.  An automated alert was distributed~\cite{I3_GCN_alert} to observers 43 seconds later, providing an initial estimate of the direction and energy of the event. A sequence of refined reconstruction algorithms was automatically started at the same time, using the full event information. A representation of this neutrino event with the best-fitting reconstructed direction is shown in Figure~\ref{fig:eview}.
Monitoring data from IceCube indicate that the observatory was functioning normally at the time of the event.

\begin{sidewaysfigure}
\begin{center}
  \begin{subfigure}[b]{0.45\textwidth}
    \caption{}
    \includegraphics[width=\textwidth]{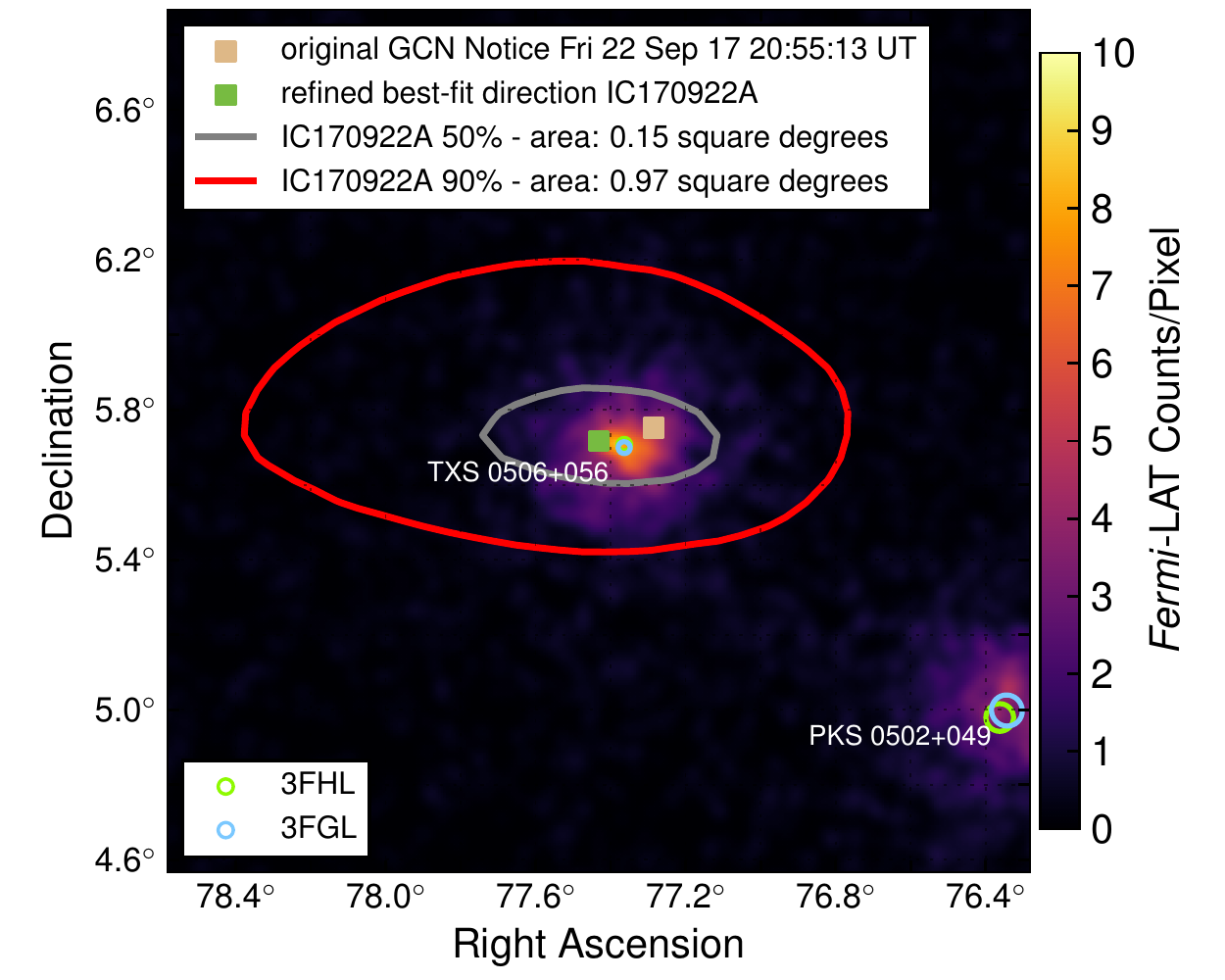}
  \end{subfigure}
  \begin{subfigure}[b]{0.45\textwidth}
    \caption{}
    \includegraphics[width=\textwidth]{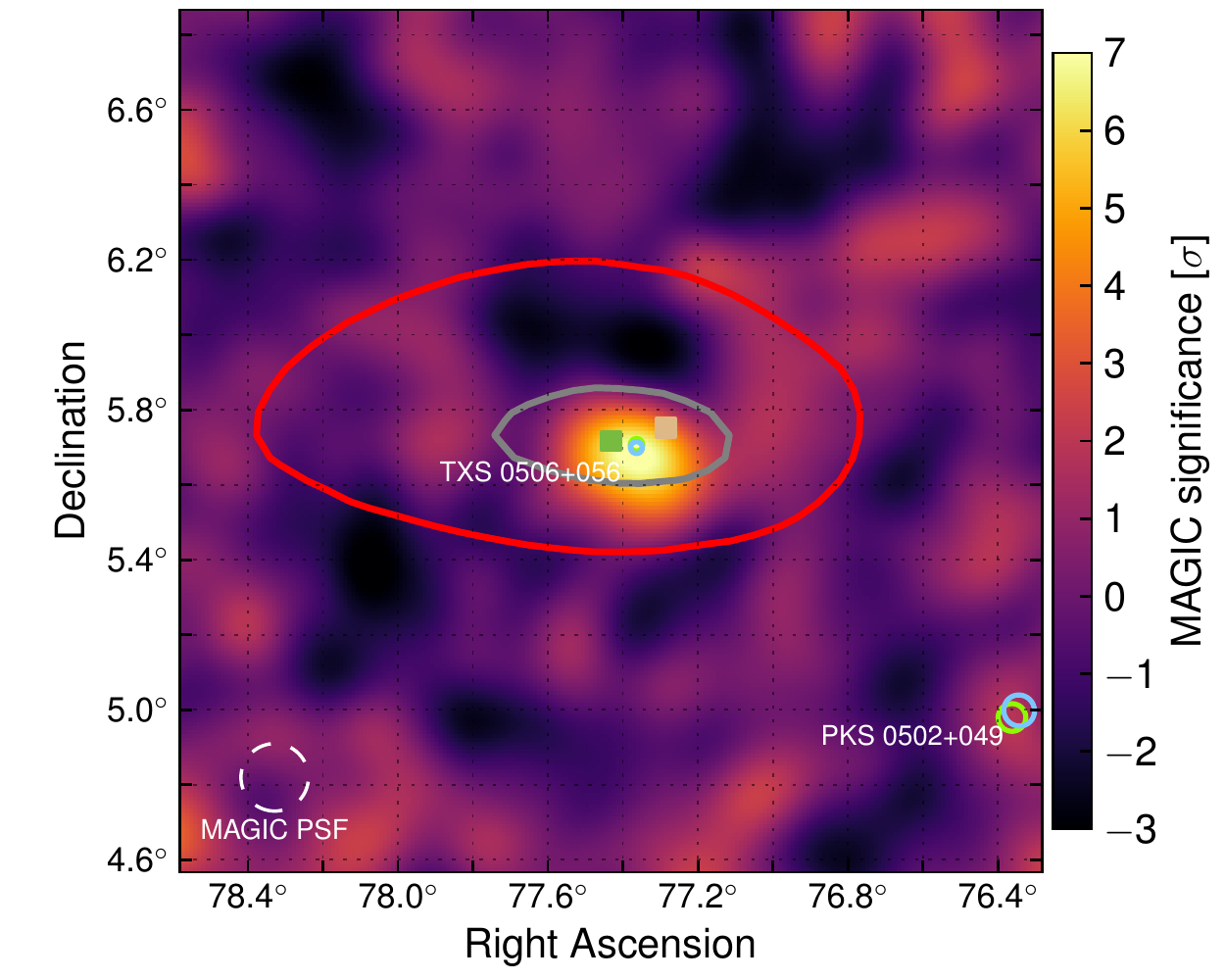}
  \end{subfigure}
\end{center}
\caption{{\bf \Fermi-LAT and MAGIC observations of IceCube-170922A's location.}
  Sky position of IceCube-170922A in J2000 equatorial coordinates overlaying
  the \g-ray counts from \Fermi-LAT above 1 GeV (A) and the
  signal significance as observed by MAGIC (B) in this region.
  The tan square indicates the position reported in the initial alert and the
  green square indicates the final best-fitting position from follow-up reconstructions~\cite{GCN21916}.
  Gray and red curves show the 50\% and 90\% neutrino containment regions, respectively, including
  statistical and systematic errors.  \Fermi-LAT data are shown as a photon counts map in 9.5 years of data
  in units of counts per pixel, using detected photons with energy of 1 to 300~GeV in a 2$^\circ$ by 2$^\circ$ region around TXS0506+056.
  The map has a pixel size of 0.02$^\circ$ and was smoothed with a 0.02~degree-wide Gaussian kernel.
  MAGIC data are shown as signal significance for \g-rays above 90~GeV.
  Also shown are the locations of a \g-ray source observed by \Fermi-LAT as given in the
  \Fermi-LAT Third Source Catalog (3FGL)~\cite{Acero:2015hja} and the Third Catalog of Hard \Fermi-LAT Sources (3FHL)~\cite{TheFermi-LAT:2017pvy} source catalogs, including
  the identified positionally coincident 3FGL object TXS~0506+056.  For \Fermi-LAT catalog objects, marker sizes indicate the 95\% C.L. positional uncertainty of the source.}
\label{fig:skymap}
\end{sidewaysfigure}

A Gamma-ray Coordinates Network (GCN) Circular~\cite{GCN21916} was issued $\sim$ 4~h after the initial notice including the refined directional information (offset 0.14$^{\circ}$ from the initial direction, see Figure~\ref{fig:skymap}).
Subsequently, further studies were performed to determine the uncertainty of the directional reconstruction arising from statistical and systematic effects, leading to a best-fitting right ascension (RA) $77.43^{+0.95}_{-0.65}$  and declination (Dec) $+5.72^{+0.50}_{-0.30}$  (degrees, J2000 equinox, 90\% containment region). The alert was later reported to be in positional coincidence with the known \g-ray blazar TXS~0506+056~\cite{2017Atel_Fermi}, which is located at RA 77.36$^{\circ}$ and Dec +5.69$^{\circ}$ (J2000)~\cite{Lanyi:2010}, 0.1$^{\circ}$ from the arrival direction of the high-energy neutrino.

The IceCube alert prompted a follow-up search by the Mediterranean neutrino telescope ANTARES (acronym for Astronomy with a Neutrino Telescope and Abyss environmental RESearch)~\cite{2011NIMPA.656...11A}. The sensitivity of ANTARES at the declination of IceCube-170922A is about one-tenth that of IceCube's~\cite{Albert:2017ohr} and no neutrino candidates were found in a $\pm$1 day period around the event time~\cite{ATel10773}.

An energy of 23.7$\pm$2.8~TeV was deposited in IceCube by the traversing muon. To estimate the parent neutrino energy, we performed simulations of the response of the detector array, considering that the muon-neutrino might have interacted outside the detector at an unknown distance. 
We assumed the best-fitting power-law energy spectrum for astrophysical high-energy muon neutrinos, ${\rm d}N/{\rm d}E \propto E^{-2.13}$~\cite{Aartsen:2016xlq} where $N$ is the number of neutrinos as a function of energy $E$.  The simulations yielded a most probable neutrino energy
of 290~TeV, with a 90\% confidence level (C.L.) lower limit of 183~TeV, depending only weakly on the assumed astrophysical energy spectrum \citesm{}.

The vast majority of neutrinos detected by IceCube arise from cosmic-ray interactions within Earth's atmosphere.  Although atmospheric neutrinos are dominant at energies below 100~TeV, their spectrum falls steeply with energy, allowing astrophysical neutrinos to be more easily identified at higher energies.
The muon-neutrino astrophysical spectrum, together with simulated data, was used to calculate the
probability that a neutrino at the observed track energy and zenith angle in IceCube is of astrophysical origin.
This probability, the so-called signalness of the event~\cite{Aartsen:2016lmt}, was reported to be 56.5\%~\cite{I3_GCN_alert}. Although IceCube can robustly identify astrophysical neutrinos at PeV energies, for individual neutrinos at several hundred TeV, an atmospheric
origin cannot be excluded. Electromagnetic observations are valuable to assess the possible association of a single neutrino to an astrophysical source. 

Following the alert, IceCube performed a complete analysis of relevant data prior to 31 October 2017. Although no additional excess of neutrinos was found from the direction of TXS~0506+056 near the time of the alert, there are indications at the 3$\sigma$ level of high-energy neutrino emission from that direction in data prior to 2017, as discussed in a companion paper~\cite{TXS0506_PS_INPREP}.

\section*{High-energy \g-ray observations of TXS~0506+056}

On 28 September 2017, the \Fermi~Large Area Telescope (LAT) Collaboration reported that the direction of origin of IceCube-170922A  was consistent with a known \g-ray source in a state of enhanced emission \cite{2017Atel_Fermi}.
\Fermi-LAT is a pair-conversion telescope aboard the {\it Fermi Gamma-ray Space Telescope} sensitive to \g-rays with energies from 20\,MeV to greater than 300\,GeV~\cite{Atwood}. Since August 2008, it has operated continuously, primarily in an all-sky survey mode. Its wide field of view of $\sim$2.4 steradian provides coverage of the entire \g-ray sky every 3 hours.  The search for possible counterparts to IceCube-170922A was part of the \Fermi-LAT collaboration’s routine multi-wavelength/multi-messenger program. 

Inside the error region of the neutrino event, a positional coincidence was found with a previously cataloged \g-ray source, 0.1$^\circ$ from the best-fitting neutrino direction. TXS~0506+056 is a blazar of BL Lacertae (BL Lac) type. Its redshift of $z=0.3365 \pm 0.0010$ was measured only recently based on the optical emission spectrum in a study triggered by the observation of IceCube-170922A~\cite{Paiano:2018qeq}.

TXS~0506+056 is a known \Fermi-LAT \g-ray source, appearing in three catalogs of \Fermi\ sources~\cite{Acero:2015hja, 2016ApJS..222....5A, TheFermi-LAT:2017pvy} at energies above 0.1, 50, and 10\,GeV, respectively. 
An examination of the \Fermi~ All-Sky Variability Analysis (FAVA)~\cite{FAVA} photometric light curve for this object
showed that TXS~0506+056 had brightened considerably in the GeV band starting in April 2017~\cite{2017Atel_Fermi}.
Independently, a \g-ray flare was also found by \Fermi's Automated Science Processing (ASP, \citesm). Such flaring is not unusual for a BL Lac object, and would not have been followed up as extensively if the neutrino were not detected.

\begin{sidewaysfigure}
\centering
\includegraphics[width=0.999\textwidth]{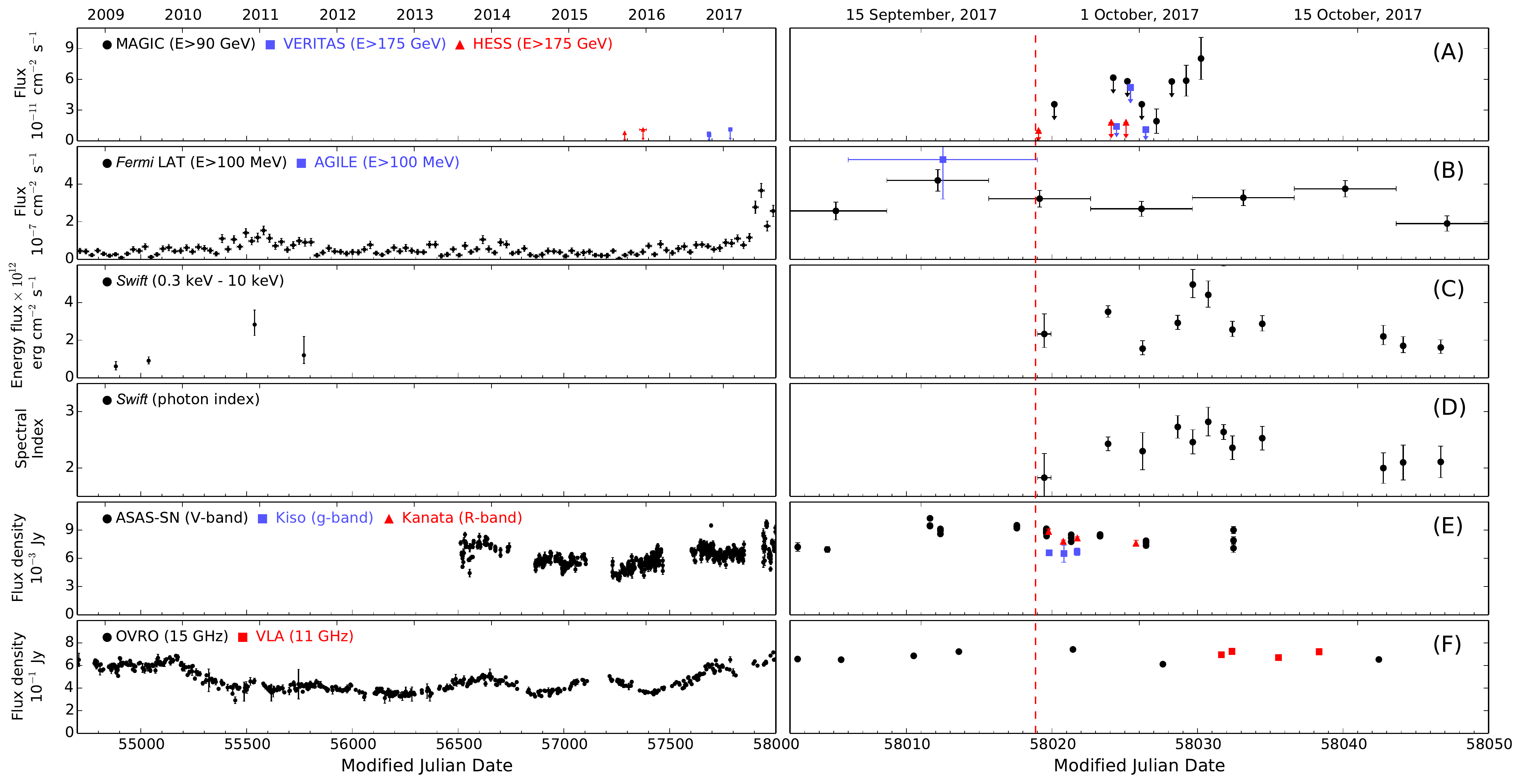}
\caption{{\bf Time-dependent multi-wavelength observations of TXS~0506+056 before and after IceCube-170922A.} Significant variability of the electromagnetic emission can be observed in all displayed energy bands, with the source being in a high emission state around the time of the neutrino alert. From top to bottom: (A) VHE \g-ray observations by MAGIC, H.E.S.S. and VERITAS; (B) high-energy \g-ray observations by \Fermi-LAT and {\it AGILE}; (C and D) x-ray observations by \swift\ XRT; (E) optical light curves from ASAS-SN,  Kiso/KWFC, and Kanata/HONIR; and (F) radio observations by OVRO and VLA. The red dashed line marks the detection time of the neutrino IceCube-170922A. The left set of panels shows measurements between MJD~54700 (22 August, 2008) and MJD~58002 (6 September, 2017). The set of panels on the right shows an expanded scale for time range MJD~58002 $-$ MJD~58050 (24 October, 2017). The \Fermi-LAT light curve is binned in 28~day bins on the left panel, while finer 7~day bins are used on the expanded panel. A VERITAS limit from MJD 58019.40 (23 September, 2017) of $2.1 \times 10^{-10}$~cm$^{-2}$~s$^{-1}$ is off the scale of the plot and not shown.}

\label{fig:CombinedLC}
\end{sidewaysfigure}

Figure~\ref{fig:CombinedLC} shows the \Fermi-LAT light curve and the detection time of the neutrino alert. 
The light curve of TXS~0506+056 from August 2008 to October 2017 was calculated in bins of 28 days for the energy range above 0.1~GeV. An additional light curve with 7-day bins was calculated for the period around the time of the neutrino alert. The \g-ray flux of TXS~0506+056 in each time bin was determined through a simultaneous fit of this source and the other \Fermi-LAT sources in a  10$^\circ$ by 10$^\circ$ region of interest along with the Galactic and isotropic diffuse backgrounds, 
using a maximum likelihood technique~\citesm{}. The integrated \g-ray flux of TXS~0506+056 for $E$~$>$~0.1~GeV, averaged over all \Fermi-LAT observations spanning 9.5 years, is $(7.6 \pm 0.2) \times 10^{-8}$ cm$^{-2}$ s$^{-1}$. The highest flux observed in a single 7-day light curve bin was $(5.3 \pm 0.6) \times 10^{-7}$ cm$^{-2}$ s$^{-1}$, measured in the week 4 to 11 July 2017. Strong flux variations were observed during the \g-ray flare, the most prominent being a flux increase from $(7.9 \pm 2.9) \times 10^{-8}$ cm$^{-2}$ s$^{-1}$ in the week 8 to 15 August 2017 to $(4.0 \pm 0.5) \times 10^{-7}$ cm$^{-2}$ s$^{-1}$ in the week 15 to 22 August 2017. 

The Astro-Rivelatore Gamma a Immagini Leggero ({\it AGILE}) \g-ray telescope~\cite{Tavani:2008sp} confirmed the elevated level of \g-ray emission at energies above 0.1~GeV from TXS~0506+056 in a 13-day window (10 to 23 September 2017). The {\it AGILE} measured flux of $(5.3 \pm 2.1) \times 10^{-7}$ cm$^{-2}$ s$^{-1}$ is consistent with the \Fermi-LAT observations in this time period. 

High-energy \g-ray observations are shown in Figures~\ref{fig:CombinedLC} and \ref{fig:txs0506_sed}. Details on the \Fermi-LAT and {\it AGILE} analyses can be found in \citesm.

\section*{Very-high-energy \g-ray observations of TXS~0506+056}
Following the announcement of IceCube-170922A, TXS~0506+056 was observed by several ground based Imaging Atmospheric Cherenkov Telescopes (IACTs). A total of 1.3 hours of observations in the direction of the blazar TXS~0506+056 were taken using the High-Energy Stereoscopic System (H.E.S.S.)~\cite{2006A&A...457..899A}, located in Namibia, on 23 September, 2017 (Modified Julian Date (MJD) 58019), $\sim$4 hours after the circulation of the neutrino alert. A 1-hour follow-up observation of the neutrino alert under partial cloud coverage was performed using the Very Energetic Radiation Imaging Telescope Array System (VERITAS) \g-ray telescope array~\cite{VTS}, located in Arizona, USA, later on the same day, $\sim$12 hours after the IceCube detection. Both telescopes made additional observations on subsequent nights but neither detected \g-ray emission from the source [see Figure~\ref{fig:CombinedLC} and \citesm]. Upper limits at 95\% C.L. on the \g-ray flux were derived accordingly (assuming the measured spectrum, see below): $7.5 \times 10^{-12}$ cm$^{-2}$ s$^{-1}$ during the H.E.S.S. observation period, and $1.2 \times 10^{-11}$ cm$^{-2}$ s$^{-1}$ during the VERITAS observations, both for energies $E>$175~GeV. 

The Major Atmospheric Gamma Imaging Cherenkov (MAGIC) Telescopes~\cite{Aleksic:2014lkm} observed  TXS~0506+056 for 2~hours on 24 September 2017 (MJD 58020) under non-optimal weather conditions and then for a period of 13 hours from 28 September to 4 October 2017 (MJD 58024--58030) under good conditions. MAGIC consists of two 17~m telescopes, located at the Roque de los Muchachos Observatory on the Canary Island of La Palma (Spain). 

No \g-ray emission from TXS~0506+056 was detected in the initial MAGIC observations on 24 September 2017, and an upper limit was derived on the flux above 90~GeV of $3.6 \times 10^{-11}$ cm$^{-2}$ s$^{-1}$at 95\% C.L. (assuming a spectrum $dN/dE \propto E^{-3.9}$). However, prompted by the \Fermi-LAT detection of enhanced \g-ray emission, MAGIC performed another 13 hours of observations of the region starting 28 September 2017. Integrating the data, MAGIC detected a significant very-high-energy (VHE) \g-ray signal~\cite{2017ATel10817....1M} corresponding to 
374 $\pm$ 62 excess photons, with observed energies up to about 400~GeV. This represents a 6.2 $\sigma$ excess over expected background levels~\citesm{}.
The day-by-day light curve of TXS~0506+056 for energies above 90~GeV is shown in Figure~\ref{fig:CombinedLC}.
The probability that a constant flux is consistent with the data is less than 1.35\%. The measured differential photon spectrum (Figure~\ref{fig:txs0506_sed}) can be described over the energy range of 80 to 400 GeV by a simple power law ${\rm d}N/{\rm d}E \propto E^{\gamma}$ with a spectral index \g = $-3.9\pm$0.4 and a flux normalization of (2.0 $\pm$ 0.4) $\times 10^{-10}$~ TeV$^{-1}$~cm$^{-2}$~s$^{-1}$ at $E$ = 130~GeV. Uncertainties are statistical only.  
The estimated  systematic uncertainties are $<$ 15\% in the energy scale, 11 to 18\% in the flux normalization and $\pm$0.15 for the power-law slope of the energy spectrum~\cite{Aleksic:2014lkm}. Further observations after 4 October 2017 were prevented by the full Moon.

An upper limit to the redshift of TXS~0506+056 can be inferred from VHE \g-ray observations using limits on the attenuation of the VHE flux due to interaction with the EBL. Details on the method are available in~\citesm{}. The obtained upper limit ranges from 0.61 to 0.98 at a 95\% C.L., depending on the EBL model used. These upper limits are consistent with the measured redshift of $z = 0.3365$~\cite{Paiano:2018qeq}. 

No \g-ray source above 1~TeV at the location of TXS~0506+056 was found in survey data of the High Altitude Water Cherenkov (HAWC) \g-ray observatory~\cite{CrabHAWC}, either close to the time of the neutrino alert, or in archival data taken since November 2014~\citesm{}. 

VHE \g-ray observations are shown in Figures~\ref{fig:CombinedLC} and \ref{fig:txs0506_sed}. All measurements are consistent with the observed flux from MAGIC, considering the differences in exposure, energy range and observation periods. 

\section*{Radio, optical and x-ray observations}
The Karl G. Jansky Very Large Array (VLA)~\cite{2011ApJ...739L...1P}  observed TXS~0506+056 starting 2 weeks after the alert in several radio bands from 2 to 12~GHz~\cite{2017ATel10861....1T}, detecting significant radio flux variability and some spectral variability of this source. The source is also in the long-term blazar monitoring program of the Owens Valley Radio Observatory (OVRO) 40-m telescope at 15~GHz~\cite{2011ApJS..194...29R}. The light curve shows a gradual increase in radio emission during the 18 months preceding the neutrino alert.   

Optical observations were performed by the All-Sky Automated Survey for Supernovae (ASAS-SN)~\cite{2017PASP..129j4502K}, the Liverpool telescope~\cite{2004SPIE.5489..679S}, the Kanata Telescope~\cite{2014SPIE.9147E..4OA}, the Kiso Schmidt Telescope~\cite{2012SPIE.8446E..6LS}, the high resolution spectrograph (HRS) on the Southern African Large Telescope (SALT)~\cite{2014SPIE.9147E..6TC}, the Subaru telescope Faint Object Camera and Spectrograph (FOCAS)~\cite{2002PASJ...54..819K} and the X-SHOOTER instrument on the Very Large Telescope (VLT)~\cite{2011A&A...536A.105V}. The {\it V} band flux of the source is the highest observed in recent years, and the spectral energy distribution has shifted towards blue wavelengths. Polarization was detected by Kanata in the {\it R} band at the level of~7\%. Redshift determination for BL Lac objects is difficult owing to the non-thermal continuum from the nucleus outshining the spectral lines from the host galaxies. Attempts were made using optical spectra from the Liverpool, Subaru and VLT telescopes to measure the redshift of TXS~0506+056, but only limits could be derived, [see, e.g., \cite{2017ATel10840....1C}]. The redshift of TXS0506+056 was later determined to be $z=0.3365 \pm 0.0010$ using the Gran Telescopio Canarias~\cite{Paiano:2018qeq}.

X-ray observations were made by the x-Ray Telescope (XRT) on the {\it Neil Gehrels Swift Observatory} (0.3 to 10~keV)~\cite{xrt2005}, {\it MAXI} Gas Slit Camera (GSC) (2 to 10~keV)~\cite{2009PASJ...61..999M}, Nuclear Spectroscopic Telescope Array (\nustar) (3 to 79~keV)~\cite{Harrison:2013md} and the INTErnational Gamma-Ray Astrophysics Laboratory ({\it INTEGRAL}) (20 to 250~keV)~\cite{Winkler:2003nn}, with detections by \swift\ and \nustar.  In a 2.1\,square degree region around the neutrino alert, \swift\ identified nine x-ray sources, including TXS~0506+056. 

\begin{figure}[htp!]
\begin{center}
\includegraphics[width=\linewidth]{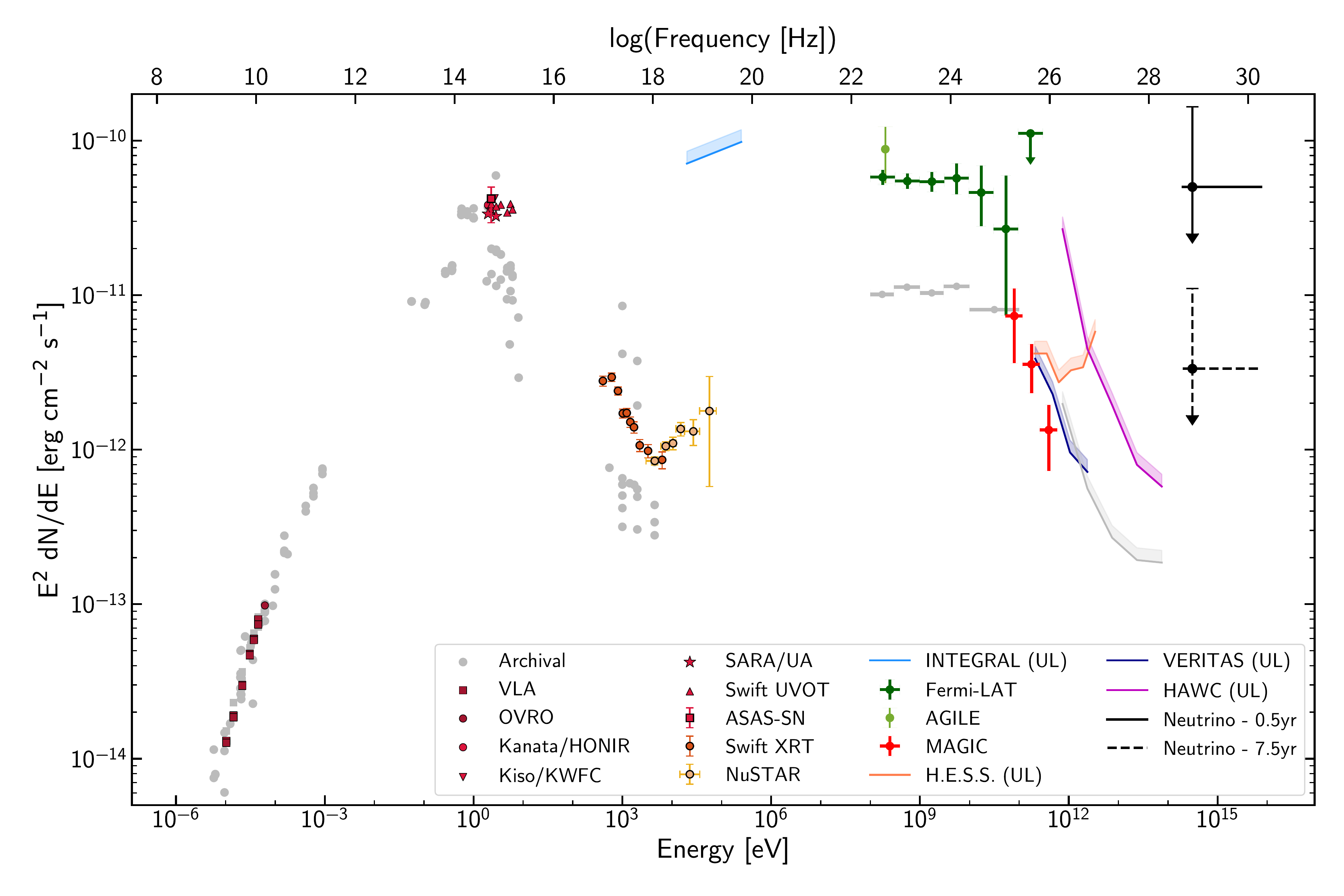}
\caption{{\bf Broadband spectral energy distribution for the blazar TXS~0506+056.} The SED is based on observations obtained within 14 days of the detection of the IceCube-170922A event.
The ${\rm E}^2 {\rm dN}/{\rm dE}$ vertical axis is equivalent to a $\nu {\rm F}_\nu$ scale. 
Contributions are provided by the following instruments: VLA~\cite{2017ATel10861....1T}, 
OVRO~\cite{2011ApJS..194...29R}, Kanata Hiroshima Optical and Near-InfraRed camera (HONIR)~\cite{2017ATel10844....1Y},
Kiso and the Kiso Wide Field Camera (KWFC)~\cite{2012SPIE.8446E..6LS},
Southeastern Association for Research in Astronomy Observatory (SARA/UA)~\cite{2017ATel10831....1K},
ASAS-SN~\cite{2017ATel10794....1F},
{\it Swift} Ultraviolet and Optical Telescope (UVOT) and XRT~\cite{2017ATel10792....1K}, 
{\it NuSTAR}~\cite{2017ATel10845....1F}, 
{\it INTEGRAL}~\cite{GCN21917},
{\it AGILE}~\cite{2017ATel10801....1L}, 
\Fermi-LAT~\cite{2017Atel_Fermi}, 
MAGIC~\cite{2017ATel10817....1M}, 
VERITAS~\cite{2017ATel10833....1M}, 
H.E.S.S.~\cite{2017ATel10787....1D} 
and HAWC~\cite{2017ATel10802....1M}. Specific observation dates and times are provided in \citesm. Differential flux upper limits (shown as colored bands and indicated as ``UL" in the legend) are quoted at the 95\% C.L. while markers indicate significant detections. Archival observations are shown in gray to illustrate the historical flux level of the blazar in the radio-to-keV range as retrieved from the ASDC SED Builder~\cite{2011arXiv1103.0749S}, and in the \g-ray band as listed in the \Fermi-LAT 3FGL catalog~\cite{Acero:2015hja} and from an analysis of 2.5 years
of HAWC data. The \g-ray observations have not been corrected for absorption owing to the EBL. SARA/UA, ASAS-SN, and Kiso/KWFC observations have not been corrected for Galactic attenuation. The electromagnetic SED displays a double-bump structure, one peaking in the optical-ultraviolet range and the second one in the GeV range, which is characteristic of the non-thermal emission from blazars. Even within this 14-day period, there is variability observed in several of the energy bands shown (see Figure~\ref{fig:CombinedLC}) and the data are not all obtained simultaneously. Representative $\nu_{\mu}+\overline{\nu}_{\mu}$ neutrino flux upper limits that produce on average one detection like IceCube-170922A over a period of 0.5 (solid black line) and 7.5 years (dashed black line) are shown assuming a spectrum of $dN/dE \propto E^{-2}$ at the most probable neutrino energy (311~TeV).} 
\label{fig:txs0506_sed}
\end{center}
\end{figure}

\swift\ monitored the x-ray flux from TXS~0506+056 for 4 weeks after the alert, starting 23 September 2017 00:09:16 UT, finding clear evidence for spectral variability (see Figure~\ref{fig:CombinedLC}D). The strong increase in flux observed at VHE energies over several days up until MJD 58030 (4 October 2017) correlates well with an increase in the x-ray emission in this period of time. 
The spectrum of TXS~0506+056 observed in the week after the flare is compatible with the sum of two power-law spectra, a soft spectrum with index $-2.8\pm 0.3$ in the soft x-ray band covered by \swift\ XRT and a hard spectrum with index $-1.4\pm 0.3$ in the hard x-ray band covered by \nustar~\citesm{}. Extrapolated to 20 MeV, the \nustar\ hard-spectrum component connects smoothly to the plateau (index $-2$) component observed by the \Fermi-LAT between 0.1 and 100~GeV and the soft VHE \g-ray component observed by MAGIC (compare Fig.~\ref{fig:txs0506_sed}). Taken together, these observations provide a mostly complete, contemporaneous picture of the source emissions from 0.3~keV to 400~GeV, more than nine orders of magnitude in photon energy.

Figures~\ref{fig:CombinedLC} and \ref{fig:txs0506_sed} summarize 
the multi-wavelength 
light curves and the changes in the broadband spectral energy distribution (SED), compared to archival observations. Additional details about the radio, optical, and x-ray observations can be found in \citesm. 
 
\section*{Chance coincidence probability}
Data obtained from multi-wavelength observations
of TXS~0506+056 can be used to constrain the blazar-neutrino chance coincidence probability. This coincidence probability is a measure of the likelihood that a neutrino alert like IceCube-170922A is correlated by chance with a flaring blazar, considering the large number of known \g-ray sources and the modest number of neutrino alerts.

Given the large number of potential neutrino source classes available, no a priori blazar-neutrino coincidence model was selected ahead of the alert.  After the observation, however,
several correlation scenarios were considered and tested to quantify the a posteriori significance of the observed coincidence. Testing multiple models is important as the specific assumptions about the correlation between neutrinos and \g~rays have an impact on the chance coincidence probability. In each case, the probability to obtain, by chance, a degree of correlation at least as high as that observed for IceCube-170922A was calculated using simulated neutrino alerts and the light curves of \Fermi-LAT \g-ray sources. 
Given the continuous all-sky monitoring of the \Fermi-LAT since 2008, all tests utilized 28-day binned \g-ray light curves above 1\,GeV from 2257 extragalactic \Fermi-LAT sources, derived in the same manner as used for the analysis of TXS~0506+056 \g-ray data.

To calculate the chance probabilities, a likelihood ratio test is used that allows different models of blazar-neutrino flux correlation to be evaluated in a consistent manner. All models assume that at least some of the observed \g-ray flux is produced in the same hadronic interactions that would produce high-energy neutrinos within the source. Our first model assumes that the neutrino flux is linearly correlated with the high-energy \g-ray energy flux~\cite{Aartsen:2016lir}. In this scenario, neutrinos are more likely to be produced during periods of bright, hard \g-ray emission. In the second model, the neutrino flux is modeled as strongly tied to variations in the observed \g-ray flux, regardless of the average flux of \g-rays.
Here, a weak or a strong \g-ray source is equally likely to be a neutrino source if the neutrino is temporally
correlated with variability in the \g-ray light curve.
Third, we consider a correlation of the neutrino flux with the VHE \g-ray flux. 
Because hadronic acceleration up to a few PeV is required to explain the detected neutrino energy, VHE \g-ray sources are potential progenitors.
Full details and results from these analyses are presented in \citesm.

The neutrino IceCube-170922A was found to arrive in a period of flaring activity in high-energy \g-rays.
Prior to IceCube-170922A, nine public alerts had been issued by the IceCube real-time system. Additionally, 41 archival events have been identified among the IceCube data recorded since 2010, before the start of the real-time program in April 2016, which would have caused alerts if the real-time alert system had been in place. These events were also tested for coincidence with the \g-ray data.  

Chance coincidence of the neutrino with the flare of TXS~0506+056 is disfavored at the 3$\sigma$ level in any scenario where neutrino production is linearly correlated with \g-ray production or with \g-ray flux variations. This includes look-elsewhere corrections for all 10 alerts issued previously by IceCube and the 41 archival events.
One of the neutrino events that would have been sent as an alert and had a good angular resolution ($<$5$^\circ$) is in a spatial correlation with the \g-ray blazar 3FGL~J1040.4+0615. However, this source was not in a particularly bright emission state at the detection time of the corresponding neutrino. Therefore, a substantially lower test statistic would be obtained in the chance correlation tests defined in this paper~\citesm{}. 

We have investigated how typical the blazar TXS~0506+056 is among those blazars that might have given rise to a coincident observation similar to the one reported here. 
A simulation that assumes that the neutrino flux is linearly correlated with the blazar \g-ray energy flux shows that in $14\%$ of the signal realizations we would find a neutrino coincident with a similarly bright \g-ray state as that observed for TXS~0506+056~\citesm{}. 
The detection of a single neutrino does not allow us to probe the details of neutrino production models or measure the neutrino-to-\g-ray production ratio.
Further observations will be needed to unambiguously establish a correlation between high-energy neutrinos and blazars, as well as to understand the emission and acceleration mechanism in the event of a correlation.

\section*{Discussion}
Blazars have often been suggested as potential sources of high-energy neutrinos. The calorimetric high-energy output of certain candidate blazars is high enough to explain individual observed IceCube events at 100~TeV to 1~PeV energies~\cite{2014A&A...566L...7K}. Spatial coincidences between catalogs of blazars and neutrinos have been examined in \cite{Padovani:2016wwn}, while \cite{Kadler:2016ygj} investigated one shower-like event with several thousand square degrees angular uncertainty observed in time coincidence with a blazar outburst. A track-like event, IceCube-160731, has been previously connected to a flaring \g-ray source~\cite{Lucarelli:2017hhh}. However, the limited evidence for a flaring source in the multi-wavelength coverage did not permit an identification of the source type of the potential counterpart~\cite{Lucarelli:2017hhh}.

Owing to the precise direction of IceCube-170922A, combined with extensive multi-wavelength observations, a chance correlation between a high-energy neutrino and the potential counterpart can be rejected at the 3$\sigma$ level. Considering the association between IceCube-170922A and TXS~0506+056, \g-ray blazars are strong candidate sources for at least a fraction of the observed astrophysical neutrinos. Earlier studies of the cross-correlation between IceCube events and the \g-ray blazar population observed by \Fermi-LAT demonstrated that these blazars can only produce a fraction of the observed astrophysical neutrino flux above 10~TeV~\cite{Aartsen:2016lir}.  
Although these limits constrain the contribution from blazars to the diffuse neutrino background, the potential association of one or two high-energy neutrinos to blazars over the total observing time of IceCube is fully compatible with the constraint.

Adopting cosmological parameters~\cite{Ade:2015xua} $H_{0} = 67.8$, $\Omega_{m}=0.308$, $\Omega_{\lambda} = 0.692$, where $H_{0}$ is the Hubble constant, $\Omega_{m}$ is the matter density and $\Omega_{\lambda}$ is the dark energy density, the observed redshift of $z = 0.3365$ implies an isotropic \g-ray luminosity between 0.1~GeV and 100~GeV of $ 1.3 \times 10^{47}$~erg s$^{-1}$ in the $\pm$2~weeks around the arrival time of the IceCube neutrino, and a luminosity of $ 2.8 \times 10^{46}$~erg s$^{-1}$, averaged over all \Fermi-LAT observations. 
Observations in the optical, x-ray, and VHE \g-ray bands show typical characteristics of blazar flares: strong variability on time scales of a few days, and an indication of a shift of the synchrotron emission peak towards higher frequencies. VHE \g-ray emission is found to change by a factor of $\sim 4$ within just 3 days. Similarly, the high-energy \g-ray energy band shows flux variations up to a factor of $\sim 5$ from 1 week to the next.

No other neutrino event that would have passed the selection criteria for a high-energy alert was observed from this source since the start of IceCube observations in May 2010. The muon neutrino fluence for which we would expect to detect one high-energy alert event with IceCube in this period of time is $2.8\times10^{-3}$~erg cm$^{-2}$. A power-law neutrino spectrum is assumed in this calculation with an index of $-2$ between 200~TeV and 7.5~PeV, the range between the 90\%~C.L. lower and upper limits for the energy of the observed neutrino (see~\citesm{} for details). 

The fluence can be expressed as an integrated energy flux if we assume a time period during which the source was emitting neutrinos. 
For a source that emits neutrinos only during the $\sim$ 6 month period corresponding to the duration of the high-energy \g-ray flare, the corresponding average integrated muon neutrino energy flux would be $1.8\times10^{-10}$~erg~cm$^{-2}$~s$^{-1}$. Alternatively, the average integrated energy flux of a source that emits neutrinos over the whole observation period of IceCube, {\it i.e.} 7.5 years, would be $1.2 \times 10^{-11}$~erg cm$^{-2}$ s$^{-1}$. These two benchmark cases are displayed in Figure~\ref{fig:txs0506_sed}. 
In an ensemble of faint sources with a summed expectation of order 1, we would anticipate observing a neutrino even if the individual expectation value is $\ll 1$. This is expressed by the downward arrows on the neutrino flux points in Figure~\ref{fig:txs0506_sed}.

The two cases discussed above correspond to average isotropic muon neutrino luminosities of $7.2 \times 10^{46}$~erg s$^{-1}$ for a source that was emitting neutrinos in the $\sim$ 6 month period of the high-energy \g-ray flare, and $4.8 \times 10^{45}$~erg s$^{-1}$ for a source that emitted neutrinos throughout the whole observation period. This is similar to the luminosity observed in \g-rays, and thus broadly consistent with hadronic source scenarios~\cite{Gaisser:1994yf}.

A neutrino flux that produces a high-energy alert event can, over time, produce many lower-energy neutrino-induced muons in IceCube. A study of neutrino emission from TXS~0506+056 prior to the high-energy \g-ray flare, based on the investigation of these lower-energy events, is reported in a companion paper~\cite{TXS0506_PS_INPREP}.

\noindent {\bf Supplementary Materials}\\
Supplementary Text\\
Table S1 — S10\\
Fig S1 — S7\\
References (69 — 116)


\bibliographystyle{Science}
\newpage
\subsection*{Acknowlegements}
{\bf MAGIC:}
We would like to thank the Instituto de Astrof\'{\i}sica de Canarias for the excellent working conditions at the Observatorio del Roque de los Muchachos in La Palma.
\\
{\bf \textit{AGILE}:}
We thank ASI personnel involved in the operations and data center of the AGILE mission.
\\
{\bf ASAS-SN:}
We thank Las Cumbres Observatory and its staff for their continued support of ASAS-SN. 
\\
{\bf VERITAS:}
We acknowledge the excellent work of the technical support staff at the Fred Lawrence Whipple Observatory and at the collaborating institutions in the construction and operation of the instrument. 

\subsubsection*{Funding}
{\bf IceCube Collaboration:}
The IceCube collaboration gratefully acknowledge the support from the following agencies and institutions: USA – U.S. National Science Foundation-Office of Polar Programs, U.S. National Science Foundation-Physics Division, Wisconsin Alumni Research Foundation, Center for High Throughput Computing (CHTC) at the University of Wisconsin–Madison, Open Science Grid (OSG), Extreme Science and Engineering Discovery Environment (XSEDE), U.S. Department of Energy–National Energy Research Scientific Computing Center, Particle astrophysics research computing center at the University of Maryland, Institute for Cyber-Enabled Research at Michigan State University, and Astroparticle physics computational facility at Marquette University; Belgium – Funds for Scientific Research (FRS-FNRS and FWO), FWO Odysseus and Big Science programmes, and Belgian Federal Science Policy Office (Belspo); Germany – Bundesministerium für Bildung und Forschung (BMBF), Deutsche Forschungsgemeinschaft (DFG), Helmholtz Alliance for Astroparticle Physics (HAP), Initiative and Networking Fund of the Helmholtz Association, Deutsches Elektronen Synchrotron (DESY), and High Performance Computing cluster of the RWTH Aachen; Sweden – Swedish Research Council, Swedish Polar Research Secretariat, Swedish National Infrastructure for Computing (SNIC), and Knut and Alice Wallenberg Foundation; Australia – Australian Research Council; Canada – Natural Sciences and Engineering Research Council of Canada, Calcul Québec, Compute Ontario, Canada Foundation for Innovation, WestGrid, and Compute Canada; Denmark – Villum Fonden, Danish National Research Foundation (DNRF); New Zealand – Marsden Fund; Japan - Japan Society for Promotion of Science (JSPS) and Institute for Global Prominent Research (IGPR) of Chiba University; Korea – National Research Foundation of Korea (NRF); Switzerland – Swiss National Science Foundation (SNSF).
\\
{\bf \Fermi-LAT collaboration:}
The \textit{Fermi}-LAT Collaboration acknowledges generous ongoing support
from a number of agencies and institutes that have supported both the
development and the operation of the LAT as well as scientific data analysis.
These include the National Aeronautics and Space Administration and the
Department of Energy in the United States, the Commissariat \`a l'Energie Atomique
and the Centre National de la Recherche Scientifique / Institut National de Physique
Nucl\'eaire et de Physique des Particules in France, the Agenzia Spaziale Italiana
and the Istituto Nazionale di Fisica Nucleare in Italy, the Ministry of Education,
Culture, Sports, Science and Technology (MEXT), High Energy Accelerator Research
Organization (KEK) and Japan Aerospace Exploration Agency (JAXA) in Japan, and
the K.~A.~Wallenberg Foundation, the Swedish Research Council and the
Swedish National Space Board in Sweden.
\\
Additional support for science analysis during the operations phase is gratefully
acknowledged from the Istituto Nazionale di Astrofisica in Italy and the Centre
National d'\'Etudes Spatiales in France. This work performed in part under DOE
Contract DE-AC02-76SF00515.
\\
{\bf MAGIC collaboration:}
The financial support of the German BMBF and MPG, the Italian INFN and INAF, the Swiss National Fund SNF, the ERDF under the Spanish MINECO (FPA2015-69818-P, FPA2012-36668, FPA2015-68378-P, FPA2015-69210-C6-2-R, FPA2015-69210-C6-4-R, FPA2015-69210-C6-6-R, AYA2015-71042-P, AYA2016-76012-C3-1-P, ESP2015-71662-C2-2-P, CSD2009-00064), and the Japanese JSPS and MEXT is gratefully acknowledged. This work was also supported by the Spanish Centro de Excelencia ``Severo Ochoa'' SEV-2012-0234 and SEV-2015-0548, and Unidad de Excelencia ``Mar\'{\i}a de Maeztu'' MDM-2014-0369, by the Croatian Science Foundation (HrZZ) Project IP-2016-06-9782 and the University of Rijeka Project 13.12.1.3.02, by the DFG Collaborative Research Centers SFB823/C4 and SFB876/C3, the Polish National Research Centre grant UMO-2016/22/M/ST9/00382 and by the Brazilian MCTIC, CNPq and FAPERJ.
\\
{\bf \textit{AGILE}:}
AGILE is an ASI space mission developed with scientific and programmatic support 
from INAF and INFN. Research partially supported through the ASI grant no. I/028/12/2.
\\
Part of this work is based on archival data, software or online services provided by 
the ASI Space Science Data Center (SSDC, previously known as ASDC).
\\
{\bf ASAS-SN:}
ASAS-SN is funded in part by the Gordon and Betty Moore Foundation through grant GBMF5490 to the Ohio State University, NSF grant AST-1515927, the Mt. Cuba Astronomical Foundation, the Center for Cosmology and AstroParticle Physics (CCAPP) at OSU, and the Chinese Academy of Sciences South America Center for Astronomy (CASSACA). A. F. was supported by the Initiative and Networking Fund of the Helmholtz Association. J. F. B is supported by NSF Grant PHY-1714479. S.D. acknowledges Project 11573003 supported by NSFC. J. L. P. is supported by FONDECYT grant 1151445 and by the Ministry of Economy, Development, and Tourism’s Millennium Science Initiative through grant IC120009, awarded to The Millennium Institute of Astrophysics, MAS. This research was made possible through the use of the AAVSO Photometric All-Sky Survey (APASS), funded by the Robert Martin Ayers Sciences Fund.
\\
{\bf HAWC:}
HAWC acknowledges the support from: the US National Science Foundation (NSF) the US Department of Energy Office of High-Energy Physics; 
the Laboratory Directed Research and Development (LDRD) program of Los Alamos National Laboratory; 
Consejo Nacional de Ciencia y Tecnolog\'{\i}a (CONACyT), M{\'e}xico (grants 271051, 232656, 260378, 179588, 239762, 254964, 271737, 258865, 243290, 132197, 281653)(C{\'a}tedras 873, 1563), Laboratorio Nacional HAWC de rayos gamma; 
L'OREAL Fellowship for Women in Science 2014; 
Red HAWC, M{\'e}xico; 
DGAPA-UNAM (grants IG100317, IN111315, IN111716-3, IA102715, 109916, IA102917, IN112218); 
VIEP-BUAP; 
PIFI 2012, 2013, PROFOCIE 2014, 2015; 
the University of Wisconsin Alumni Research Foundation; 
the Institute of Geophysics, Planetary Physics, and Signatures at Los Alamos National Laboratory; 
Polish Science Centre grant DEC-2014/13/B/ST9/945; 
Coordinaci{\'o}n de la Investigaci{\'o}n Cient\'{\i}fica de la Universidad Michoacana. Thanks to Scott Delay, Luciano D\'{\i}az and Eduardo Murrieta for technical support.
\\
{\bf H.E.S.S.:}
The support of the Namibian authorities and of the University of Namibia in facilitating the construction and operation of H.E.S.S. is gratefully acknowledged, as is the support by the German Ministry for Education and Research (BMBF), the Max Planck Society, the German Research Foundation (DFG), the Helmholtz Association, the Alexander von Humboldt Foundation, the French Ministry of Higher Education, Research and Innovation, the Centre National de la Recherche Scientifique (CNRS/IN2P3 and CNRS/INSU), the Commissariat à l’\'energie atomique et aux énergies alternatives (CEA), the U.K. Science and Technology Facilities Council (STFC), the Knut and Alice Wallenberg Foundation, the National Science Centre, Poland grant no. 2016/22/M/ST9/00382, the South African Department of Science and Technology and National Research Foundation, the University of Namibia, the National Commission on Research, Science \& Technology of Namibia (NCRST), the Austrian Federal Ministry of Education, Science and Research and the Austrian Science Fund (FWF), the Australian Research Council (ARC), the Japan Society for the Promotion of Science and by the University of Amsterdam. We appreciate the excellent work of the technical support staff in Berlin, Zeuthen, Heidelberg, Palaiseau, Paris, Saclay, Tübingen and in Namibia in the construction and operation of the equipment. This work benefited from services provided by the H.E.S.S. Virtual Organisation, supported by the national resource providers of the EGI Federation. 
\\
{\bf \textit{INTEGRAL:}}
INTEGRAL is an ESA space mission, with its instruments and science data center funded by the ESA member states (specifically the PI countries: Denmark, France, Germany, Italy, Switzerland, Spain), and with additional participation of Russia and the USA. The INTEGRAL SPI instrument was provided through Co-PI institutes IRAP (Toulouse/France) and MPE (Garching/Germany), the SPI project was coordinated and managed by CNES (Toulouse/France). The INTEGRAL IBIS instrument was provided through Co-PI institutes IAPS (Rome/Italy) and CEA (Saclay/France). The SPI-ACS detector system has been provided by MPE Garching/Germany. The SPI team is grateful to ASI, CEA, CNES, DLR, ESA, INTA, NASA and OSTC for their support. The Italian INTEGRAL team acknowledges the support of ASI/INAF agreement n. 2013-025-R.1. JR acknowledges support from the European Union’s Horizon 2020 Programme under the AHEAD project (grant n. 654215). RD acknowledges the German INTEGRAL support through DLR grants 50 OG 1101 and 1601.
\\
{\bf Kanata, Kiso and Subaru observing teams:}
Observations with the Kanata and Kiso Schmidt telescopes were supported by the Optical and Near-infrared Astronomy Inter-University Cooperation Program and the Grants-in-Aid of the Ministry of Education, Science, Culture, and Sport JP23740143, JP25800103, JP16H02158, JP17K14253, JP17H04830, JP26800103, JP24103003. This work was also based in part on data collected at Subaru Telescope, which is operated by the National Astronomical Observatory of Japan.
\\
{\bf Kapteyn:}
The Jacobus Kapteyn telescope is operated at the Observatorio del Roque de los Muchachos on the Spanish island of La Palma by the SARA
consortium, whose member institutions (listed at http://saraobservatory.org) fund its operation. Refitting for remote operations and
instrumentation were funded by the National Science Foundation under grant 1337566 to Texas A\&M University—Commerce. 
\\
{\bf Liverpool telescope:}
The Liverpool Telescope is operated on the island of La Palma by Liverpool John Moores University in the Spanish Observatorio del Roque de los Muchachos of the Instituto de Astrofisica de Canarias with financial support from the UK Science and Technology Facilities Council.
\\
{\bf \textit{Swift}$/$\nustar:}
AK and and DFC acknowledge support from the National Aeronautics and Space Administration Swift Guest Investigator Program under grant NNX17AI95G.
The Swift team at the Mission Operations Center (MOC) at Penn State acknowledges support from NASA contract NAS5-00136. Swift is supported at the University of Leicester by the UK Space Agency. 
\\
{\bf VERITAS:}
This research is supported by grants from the U.S. Department of Energy Office of Science, the U.S. National Science Foundation and the Smithsonian Institution, and by NSERC in Canada. 
\\
{\bf VLA/17B-403}
The National Radio Astronomy Observatory (NRAO) is a facility of the National Science Foundation operated under cooperative agreement by Associated Universities, Inc. AJT is supported by a Natural Sciences and Engineering Research Council of Canada (NSERC) Post-Graduate Doctoral Scholarship (PGSD2-490318-2016). AJT and GRS are supported by NSERC Discovery Grants (RGPIN-402752-2011 and RGPIN-06569-2016). JCAMJ is the recipient of an Australian Research Council Future Fellowship (FT140101082). 

\subsubsection*{Author contributions}
{\bf IceCube:}
The IceCube Collaboration designed, constructed and now operates the the IceCube Neutrino Observatory. Data processing and calibration, Monte Carlo simulations of the detector and of theoretical models, and data analyses were performed by a large number of collaboration members, who also discussed and approved the scientific results presented here. It was reviewed by the entire collaboration before publication, and all authors approved the final version of the manuscript.
\\
{\bf \Fermi-LAT:}
The Fermi-LAT contact authors and internal reviewers are SB, AF, YT, KB, EC, and MW. 
\\
{\bf MAGIC:}
EB is the MAGIC multi-messenger contact and PI of the neutrino follow-up program. KS is co-convener of the MAGIC transient working group.
LF and MP are the main analyzers of the MAGIC data presented here. AM and EP derived an upper limit to the redshift inferred from MAGIC data. 
\\
{\bf \textit{AGILE}:}
All co-authors contributed to the scientific results presented in the paper. FL and MT wrote the part of the manuscript related to the \textit{AGILE} results. 
\\
{\bf ASAS-SN:}
AF, BJS and SH installed an automatic follow up to IceCube triggers which provided additional early data on this event. KZS, CSK, JFB, TAT, TWSH, SD, JLP and BJS built the telescopes and developed the data processing pipelines.
\\
{\bf HAWC:}
TW is convener of the HAWC extragalactic working group. MH is the HAWC multi-messenger contact. IT, RL and IMC are the main analysers of the HAWC data presented here. 
\\
{\bf H.E.S.S.:}
AT is convener of the H.E.S.S. extragalactic working group. FS is the H.E.S.S. multi-messenger contact and PI of the neutrino follow-up program. CH is the main analyser of the H.E.S.S. data presented here. SO provided a cross-check of the presented analysis.
\\
{\bf \textit{INTEGRAL:}}
VS has performed the INTEGRAL analysis presented here. CF is the PI of INTEGRAL Science Data Center. RD is the co-PI of the SPI instrument. EK is INTEGRAL Project Scientist. PL and PU are co-PIs of INTEGRAL/IBIS instrument. SM is responsible for the INTEGRAL/IBAS. All of the collaborators provided contribution to the text. 
\\
{\bf Kanata, Kiso and Subaru observing teams:}
YTT and YU developed the follow-up strategy to search for IceCube counterparts. TN and MK conducted the near-infrared imaging and polarimetric observations of the TXS~0506+056 using the HONIR instrument on the Kanata telescope, which were processed by the data reduction system developed by RI. The reduced images were mainly examined by HM and HN. MY reduced the polarimetric data. KSK supervised all of the above.
TM conducted the optical imaging observations of TXS~0506+056 with the KWFC instrument on the Kiso Schmidt telescope. He also reduced the data.
These two observations contribute to Figures 3 and 4.
YM conducted optical spectroscopic observations of the TXS~0506+056 with the FOCAS spectrograph on the 8.2~m Subaru telescope. The data are reduced and examined by MY, TM, and KO. The spectra are shown in Figure S5.
\\
{\bf Kapteyn:}
WCK obtained and reduced the optical observations at the Kapteyn telescope.
\\
{\bf Liverpool telescope:} 
IS and CC obtained, reduced and analysed the Liverpool Telescpe spectra presented in this paper.
\\
{\bf \textit{Swift}$/$\nustar:}
AK led reduction of Swift XRT data, and JJD led reduction of NuSTAR data. DBF carried out the joint Swift XRT + NuSTAR analysis, and AK managed author contributions to this section. 
\\
{\bf VERITAS:}
The construction, operation, and maintenance of the VERITAS telescopes, as well as the tools to analyze the VERITAS data, are the work of the the VERITAS Collaboration as a whole. The VERITAS Collaboration contacts for this paper are MS and DAW. 
\\
{\bf VLA/17B-403:}
GRS wrote the Director's Discretionary Time proposal whose VLA data are used in this paper (VLA/17B-403). AJT performed the VLA data reduction and analyses in consultation with the rest of the team. AJT and GRS wrote the VLA-related text in consultation with the rest of the team. GRS contributed to significant copy editing of the entire paper.

\subsubsection*{Competing interests}
{\bf IceCube:}
There are no competing interests to declare.
\\
{\bf \Fermi-LAT:}
There are no competing interests to declare.
\\
{\bf MAGIC:}
There are no competing interests to declare.
\\
{\bf \textit{AGILE}:}
There are no competing interests to declare.
\\
{\bf ASAS-SN:}
There are no competing interests to declare.
\\
{\bf HAWC:}
There are no competing interests to declare.
\\
{\bf H.E.S.S.:}
There are no competing interests to declare.
\\
{\bf \textit{INTEGRAL:}}
There are no competing interests to declare.
\\
{\bf Liverpool Telescope:}   
There are no competing interests to declare.
\\
{\bf Kanata, Kiso and Subaru observing teams:}
There are no competing interests to declare.
\\
{\bf Kapteyn:}
There are no competing interests to declare.
\\
{\bf \textit{Swift}$/$\nustar:}
There are no competing interests to declare.
\\
{\bf VERITAS:}
There are no competing interests to declare.
\\
{\bf VLA/17B-403:}
There are no competing interests to declare.

\subsubsection*{Data and materials availability}
{\bf IceCube:}
All IceCube data related to the the results presented in this paper are provided in \cite{supplementary_material}.
\\
{\bf \Fermi-LAT:}
Fermi-LAT data are available from the Fermi Science Support Center \\
(http://fermi.gsfc.nasa.gov/ssc) and 
https://www-glast.stanford.edu/pub\_data/1483/.
\\
{\bf MAGIC:}
The MAGIC collaboration routinely releases published graphics, tables and analysis results on a database linked from the official web pages. For this paper we will additionally release underlying data in the form of a list of the MAGIC observed events, after reconstruction and analysis cuts, and simplified instrument response functions. 
\\
Both resources (published analysis results as well as data releases for specific publications, like in this case) are accessible at the MAGIC public data website: \\ https://magic.mpp.mpg.de/public/public-data/ 
\\
We note that the format we choose for our data release, aims to make MAGIC data accessible to the wider community, without requiring the use of internal proprietary software.
\\
To exactly reproduce the published results, advanced software programs are necessary together with training on how to use them. Access to all data, analysis software and necessary training will be made available on request.
\\
{\bf \textit{AGILE}:}
Data and materials availability: All the \textit{AGILE} data used in this paper are public and available from the \textit{AGILE} Data Center webpage at the following URL: http://www.ssdc.asi.it \\
/mmia/index.php?mission=agilemmia. The \textit{AGILE} data analysis software and calibrations are also public and available at the URL: http://agile.ssdc.asi.it/publicsoftware.html.
\\
{\bf ASAS-SN:}
ASAS-SN light curves are available through the ASAS-SN sky patrol https://asas-sn.osu.edu
\\
{\bf HAWC:}
Data used for the presented results are available at: \\
https://data.hawc-observatory.org/datasets/ic170922/index.php.
\\
{\bf H.E.S.S.:}
Data used for the presented results are available at https://www.mpi-hd.mpg.de/hfm \\
/HESS/pages/publications/auxiliary/auxinfo\_TXS0506.html
\\
{\bf \textit{INTEGRAL:}}
Data and software used for the presented results are available at \\
https://www.isdc.unige.ch/.
\\
{\bf Kanata, Kiso and Subaru observing teams:}
All the raw data taken with the Kiso, Kanata, and Subaru telescopes are available in the SMOKA, which is operated by the Astronomy Data Center, National Astronomical Observatory of Japan. The Subaru data were taken in the open-use program S16B-071I.
\\
{\bf Kapteyn:}
The numerical magnitude/flux data to produced the published results here are available in at http://www.astronomerstelegram.org/?read=10831
\\
{\bf Liverpool Telescope:}    
Data used for the presented results are available in the telescope archive at http://telescope.livjm.ac.uk.
\\
{\bf \textit{Swift}$/$\nustar:}
Link to the archived Swift data: \\
http://www.swift.ac.uk/archive/obs.php \\
The initial tiling observations (shortly after the IceCube trigger) targetIDs: 10308-10326.
Monitoring ObsIDs used for this analysis (from Sept 23 to Oct 23): 00010308001, 00083368001-006, 00010308008-013.
\\
\noindent Link to the archived NuSTAR data: \\
https://heasarc.gsfc.nasa.gov/FTP/nustar/data/obs/03/9//90301618002/ \\
ObsID used for this analysis: 90301618002.
\\
{\bf VERITAS:}
Data used for the presented results is available at
https://veritas.sao.arizona.edu/veritas-science/veritas-results-mainmenu-72/490-ic-result
\\
{\bf VLA/17B-403 team:}
The VLA/17B-403 team would like to thank the NRAO for granting us DDT VLA time to observe this source and the NRAO staff for rapidly executing our observations. The original data from the VLA observations presented in this paper are publicly available through the NRAO Science Data Archive (https://archive.nrao.edu/archive/advquery.jsp, using the Project Code 17B-403).

\newpage
\renewcommand\thetable{S\arabic{table}}
\renewcommand\thefigure{S\arabic{figure}}
\renewcommand\theequation{S\arabic{equation}}
\renewcommand\thepage{S\arabic{page}}
\setcounter{table}{0}
\setcounter{figure}{0}
\setcounter{equation}{0}
\setcounter{page}{1}

\begin{center}

{\Large
Supplementary Materials for:\\
Multi-messenger observations of a flaring blazar\\
coincident with high-energy neutrino IceCube-170922A}
\end{center}

\subsection*{IceCube Collaboration$^\ddagger$:}
M.~G.~Aartsen$^{16}$,
M.~Ackermann$^{52}$,
J.~Adams$^{16}$,
J.~A.~Aguilar$^{12}$,
M.~Ahlers$^{20}$,
M.~Ahrens$^{44}$,
I.~Al~Samarai$^{25}$,
D.~Altmann$^{24}$,
K.~Andeen$^{34}$
% ...
\\ 
% ...
$^{12}$ Universit\'e Libre de Bruxelles, Science Faculty CP230, B-1050 Brussels, Belgium \\
% ...
$^{16}$ Dept.~of Physics and Astronomy, University of Canterbury, Private Bag 4800, Christchurch, New Zealand \\
% ...
$^{20}$ Niels Bohr Institute, University of Copenhagen, DK-2100 Copenhagen, Denmark \\
% ...
$^{24}$ Erlangen Centre for Astroparticle Physics, Friedrich-Alexander-Universit\"at Erlangen-N\"urnberg, D-91058 Erlangen, Germany \\
$^{25}$ D\'epartement de physique nucl\'eaire et corpusculaire, Universit\'e de Gen\`eve, CH-1211 Gen\`eve, Switzerland \\
% ...
$^{34}$ Department of Physics, Marquette University, Milwaukee, WI, 53201, USA \\
% ...
$^{44}$ Oskar Klein Centre and Dept.~of Physics, Stockholm University, SE-10691 Stockholm, Sweden \\
% ...
$^{52}$ DESY, D-15738 Zeuthen, Germany \\
% ...
$^\ddagger$E-mail: analysis@icecube.wisc.edu

\subsection*{\Fermi-LAT collaboration:}
S.~Abdollahi$^{1}$, 
M.~Ajello$^{2}$, 
R.~Angioni$^{3}$, 
L.~Baldini$^{4}$, 
J.~Ballet$^{5}$, 
G.~Barbiellini$^{6,7}$, 
D.~Bastieri$^{8,9}$, 
K.~Bechtol$^{10}$, 
R.~Bellazzini$^{11}$, 
B.~Berenji$^{12}$
% ...
\\
$^{1}$ Department of Physical Sciences, Hiroshima University, Higashi-Hiroshima, Hiroshima 739-8526, Japan \\
$^{2}$ Department of Physics and Astronomy, Clemson University, Kinard Lab of Physics, Clemson, SC 29634-0978, USA \\
$^{3}$ Max-Planck-Institut f\"ur Radioastronomie, Auf dem H\"ugel 69, D-53121 Bonn, Germany \\
$^{4}$ Universit\`a di Pisa and Istituto Nazionale di Fisica Nucleare, Sezione di Pisa I-56127 Pisa, Italy \\
$^{5}$ Laboratoire AIM, CEA-IRFU/CNRS/Universit\'e Paris Diderot, Service d'Astrophysique, CEA Saclay, F-91191 Gif sur Yvette, France \\
$^{6}$ Istituto Nazionale di Fisica Nucleare, Sezione di Trieste, I-34127 Trieste, Italy \\
$^{7}$ Dipartimento di Fisica, Universit\`a di Trieste, I-34127 Trieste, Italy \\
$^{8}$ Istituto Nazionale di Fisica Nucleare, Sezione di Padova, I-35131 Padova, Italy \\
$^{9}$ Dipartimento di Fisica e Astronomia ``G. Galilei'', Universit\`a di Padova, I-35131 Padova, Italy \\
$^{10}$ Large Synoptic Survey Telescope, 933 North Cherry Avenue, Tucson, AZ 85721, USA \\
$^{11}$ Istituto Nazionale di Fisica Nucleare, Sezione di Pisa, I-56127 Pisa, Italy \\
$^{12}$ California State University, Los Angeles, Department of Physics and Astronomy, Los Angeles, CA 90032, USA \\
% ...

\subsection*{MAGIC collaboration$^{\|}$:}
M.~L.~Ahnen$^{1}$, 
S.~Ansoldi$^{2,20}$, 
L.~A.~Antonelli$^{3}$, 
C.~Arcaro$^{4}$, 
D.~Baack$^{5}$, 
A.~Babi\'c$^{6}$, 
B.~Banerjee$^{7}$, 
P.~Bangale$^{8}$, 
U.~Barres~de~Almeida$^{8,\: 9}$, 
J.~A.~Barrio$^{10}$
% ...
\\
$^{1}$ ETH Zurich, CH-8093 Zurich, Switzerland \\
$^{2}$ Universit\`a di Udine, and INFN Trieste, I-33100 Udine, Italy \\
$^{3}$ National Institute for Astrophysics (INAF), I-00136 Rome, Italy \\
$^{4}$ Universit\`a di Padova and INFN, I-35131 Padova, Italy \\
$^{5}$ Technische Universit\"at Dortmund, D-44221 Dortmund, Germany \\
$^{6}$ Croatian MAGIC Consortium: University of Rijeka, 51000 Rijeka, University of Split - FESB, 21000 Split, University of Zagreb - FER, 10000 Zagreb, University of Osijek, 31000 Osijek and Rudjer Boskovic Ins
titute, 10000 Zagreb, Croatia \\
$^{7}$ Saha Institute of Nuclear Physics, HBNI, 1/AF Bidhannagar, Salt Lake, Sector-1, Kolkata 700064, India \\
$^{8}$ Max-Planck-Institut f\"ur Physik, D-80805 M\"unchen, Germany \\
$^{9}$ now at Centro Brasileiro de Pesquisas F\'isicas (CBPF), 22290-180 URCA, Rio de Janeiro (RJ), Brasil \\
$^{10}$ Unidad de Part\'iculas y Cosmolog\'ia (UPARCOS), Universidad Complutense, E-28040 Madrid, Spain \\
% ...
$^{20}$ Japanese MAGIC Consortium: ICRR, The University of Tokyo, 277-8582 Chiba, Japan; Department of Physics, Kyoto University, 606-8502 Kyoto, Japan; Tokai University, 259-1292 Kanagawa, Japan; The University
 of Tokushima, 770-8502 Tokushima, Japan \\
% ...

\subsection*{\textit{\textbf AGILE}:}
F.~Lucarelli$^{1,\: 2}$,
M.~Tavani$^{3,\: 4,\: 5}$,
G.~Piano$^{3}$,
I.~Donnarumma$^{6}$,
C.~Pittori$^{1,\: 2}$,
F.~Verrecchia$^{1,\: 2}$,
G.~Barbiellini$^{7}$,
A.~Bulgarelli$^{8}$,
P.~Caraveo$^{9}$,
P.~W.~Cattaneo$^{10}$,
S.~Colafrancesco$^{11,\: 2}$,
E.~Costa$^{3,\: 6}$,
G.~Di~Cocco$^{8}$,
A.~Ferrari$^{12}$,
F.~Gianotti$^{8}$,
A.~Giuliani$^{9}$,
P.~Lipari$^{13}$,
S.~Mereghetti$^{9}$,
A.~Morselli$^{14}$,
L.~Pacciani$^{3}$,
F.~Paoletti$^{15,\: 3}$,
N.~Parmiggiani$^{8}$,
A.~Pellizzoni$^{16}$,
P.~Picozza$^{17}$,
M.~Pilia$^{16}$,
A.~Rappoldi$^{10}$,
A.~Trois$^{16}$,
S.~Vercellone$^{18}$,
V. Vittorini$^{3}$,
\\
$^{1}$ ASI Space Science Data Center (SSDC), Via del Politecnico snc, I-00133 Roma, Italy \\
$^{2}$ INAF--OAR, via Frascati 33, I-00078 Monte Porzio Catone (Roma), Italy \\
$^{3}$ INAF/IAPS--Roma, Via del Fosso del Cavaliere 100, I-00133 Roma, Italy \\
$^{4}$ Univ. ``Tor Vergata'', Via della Ricerca Scientifica 1, I-00133 Roma, Italy \\
$^{5}$ Gran Sasso Science Institute, viale Francesco Crispi 7, I-67100 L`Aquila, Italy \\
$^{6}$ Agenzia Spaziale Italiana (ASI), Via del Politecnico snc, I-00133 Roma, Italy \\
$^{7}$ Dipartimento di Fisica, UniversitaÌ di Trieste and INFN, via Valerio 2, I-34127 Trieste, Italy \\
$^{8}$ INAF/IASF--Bologna, Via Gobetti 101, I-40129 Bologna, Italy \\
$^{9}$ INAF/IASF--Milano, via E.Bassini 15, I-20133 Milano, Italy \\
$^{10}$ INFN--Pavia, Via Bassi 6, I-27100 Pavia, Italy \\
$^{11}$ University of Witwatersrand, Johannesburg, South Africa \\
$^{12}$ CIFS, c/o Physics Department, University of Turin, via P. Giuria 1, I-10125 Torino, Italy \\
$^{13}$ INFN--Roma Sapienza, Piazzale Aldo Moro 2, 00185 Roma, Italy \\
$^{14}$ INFN--Roma Tor Vergata, via della Ricerca Scientifica 1, 00133 Roma, Italy \\
$^{15}$ East Windsor RSD, 25a Leshin Lane, Hightstown, NJ 08520, USA \\
$^{16}$ INAF -- Osservatorio Astronomico di Cagliari, via della Scienza 5, I-09047 Selargius (CA), Italy \\
$^{17}$ INFN--Roma Tor Vergata, via della Ricerca Scientifica 1, I-00133 Roma, Italy \\
$^{18}$ INAF -- Oss. Astron. di Brera, Via E. Bianchi 46, I-23807 Merate (LC), Italy \\

\subsection*{ASAS-SN:}
A.~Franckowiak$^{1}$,
K.~Z.~Stanek$^{2}$,
C.~S.~Kochanek$^{2, \: 3}$,
J.~F.~Beacom$^{3, \: 4, \: 2}$,
T.~A.~Thompson$^{2}$,
T. ~W.-S.~Holoien$^{5}$,
S.~Dong$^{6}$,
J. L. Prieto$^{7, \: 8}$,
B. J. Shappee$^{9}$,
S. Holmbo$^{10}$,
\\
$^{1}$ DESY, D-15738 Zeuthen, Germany \\
$^{2}$ Department of Astronomy, The Ohio State University, 140 West 18th Avenue Columbus, OH 43210 USA \\
$^{3}$ Center for Cosmology and Astroparticle Physics, The Ohio State University, 191 W. Woodruff Avenue, Columbus, OH 43210, USA \\
$^{4}$ Department of Physics, The Ohio State University, 191 W. Woodruff Avenue, Columbus, OH 43210 USA \\
$^{5}$ The Observatories of the Carnegie Institution for Science, 813 Santa Barbara St., Pasadena, CA 91101, USA \\ 
$^{6}$ Kavli Institute for Astronomy and Astrophysics, Peking University, 5 Yiheyuanlu, Haidian District
Beijing, China 100871 \\
$^{7}$ N\'ucleo de Astronom\'ia de la Facultad de Ingenier\'ia y Ciencias, Universidad Diego Portales, Av. Ej\'ercito 441, Santiago, Chile \\
$^{8}$ Millennium Institute of Astrophysics, Santiago, Chile \\
$^{9}$ Institute for Astronomy, University of Hawai'i, 2680 Woodlawn Drive, Honolulu, HI 96822, USA \\
$^{10}$ Department of Physics and Astronomy, Aarhus University, Ny Munkegade 120, 8000 Aarhus C, Denmark \\

\subsection*{HAWC:}
A.~U.~Abeysekara$^{1}$,
A.~Albert$^{2}$,
R.~Alfaro$^{3}$,
C.~Alvarez$^{4}$,
R.~Arceo$^{4}$,
J.~C.~Arteaga-Velázquez$^{5}$,
D.~Avila~Rojas$^{3}$,
H.~A.~Ayala~Solares$^{6}$,
A.~Becerril$^{3}$,
E.~Belmont-Moreno$^{3}$
% ...
\\
$^{1}$ Department of Physics and Astronomy, University of Utah, Salt Lake City, UT, USA \\
$^{2}$ Physics Division, Los Alamos National Laboratory, Los Alamos, NM, USA  \\
$^{3}$ Instituto de F\'{i}sica, Universidad Nacional Autónoma de México, Ciudad de Mexico, Mexico \\
$^{4}$ Universidad Autónoma de Chiapas, Tuxtla Gutiérrez, Chiapas, Mexico \\
$^{5}$ Universidad Michoacana de San Nicolás de Hidalgo, Morelia, Mexico \\
$^{6}$ Department of Physics, Pennsylvania State University, University Park, PA, USA  \\
% ...

\subsection*{H.E.S.S.$^{\dagger\dagger}$:}
H.~Abdalla$^{1}$,
E.O.~Ang\"uner$^{2}$,
C.~Armand$^{3}$,
M.~Backes$^{15}$,
Y.~Becherini$^{4}$,
D.~Berge$^{5}$, 
M.~B\"ottcher$^{1}$,
C.~Boisson$^{6}$,
J.~Bolmont$^{7}$
% ...
\\
$^{1}$ Centre for Space Research, North-West University, Potchefstroom 2520, South Africa \\
$^{2}$ Aix Marseille Universit\'e, CNRS/IN2P3, CPPM, Marseille, France \\
$^{3}$ Laboratoire d'Annecy de Physique des Particules, Univ. Grenoble Alpes, Univ. Savoie Mont Blanc, CNRS, LAPP, 74000 Annecy, France \\
$^{4}$ Department of Physics and Electrical Engineering, Linnaeus University,  351 95 V\"axj\"o, Sweden \\
$^{5}$ DESY, D-15738 Zeuthen, Germany \\
$^{6}$ LUTH, Observatoire de Paris, PSL Research University, CNRS, Universit\'e Paris Diderot, 5 Place Jules Janssen, 92190 Meudon, France \\
$^{7}$ Sorbonne Universit\'e, Universit\'e Paris Diderot, Sorbonne Paris Cit\'e, CNRS/IN2P3, Laboratoire de Physique Nucl\'eaire et de Hautes Energies, LPNHE, 4 Place Jussieu, F-75252 Paris, France \\
% ...
$^{15}$ University of Namibia, Department of Physics, Private Bag 13301, Windhoek, Namibia \\
% ...
$^{\dagger\dagger}$E-mail: contact.hess@hess-experiment.eu \\

\subsection*{\textit{\textbf INTEGRAL}$^{\ddagger\ddagger}$:}
V.~Savchenko$^{1}$, 
C.~Ferrigno$^{1}$, 
A.~Bazzano$^{2}$, 
R.~Diehl$^{3}$, 
E.~Kuulkers$^{4}$, 
P.~Laurent$^{5,6}$, 
S.~Mereghetti$^{7}$, 
L.~Natalucci$^{2}$, 
F.~Panessa$^{2}$, 
J.~Rodi$^{2}$, 
P.~Ubertini$^{2}$
\\
 $^{1}$ISDC, Department of astronomy, University of Geneva, chemin d'\'Ecogia, 16 CH-1290 Versoix, Switzerland \\
 $^{2}$INAF-Institute for Space Astrophysics and Planetology, Via Fosso del Cavaliere 100, 00133-Rome, Italy \\
 $^{3}$Max-Planck-Institut f\"{u}r Extraterrestrische Physik, Garching, Germany \\
 $^{4}$European Space Research and Technology Centre (ESA/ESTEC), Keplerlaan 1, 2201 AZ Noordwijk, The Netherlands \\
 $^{5}$APC, AstroParticule et Cosmologie, Universit\'e Paris Diderot, CNRS/IN2P3, CEA/Irfu, Observatoire de Paris Sorbonne Paris Cit\'e, 10 rue Alice Domont et L\'eonie Duquet, 75205 Paris Cedex 13, France. \\
$^{6}$DSM/Irfu/Service d'Astrophysique, Bat. 709 Orme des Merisiers CEA Saclay, 91191 Gif-sur-Yvette Cedex, France \\
$^{7}$INAF, IASF-Milano, via E.Bassini 15, I-20133 Milano, Italy \\ 
$^{\ddagger\ddagger}$E-mail: Volodymyr.Savchenko@unige.ch \\

\subsection*{Kanata, Kiso and Subaru observing teams$^{\|\|}$:}
T.~Morokuma$^{1}$,
K.~Ohta$^{2}$,
Y.~T.~Tanaka$^{3}$,
H.~Mori$^{4}$,
M.~Yamanaka$^{3}$,
K.~S.~Kawabata$^{3}$,
Y.~Utsumi$^{5}$,
T.~Nakaoka$^{4}$,
M.~Kawabata$^{4}$,
H.~Nagashima$^{4}$,
M. Yoshida$^{6}$,
Y. Matsuoka$^{7}$,
R. Itoh$^{8}$
\\
$^{1}$ Institute of Astronomy, Graduate School of Science, The University of Tokyo, 2-21-1 Osawa, Mitaka, Tokyo 181-0015, Japan\\
$^{2}$ Department of Astronomy, Graduate School of Science, Kyoto University, Sakyo-ku, Kyoto, Kyoto 606-8502, Japan\\
$^{3}$ Hiroshima Astrophysical Science Center, Hiroshima University, 1-3-1 Kagamiyama, Higashi-Hiroshima, Hiroshima 739-8526, Japan\\
$^{4}$ Department of Physical Science, Hiroshima University, 1-3-1 Kagamiyama, Higashi-Hiroshima, Hiroshima 739-8526, Japan\\
$^{5}$ Kavli Institute for Particle Astrophysics and Cosmology, SLAC National Accelerator Laboratory, Stanford University, 2575 Sand Hill Road, Menlo Park, CA  94025, USA\\
$^{6}$ Subaru Telescope, National Astronomical Observatory of Japan, National Institutes of Natural Sciences, 650 North A’ohoku Place, Hilo, HI 96720, USA\\
$^{7}$ Research Center for Space and Cosmic Evolution, Ehime University, Matsuyama, Ehime 790-8577, Japan\\
$^{8}$ Department of Physics, Tokyo Institute of Technology, 2-12-1 Ohokayama, Meguro, Tokyo 152-8551, Japan \\
$^{\|\|}$E-mail: tmorokuma@ioa.s.u-tokyo.ac.jp
\\
\subsection*{Kapteyn$^{\P\P}$:}
W.~Keel$^{1}$,	
\\
$^{1}$ Department of Physics and Astronomy, University of Alabama, Box 870324, Tuscaloosa, AL 35487, USA \\
$^{\P\P}$E-mail:keel@ua.edu
\\
\subsection*{Liverpool telescope:}
C.~Copperwheat$^{1}$,
I.~Steele $^{1}$,
\\
$^{1}$ Astrophysics Research Institute, Liverpool John Moores University, IC2, Liverpool Science Park, Liverpool L3 5RF, UK \\
\\
\subsection*{\textit{\textbf Swift}$/$\nustar:}
S.~B.~Cenko$^{1,\: 2}$,
D.~F.~Cowen$^{3,\: 4,\: 5}$,
J.~J.~DeLaunay$^{3,\: 4}$, 
P.~A.~Evans$^{6}$,
D.~B.~Fox$^{4,\: 5,\: 7}$,
A.~Keivani$^{3,\: 4}$,
J.~A.~Kennea$^{5}$,
F.~E.~Marshall$^{8}$,
J.~P.~Osborne$^{6}$,
M.~Santander$^{9}$,
A.~Tohuvavohu$^{5}$,
C.~F.~Turley$^{3,\: 4}$,
\\
$^{1}$ Astrophysics Science Division, NASA Goddard Space Flight Center, Mail Code 661, Greenbelt, MD 20771, USA \\
$^{2}$ Joint Space-Science Institute, University of Maryland, College Park, MD 20742, USA \\
$^{3}$ Department of Physics, Pennsylvania State University, University Park, PA 16802, USA \\
$^{4}$ Center for Particle \& Gravitational Astrophysics, Institute for Gravitation and the Cosmos, Pennsylvania State University, University Park, PA 16802, USA \\
$^{5}$ Department of Astronomy \& Astrophysics, Pennsylvania State University, University Park, PA 16802, USA \\
$^{6}$ University of Leicester, X-ray and Observational Astronomy Research Group, Leicester Institute for Space and Earth Observation, Department of Physics \& Astronomy, University Road, Leicester, LE1 7RH, UK 
\\
$^{7}$ Center for Theoretical \& Observational Cosmology, Institute for Gravitation and the Cosmos,  Pennsylvania State University, University Park, PA 16802, USA \\
$^{8}$ NASA Goddard Space Flight Center, Mail Code 660.1, Greenbelt, MD 20771, USA \\
$^{9}$ Department of Physics and Astronomy, University of Alabama, Tuscaloosa, AL 35487, USA \\

\subsection*{VERITAS:}
A.~U.~Abeysekara$^{1}$,
A.~Archer$^{2}$,
W.~Benbow$^{3}$,
R.~Bird$^{4}$,
A.~Brill$^{5}$,
R.~Brose$^{6,7}$,
M.~Buchovecky$^{4}$,
J.~H.~Buckley$^{2}$,
V.~Bugaev$^{2}$,
J.~L.~Christiansen$^{8}$,
M.~P.~Connolly$^{9}$,
W.~Cui$^{10,11}$,
M.~K.~Daniel$^{3}$,
M.~Errando$^{2}$,
A.~Falcone$^{12}$,
Q.~Feng$^{13}$,
J.~P.~Finley$^{10}$,
L.~Fortson$^{14}$,
A.~Furniss$^{15}$,
O.~Gueta$^{7}$,
M.~H\"utten$^{7}$,
O.~Hervet$^{16}$,
G.~Hughes$^{3}$,
T.~B.~Humensky$^{5}$,
C.~A.~Johnson$^{16}$,
P.~Kaaret$^{17}$,
P.~Kar$^{1}$,
N.~Kelley-Hoskins$^{7}$,
M.~Kertzman$^{18}$,
D.~Kieda$^{1}$,
M.~Krause$^{7}$,
F.~Krennrich$^{19}$,
S.~Kumar$^{20}$,
M.~J.~Lang$^{9}$,
T.~T.Y.~Lin$^{13}$,
G.~Maier$^{7}$,
S.~McArthur$^{10}$,
P.~Moriarty$^{9}$,
R.~Mukherjee$^{21}$,
D.~Nieto$^{5}$,
S.~O'Brien$^{22}$,
R.~A.~Ong$^{4}$,
A.~N.~Otte$^{23}$,
N.~Park$^{24}$,
A.~Petrashyk$^{5}$,
M.~Pohl$^{6,7}$,
A.~Popkow$^{4}$,
E.~Pueschel$^{7}$,
J.~Quinn$^{22}$,
K.~Ragan$^{13}$,
P.~T.~Reynolds$^{25}$,
G.~T.~Richards$^{23}$,
E.~Roache$^{3}$,
C.~Rulten$^{14}$,
I.~Sadeh$^{7}$,
M.~Santander$^{26}$,
S.~S.~Scott$^{16}$,
G.~H.~Sembroski$^{10}$,
K.~Shahinyan$^{14}$,
I.~Sushch$^{7}$,
S.~Tr\'epanier$^{13}$,
J.~Tyler$^{13}$,
V.~V.~Vassiliev$^{4}$,
S.~P.~Wakely$^{24}$,
A.~Weinstein$^{19}$,
R.~M.~Wells$^{19}$,
P.~Wilcox$^{17}$,
A.~Wilhelm$^{6,7}$,
D.~A.~Williams$^{16}$,
B.~Zitzer$^{13}$
\\
$^{1}$ Department of Physics and Astronomy, University of Utah, Salt Lake City, UT 84112, USA \\
$^{2}$Department of Physics, Washington University, St. Louis, MO 63130, USA \\
$^{3}$ Fred Lawrence Whipple Observatory, Harvard-Smithsonian Center for Astrophysics, Amado, AZ 85645, USA \\
$^{4}$ Department of Physics and Astronomy, University of California, Los Angeles, CA 90095, USA \\
$^{5}$ Physics Department, Columbia University, New York, NY 10027, USA \\
$^{6}$ Institute of Physics and Astronomy, University of Potsdam, 14476 Potsdam-Golm, Germany \\
$^{7}$ DESY, Platanenallee 6, 15738 Zeuthen, Germany \\
$^{8}$ Physics Department, California Polytechnic State University, San Luis Obispo, CA 94307, USA \\
$^{9}$ School of Physics, National University of Ireland Galway, University Road, Galway, Ireland \\
$^{10}$ Department of Physics and Astronomy, Purdue University, West Lafayette, IN 47907, USA \\
$^{11}$ Department of Physics and Center for Astrophysics, Tsinghua University, Beijing 100084, China. \\
$^{12}$ Department of Astronomy and Astrophysics, 525 Davey Lab, Pennsylvania State University, University Park, PA 16802, USA \\
$^{13}$ Physics Department, McGill University, Montreal, QC H3A 2T8, Canada \\
$^{14}$ School of Physics and Astronomy, University of Minnesota, Minneapolis, MN 55455, USA \\
$^{15}$ Department of Physics, California State University - East Bay, Hayward, CA 94542, USA \\
$^{16}$ Santa Cruz Institute for Particle Physics and Department of Physics, University of California, Santa Cruz, CA 95064, USA \\
$^{17}$ Department of Physics and Astronomy, University of Iowa, Van Allen Hall, Iowa City, IA 52242, USA \\
$^{18}$ Department of Physics and Astronomy, DePauw University, Greencastle, IN 46135-0037, USA \\
$^{19}$ Department of Physics and Astronomy, Iowa State University, Ames, IA 50011, USA \\
$^{20}$ Department of Physics and Astronomy and the Bartol Research Institute, University of Delaware, Newark, DE 19716, USA \\
$^{21}$ Department of Physics and Astronomy, Barnard College, Columbia University, NY 10027, USA \\
$^{22}$ School of Physics, University College Dublin, Belfield, Dublin 4, Ireland \\
$^{23}$ School of Physics and Center for Relativistic Astrophysics, Georgia Institute of Technology, 837 State Street NW, Atlanta, GA 30332-0430 \\
$^{24}$ Enrico Fermi Institute, University of Chicago, Chicago, IL 60637, USA \\
$^{25}$ Department of Physical Sciences, Cork Institute of Technology, Bishopstown, Cork, Ireland \\
$^{26}$ Department of Physics and Astronomy, University of Alabama, Tuscaloosa, AL 35487, USA \\
%

\subsection*{VLA/17B-403 team:}
A.~J.~Tetarenko$^{1}$,
A.~E.~Kimball$^{2}$,
J.~C.~A.~Miller-Jones$^{3}$,
G.~R.~Sivakoff$^{1}$
\\ 
\\
$^{1}$ Department of Physics, CCIS 4-181, University of Alberta, Edmonton, Alberta, Canada T6G 2E1 \\
$^{2}$ National Radio Astronomy Observatory, 1003 Lopezville Rd, Socorro, NM, 87801, USA \\
$^{3}$ International Centre for Radio Astronomy Research – Curtin University, GPO Box U1987, Perth, WA 6845, Australia \\

\tableofcontents

\subsection*{IceCube}
\addcontentsline{toc}{section}{\protect\numberline{}IceCube}%
IceCube is a cubic-kilometer-sized neutrino detector~\cite{Aartsen:2016nxy}
installed in the ice at the geographic South Pole, Antarctica
between depths of 1450\,m and 2450\,m.  The detector consists of
5160 digital optical modules (DOMs) attached to 86 cables (called strings),
each instrumented with 60 DOMs.
The strings are arranged in a hexagonal
pattern with 125\,m average horizontal spacing.
Each DOM consists of a glass pressure-resistant sphere containing
a photomultiplier and electronics, and operates independently producing digital signals, which
are transmitted to the surface along the string.
Detector construction was completed in 2010,
and IceCube has operated with an $\sim$99\% duty cycle since then.

IceCube does not directly observe neutrinos, but rather the secondary particles produced
in the neutrino interaction with matter.  IceCube detects these particles by observing the
Cherenkov light emitted as they travel through the ice.  The ability
to accurately determine the direction of a neutrino event recorded in IceCube is highly
dependent on the ability to reconstruct the trajectories of these secondary particles. The secondary particles
produce two distinct classes of signals within the instrumented
volume: tracks and cascades.  Track events, the primary focus of the IceCube alert system,
are produced by muons, arising primarily
from the charged-current interaction of muon-type neutrinos,
which produce tracks with lengths of the order of a few kilometers.  These tracks can be reconstructed
with a directional uncertainty less than 1\,deg, but with a large uncertainty on the neutrino energy since an
unmeasured fraction of their energy is deposited outside the instrumented volume.

\begin{figure*}
  \centering
  \includegraphics[width=0.75\textwidth]{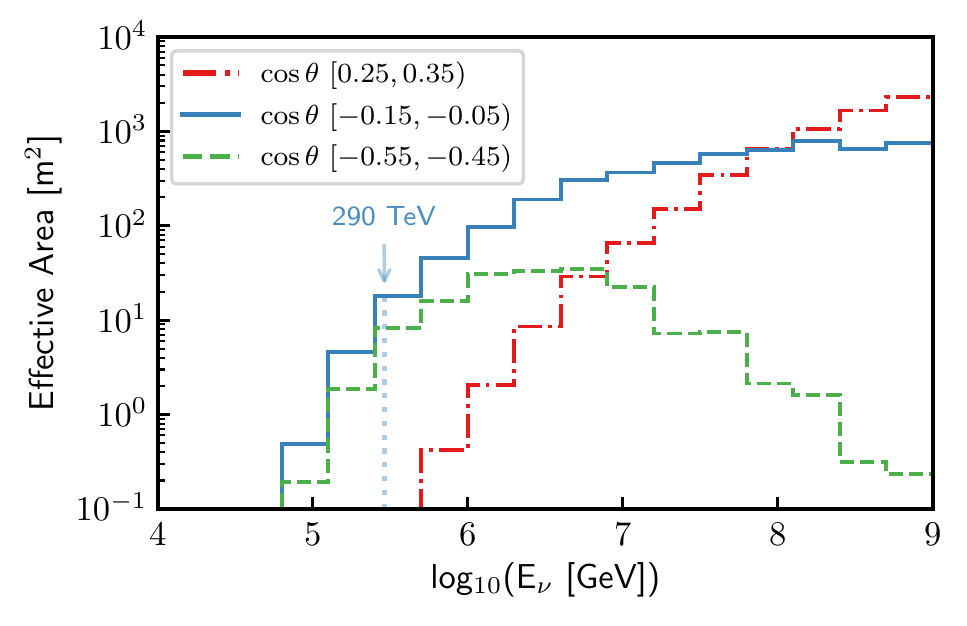}
  \caption{{\bf Neutrino effective area for the through-going track alert channel.}
Effective area for the online through-going track (``EHE'') selection in three zenith angle ranges. The zenith angle of IceCube-170922A was cos(zenith) = $-0.1$, a preferred direction for this event selection. In the range $-0.55$ to $-0.45$ ($\sim$30\,deg below the horizon) a strong absorption by the Earth at the highest neutrino energies is seen, while in the interval 0.25 to 0.35 ($\sim$20\,deg above the horizon) strong cuts on track energy are needed to suppress the background from cosmic-ray muons, limiting sensitivity below 1~PeV.  The most probable neutrino energy of 290~TeV is also shown.}
  \label{fig:ehe_eff_area}
\end{figure*}

The IceCube neutrino alerts are generated in real time by applying direction and energy estimates
to all events as the data are collected~\cite{Aartsen:2016lmt}, and notifying the
astronomical community immediately of a candidate astrophysical neutrino event.  IceCube-170922A was
generated by the ``EHE'' through-going track selection in the real-time alert system.  The event
selection was inspired by the event requirements used to search for cosmogenic/GZK neutrinos~\cite{Aartsen:2016ngq}
and was modified for online use to give a larger number of astrophysical neutrinos.  The
sensitivity of this event selection is highlighted by the effective area
for several zenith angle ranges, shown in Figure~\ref{fig:ehe_eff_area}.
The zenith angle of IceCube-170922A (cos(zenith) = $-0.1$) was in the most
sensitive zenith acceptance range, a direction where atmospheric muons are
easily blocked, but neutrino absorption in the Earth
has not depleted the high-energy neutrino flux.

\paragraph*{Calculations of systematic uncertainties for IceCube-170922A}
\addcontentsline{toc}{subsection}{\protect\numberline{}Calculations of systematic uncertainties for IceCube-170922A}%

The directional resolution of muon tracks passing through the IceCube detector is limited
by the stochastic nature of the detected light, the finite density of DOMs
where Cherenkov light is detected and the uncertainty in the optical properties
of the glacial ice~\cite{GlaciologyPaper}. We modeled the expected uncertainty due to these statistical and systematic effects by re-simulating a large sample
of candidate events similar to the observed event, and studying the distance of
their best-fitting directions from their true simulated direction.

A dedicated simulation set was generated containing muon tracks passing through the same part of the detector
as the originally observed event (closer than 30\,m from the original best-fitting track at any point within the instrumented
volume and within 2\,deg of the best-fitting direction) and with a similar energy
loss pattern (total deposited charge within $\pm 20\%$ of the original charge).
Each event was simulated using an ice model sampled from the space of ice models
compatible with the current baseline best-fitting ice model~\cite{SPICEPaper}.

Each event in this simulation set is reconstructed using the same method as is
applied to the observed event and a test statistic (TS), defined as the difference in log-likelihood ($L$) between the best-fitting
direction and the true direction is recorded as $\mathrm{TS}=2 (\log L_{\mathrm{true}}-\log L_{\mathrm{best}})$.
The 50\% and 90\% percentiles of the distribution of $\mathrm{TS}$ over this simulation
set are recorded and used to draw the 50\% and 90\% contour lines in the
reconstructed likelihood fit at the corresponding likelihood ratios.

This algorithm allows us to include the uncertainty in the modeling of  optical properties of the glacial ice into the
fit uncertainty providing a combined statistical and systematic error
(taking into account ice model systematics only). By construction, this method
is not able to shift the best-fitting direction of the reconstruction and will
include systematic bias on average only.

\paragraph*{Calculation of the neutrino energy}
\addcontentsline{toc}{subsection}{\protect\numberline{}Calculation of the neutrino energy}%
As IceCube detects the secondary muon produced in the neutrino's interaction in or near the instrumented volume, a precise determination of the neutrino energy is generally not possible for track events.  However, for high-energy muons, a robust estimation of the energy of the muon as it traverses the instrumented volume is available~\cite{2013NIMPA.703..190A,Aartsen:2013vja}.  Muons
above $\sim$1~TeV experience large stochastic energy losses due to pair production,
bremsstrahlung, and photo-nuclear interactions. These energy losses grow with muon energy and can be used to estimate the energy as the muon passes through the detector.

\begin{figure*}
\begin{center}
\begin{subfigure}[b]{0.55\textwidth}
  \caption{}
  \includegraphics[width=\textwidth]{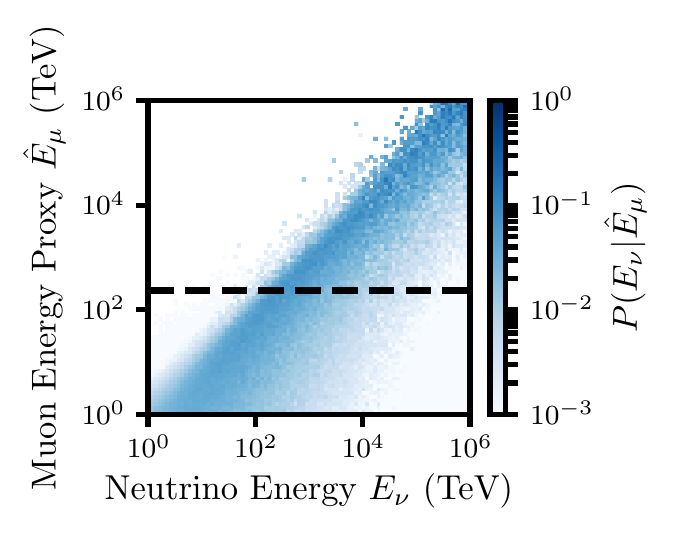}
\end{subfigure}
\begin{subfigure}[b]{0.70\textwidth}
  \caption{}
  \includegraphics[width=\textwidth]{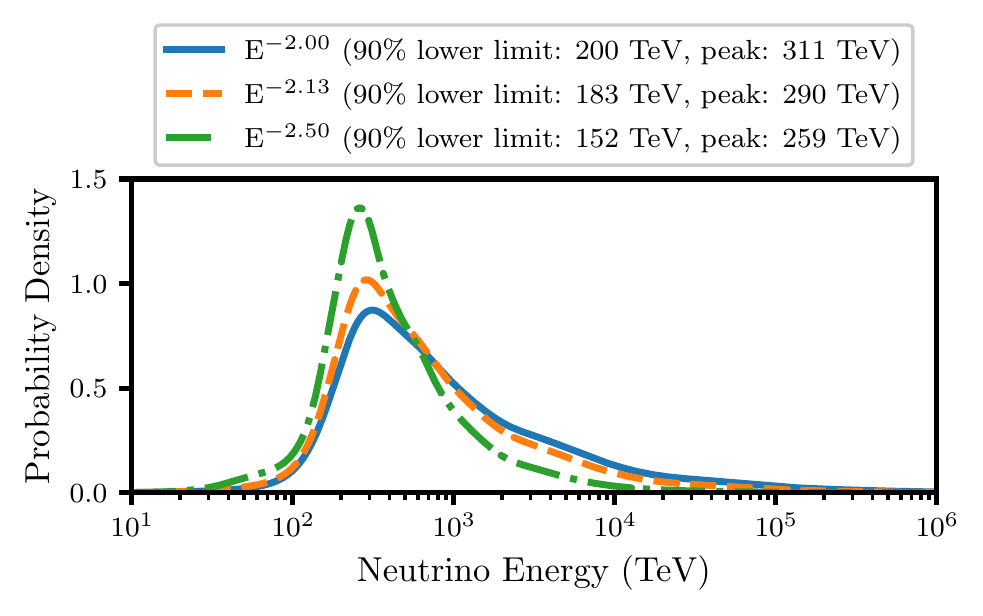}
\end{subfigure}
\end{center}
\caption{{\bf Estimate of neutrino energy for IceCube-170922A.}
Estimate of the neutrino energy of IceCube-170922A derived from an estimator of the muon
energy in the detector~\cite{2013NIMPA.703..190A}.  Note that the muon energy estimator is not equivalent to the deposited energy as the muon passed through the detector. The deposited muon energy sets a lower limit on the neutrino and muon energies. Panel~A presents the 2-D distribution
of neutrino energy vs. muon energy estimator (``Muon Energy Proxy'') from simulation.  The observed energy estimator is indicated
by a horizontal dashed black line. Assuming a prior distribution of true neutrino energies (modeled as power-law spectra with various indices),
a probability distribution of true neutrino energies for the event can be derived (Panel~B). For each neutrino spectral index, the 90\% C.L. lower limit and most probable ("peak") neutrino energies are listed.
The result is only weakly dependent on the chosen spectral index.}

\label{fig:nu_energy_like}
\end{figure*}

Figure~\ref{fig:nu_energy_like} presents the measured muon energy~\cite{2013NIMPA.703..190A} observed in simulation of neutrino track events for a wide range of neutrino energies. The exact distribution of muon energies will depend on the assumed neutrino spectral index. For the observed muon energy of the IceCube-170922A track, the most-probable neutrino energy and the 90\% C.L. lower limit can be calculated, and is shown in Figure~\ref{fig:nu_energy_like} for three spectral indices.  Using the measured spectral index of $-2.13$ ($-2.0$) for the estimated diffuse astrophysical muon neutrino spectrum~\cite{Aartsen:2016xlq}, the most-probable neutrino energy of 290\,TeV (311\,TeV), a 90\% C.L. lower limit on the neutrino energy of 183\,TeV (200\,TeV), and a 90\% C.L. upper limit on the neutrino energy of 4.3\,PeV (7.5\,PeV) are determined.

\subsection*{High-energy \g-ray observations}
\addcontentsline{toc}{section}{\protect\numberline{}High-energy \g-ray observations}%

\paragraph*{Generation of the \Fermi-LAT light curves of TXS~0506+056}
\addcontentsline{toc}{subsection}{\protect\numberline{}Generation of the \Fermi-LAT light curves of TXS~0506+056}%
The light curve is based on Pass\,8 SOURCE class photons detected in the time interval from the start of the science phase of the mission in 4 August, 2008 to 24 October, 2017.  This is the recommended class for most analyses and provides good
sensitivity for analysis of point sources.  Standard good-time intervals were selected excluding time
intervals when the field of view of the LAT intersected the Earth, and during which bright \g-ray bursts and solar flares were observed.

A binned maximum likelihood technique (binned in space and energy) was applied
using the standard \Fermi-LAT Science-Tools package version v11r05p02 available from the \Fermi~Science Support Center (FSSC)~\cite{fsscref} and the P8R2\_SOURCE\_V6 instrument response functions.  Data in the energy
range of 100\,MeV to 1\,TeV were binned into eight equally spaced logarithmic energy intervals per decade. To minimize the contamination from the \g-rays produced in Earth's upper atmosphere, a zenith angle cut of $<$\,90\,deg was applied.

A 10\,deg $\times$ 10\,deg region of interest (ROI) was selected centered on the assumed source position and binned in 0.1\,deg pixels.  The input model for the ROI included all known \g-ray sources from the \Fermi-LAT Third Source Catalog (3FGL)~\cite{Acero:2015hja}. We refined the best-fitting position of TXS~0506+056, including the additional data taken since the release of the 3FGL catalog. Similarly, we searched for additional sources in the ROI that may be significantly detected 
in the current data set, but were too faint to be included in this catalog, a standard procedure~\cite{2017arXiv170500009F}.
The model of the ROI included the isotropic and Galactic diffuse
components (gll\_iem\_ext\_v06.fits and iso\_P8R2\_SOURCE\_V6\_v06.txt).
To build the light curve, the spectral functional forms given in the
3FGL/3FHL~\cite{TheFermi-LAT:2017pvy} catalog for each source in the ROI were adopted.
For each time interval analyzed for the light curve, the flux normalization of TXS~0506+056 and of other sources within 3\,deg of it were free parameters, while the spectral shapes were fixed to their forms in the overall best-fitting model for the entire 9.5-year dataset starting in August 2008. Light curves for TXS~0506+056 were created with a time binning of 28 days over the full \Fermi-LAT observation period, and a binning of 7 days around the time of the IceCube neutrino alert.

Similar light curves, but with an energy threshold of 1\,GeV and using 28 day bins only, were compiled for all extragalactic \Fermi-LAT sources (time binning and energy threshold were chosen to reduce the required computing resources).
All sources from the four-year source catalog (3FGL) and the six-year hard source catalog (3FHL)~\cite{TheFermi-LAT:2017pvy} which are classified as extragalactic objects were included. Unclassified sources were added if they were more than 5\,deg from the Galactic equator. Sources that were marked with an analysis flag~\cite{Acero:2015hja},  were removed. In total, 2257 sources were selected.
These light curves were used in the calculation of the chance coincidence probability of the apparent neutrino-flaring blazar correlation. Very bright sources
were modeled with log-parabolic spectra.
For the light curve generation, the spectra of the sources were kept fixed to the values obtained from a fit over the total observation period. The limited statistics in the 28 day time bins do not allow fitting bin-by-bin spectral parameters for most sources.

\paragraph*{\Fermi-LAT real-time follow up pipelines }
\addcontentsline{toc}{subsection}{\protect\numberline{}\Fermi-LAT real-time follow up pipelines}%

Following the neutrino alert, the quick recognition of the coincident blazar flare was made possible by automated high-level software pipelines developed by the LAT collaboration that provide continuous monitoring of the \g-ray sky.
The Automated Science Processing (ASP)~\cite{2007AIPC..921..544C} and
the \Fermi~All-sky Variability Analysis (FAVA)~\cite{FAVA}  are model independent techniques that search for variations in \g-ray flux, on timescales from hours to one week.  The statistical
significances of the candidate sources identified by ASP and FAVA are subsequently evaluated
with the more robust maximum-likelihood technique, and further inspected by  the so-called 
Flare Advocates in the LAT collaboration, who communicate 
significant results to 
the external scientific community through the \Fermi~multi-wavelength mailing list, Astronomer's
Telegrams, and direct e-mail~\cite{2011arXiv1111.6803C}.  Significant flares seen with ASP are reported
automatically as Gamma-ray Coordinates Network notices~\cite{FermiASPGCN}.

ASP is based on the source detection algorithm PGWave~\cite{1997ApJ...483..350D}, used also in the \Fermi-LAT
catalog pipeline to identify candidate sources. PGWave applies a two-dimensional Mexican Hat wavelet filtering to
find significant clusters of photons in the LAT data from different intervals (6-hour, 1-day and 1-week)
and in three energy ranges (0.1~GeV\,--\,300~GeV, 0.1~GeV\,--\,1~GeV and 1~GeV\,--\,300~GeV).

FAVA searches for \g-ray variability on a weekly time scale in a low and high energy bands, {\it i.e.}, from $0.1$\,GeV to $0.8$\,GeV and from $0.8$\,GeV to $300$\,GeV. A photometric
technique is used to compare the weekly flux to the long-term average flux over the first four
years of the mission in a grid of regions covering the entire sky.
If the photometric technique finds a deviation from the average in at least one of the two energy bands with a significance greater than $4\sigma$, a maximum likelihood analysis is applied. This models the ROI, including background sources and the diffuse emission, taking into account the LAT PSF that is applied to accurately assess the statistical significance.

The FAVA results are updated in real time and are displayed in a public web
interface~\cite{favaref}. FAVA
has been used as a tool to quickly find potentially variable sources in the neutrino error
circle. In the time bin during which IceCube-170922A arrived, FAVA reported a significance at the position of TXS 0506+056 obtained in the likelihood analysis of $6.5\sigma$ in the low-energy band and $6.9\sigma$ in the high-energy band.

\paragraph*{\textbf{\textit{AGILE}}}
\addcontentsline{toc}{subsection}{\protect\numberline{}\textit{AGILE}}%

The \g-ray satellite {\it AGILE}~\cite{Tavani:2008sp} monitors
the sky in spinning mode in the energy range 30 MeV--30 GeV.  {\it AGILE}
detected enhanced \g-ray emission above 100~MeV from the
IceCube-170922A/TXS~0506+056 region and reported this in an Astronomer's
Telegram, issued 7 days after the neutrino detection~\cite{2017ATel10801....1L}.

A refined analysis of the data acquired with the {\it AGILE} imaging \g-ray
detector leads to significant detections from this region
on short and long timescales before and near the time of the IceCube neutrino alert, compatible with the flaring activity observed by \Fermi-LAT from TXS~0506+056.

The {\it AGILE} \g-ray flux above 100~MeV from TXS~0506+056, estimated with the
{\it AGILE} Maximum Likelihood (ML) algorithm~\cite{Bulgarelli:2012he}
in a time window of 13 days centered at MJD 58012.5 (16 September, 2017) is
found to be $(5.3 \pm 2.1) \times 10^{-7}$ cm$^{-2}$ s$^{-1}$.
The corresponding energy flux density of this {\it AGILE} observation,
scaled at 200~MeV assuming a power-law index of -2,
is $(8.8 \pm 3.5) \times 10^{-11}$ erg cm$^{-2}$ s$^{-1}$.

\subsection*{Very-high-energy \g-ray observations}
\addcontentsline{toc}{section}{\protect\numberline{}Very-high-energy \g-ray observations}%

\paragraph*{MAGIC}
\addcontentsline{toc}{subsection}{\protect\numberline{}MAGIC}%

\begin{figure*}
\centering
\includegraphics[width=0.9\textwidth]{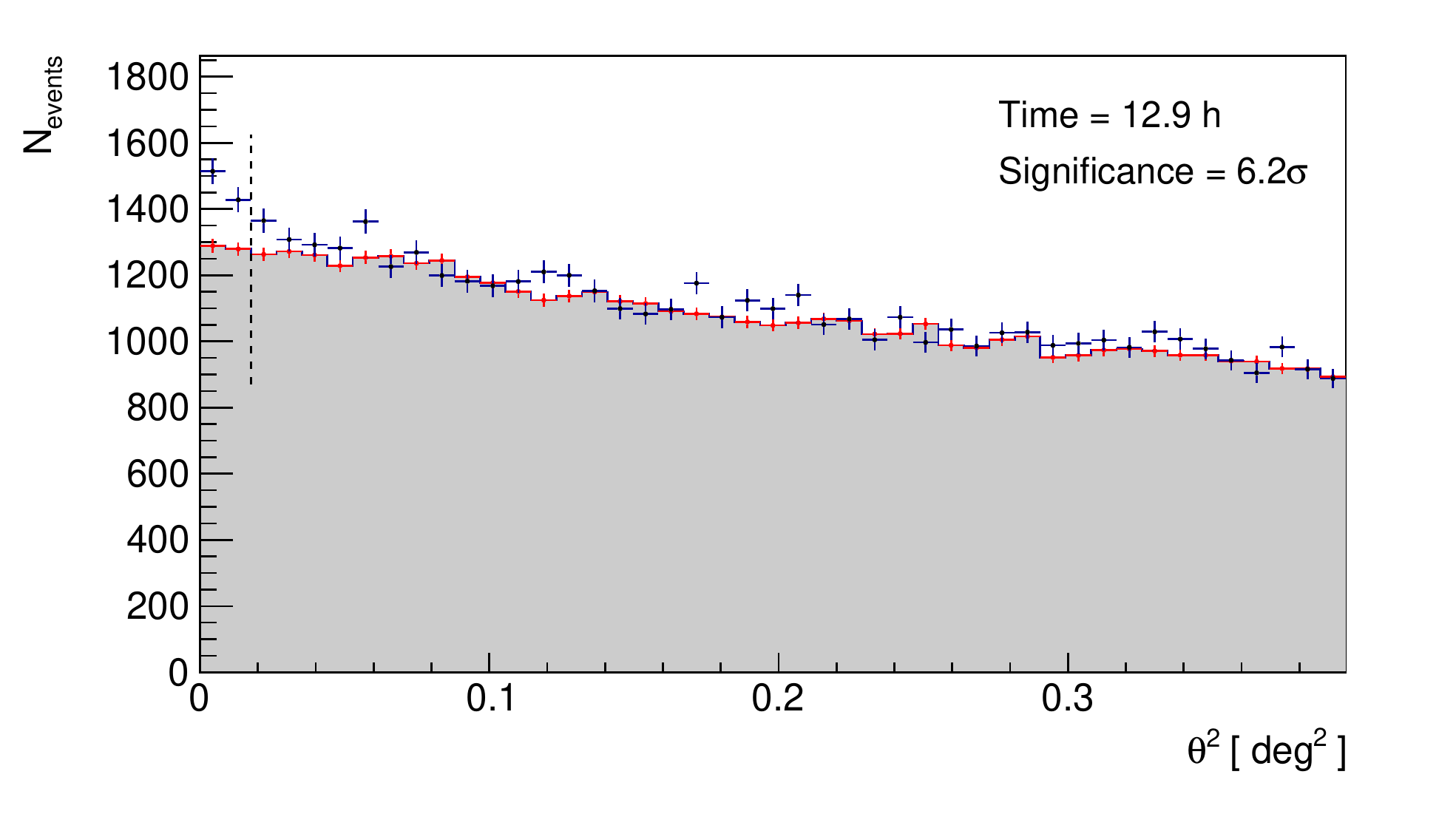}
 \caption{{\bf  Significance of the VHE \g-ray signal from TXS 0506+056 as measured by MAGIC.}
Distribution of the squared angular distance, $\theta^{2}$, between the re-constructed source position and the nominal source position (blue points) or the background estimation position (shaded area) for the direction of the blazar TXS 0506+056. Statistical
uncertainties on the number of signal (blue markers) or background (red markers) events are shown as vertical error bars. The number of excess events (N$_{\rm ex}$ = 374 $\pm$ 62) and significance~\cite{LiMa:1983} are calculated in the region from 0 to the vertical dashed line. The estimated energy threshold is 90~GeV.}
  \label{fig_th2}
\end{figure*}

MAGIC followed-up 4 of the 10 IceCube alerts that had been issued by 1 October, 2017. The properties of the events that were followed-up are listed in Table~\ref{tab_events}. For the alerts issued before September 2017 no signal was detected by MAGIC within the 50\% containment radius reported by IceCube~\cite{amongcnref}.

\begin{table}[ht]
\centering
\caption{{\bf Neutrino alerts selected as MAGIC targets.}  
For each set of targeted observations, the position, directional uncertainty and reported deposited energy
from the neutrino alert are listed.  Additionally, MAGIC zenith angle ranges and times over 
which observations were made are listed.}
\begin{tabular}{lllll}
\hline
IceCube- & 160427A & 160731A & 170321A & {\bf 170922A} \\
\hline
{\it From IceCube:}& & &  &\\
Right Ascension [deg]&240.57&214.54&98.33&{\bf 77.43}\\
Declination [deg]&	9.34&$-0.33$&	$-14.48$&{\bf 5.72}\\
 Median angular resolution [deg]& 0.60& 0.33 & 0.33 &{\bf 0.25}\\
 Deposited energy [TeV]& $\sim$140 & $<$100 & $>$120 & {\bf $>$23}\\
 \hline
{\it MAGIC data taking:}& & &  &\\
 Zenith angle range [deg]& 18 - 26&45-65&45-60 & {\bf 22-52}\\
 Effective  observation time [h] &1.85&1.3&1.0&{\bf 12.9}\\
\hline
 \end{tabular}
\label{tab_events}
\end{table}

MAGIC performed the first observations of the reported IceCube-170922A event direction for 2~hours on 24 September, 2017 (32~hours after the IceCube alert was issued), under non-optimal weather conditions. The standard MAGIC analysis framework \texttt{MARS} was used for the data analysis~\cite{Aleksic:2014lkm}. After data quality selection, 1.07~hours of observations were used to derive an upper limit on the TXS~0506+056 flux above 90 GeV of $3.56 \times 10^{-11}$ cm$^{-2}$ s$^{-1}$ at 95\% C.L.
The upper limit was calculated~\cite{2005NIMPA.551..493R} assuming a 30\% systematic uncertainty in the estimated \g-ray detection efficiency of MAGIC.
Observations were resumed on 28 September, 2017. With good observational conditions, data from 28 September, 2017 and 29 September, 2017 revealed hints of a signal, with an excess observed at the $\sim$3.5~$\sigma$ significance level. The significance of this signal grew steadily over the following nights, motivating the long exposure.
Analysis cuts optimized for Crab Nebula detection above 90~GeV were applied to the data. Integrating 12.9 hours of good quality data, MAGIC detected a clear signal with 374 $\pm$ 62 excess events at the location of the blazar TXS~0506+056 (RA: 77.36\,deg, Dec: +5.69\,deg (J2000)~\cite{Lanyi:2010}).  The combined signal is shown in Figure \ref{fig_th2}. The day-to-day results from the MAGIC observations are provided in Table \ref{tab_obs}. Observations with a detection significance of less than 2$\sigma$ are reported as flux upper limits at 95\% C.L., again including a 30\% systematic uncertainty on the detection efficiency.

The MAGIC VHE \g-ray observations can be used to determine an upper limit for the redshift of TXS~0506+056 constraining the attenuation of the VHE flux due to interaction with the EBL~\cite{Stecker1992,Hauser2001}. These redshift limits are derived from assumed properties of the intrinsic spectrum, the measured VHE spectrum from MAGIC and models for the EBL. Here, the intrinsic spectrum is assumed to be a simple power-law $dN/dE \propto E^{\gamma}$, with the index constrained to be $\gamma<-1.5$.  For each assumed redshift value, the VHE gamma-ray spectrum is evaluated including attenuation with the EBL and the expected rate of \g-ray events is calculated using the MAGIC instrument response. A likelihood test is constructed by comparing the expected event rates to those observed in the source region and in the background control regions. The likelihood ($L$) is maximized by allowing the intrinsic spectral model parameters to vary and treating uncertainties in the cosmic-ray induced background as nuisance parameters. Performing a scan in redshift, at each step the profile likelihood following~\cite{2001NIMPA.458..745R} is used to derive an upper limit to the source redshift at 95\% C.L. When the maximum likelihood value is obtained for z=0, the so-called "bounded likelihood" approach is followed, {\it i.e.} the increase in $-2\ln\mathcal{L}$ is computed relative to its value at z=0. The dominant experimental  systematics are evaluated by varying the simulated total light throughput of the instrument (including effects in the atmosphere) by $\pm$15\%. The most conservative value derived from all realizations is taken as a result for the upper limit.

For TXS~0506+056, the redshift upper limit ranges from 0.61 to 0.67 at 95\% C.L, adopting the EBL models from~\cite{Dominguez:2011a, Franceschini2008, 2010ApJ...712..238F}. More conservative are the results obtained using the models from \cite{0004-637X-827-1-6,2041-8205-781-2-L35,0004-637X-752-2-113} for which the redshift upper limit ranges from 0.83 to 0.98 at 95\% C.L. Considering a lower confidence level of 90\% C.L. the results are in better agreement, the full range of upper limits values for all EBL models considered here being 0.41 to 0.57.  Conservatively, the resulting 95\% confidence level upper limit on the source redshift is z $<$ 0.98 taking into account a 15\% systematic uncertainty on the total light throughput of the instrument~\cite{Stecker1992}.  These results are consistent with the measured redshift of z=0.3365~\cite{Paiano:2018qeq}.

\begin{table}[ht]
\centering
\caption{{\bf MAGIC nightly observations of TXS~0506+056.}
Summary of MAGIC observations for each night's observation of TXS~0506+056, including: date corresponding to the middle of the observation window; effective observation time after quality cuts; integral photon flux above 90 GeV, with flux upper limits (indicated by $<$) given at 95\% C.L.; and per-night significance.}
\label{tab_obs}
\begin{tabular}{lccc}
\hline
Date & Effective time & Flux $>$ 90 GeV  & Significance \\
MJD & [hours] & [$\mathrm{ph\,cm^{-2}\,s^{-1}}$] & $\sigma$\\
\hline
58020.16 & 1.07 & $<3.6\times 10^{-11}$ & 0 \\
58024.21 & 1.25 & $<6.2\times 10^{-11}$ & 1.8 \\
58025.18 & 2.9 & $<5.8 \times 10^{-11}$ & 1.0 \\
58026.17 & 3.0 & $<3.6 \times 10^{-11}$ & 0.95 \\
58027.18 & 2.9 & $1.9 \pm 1.2 \times 10^{-11}$ & 2.5 \\
58028.23 & 0.8 & $<5.8\times 10^{-11}$ & 1.7 \\
58029.22 & 1.3 & $5.9 \pm 1.5\times 10^{-11}$ & 4.3\\
58030.24 & 0.65 & $8.0 \pm 2.0\times 10^{-11}$ & 5.4\\
\hline
 \end{tabular}
\end{table}

\paragraph*{H.E.S.S.} 
\addcontentsline{toc}{subsection}{\protect\numberline{}H.E.S.S.}%
The High Energy Stereoscopic System (H.E.S.S.) of imaging atmospheric Cherenkov telescopes has been routinely performing follow-up observations of high-energy neutrinos detected by IceCube and ANTARES since 2015. The H.E.S.S. system has been designed to automatically react to neutrino alerts, allowing for a search for VHE \g-ray sources in coincidence with a neutrino detection on timescales that range from a few tens of seconds to several days~\cite{hess_transients_system}. In the past, several neutrino alerts were followed up by H.E.S.S.~\cite{HESSMM_ICRC2017}. All follow-up observations of IceCube alerts performed by H.E.S.S. are summarized in Table~\ref{tab:nu_followup_summary}.

\begin{table}[htp]
\begin{center}
\caption{{\bf Neutrino follow-up observations performed by H.E.S.S.} IceCube-170922A represents the IceCube alert which was followed up with the longest exposure and the shortest delay. The long exposure was taken following the \Fermi-LAT announcement of TXS~0506+056 being in a high emission state~\cite{2017Atel_Fermi}.
\label{tab:nu_followup_summary}}
\begin{tabular}{cccc}
\hline
Date 					&  Alert 		& Delay of  	& Duration of  \\
						&identifier	& observations	&observations  \\
\hline
Apr 27, 2016 & IceCube-160427A	& 2d 15h	&	2h	\\
Jul 31, 2016 & IceCube-160731A	& 16h		&	2h	\\
Nov 3, 2016 & IceCube-161103A	& 12h		&	2h	\\
{\bf Sep 22, 2017}	& {\bf IceCube-170922A}	& {\bf 4h}		& 	{\bf 3h 14m}	\\
\hline
\end{tabular}
\end{center}
\end{table}%

H.E.S.S. performed follow-up observations towards the direction of the blazar TXS~0506+056 during the nights of 22 September, 2017 and 23 September, 2017 after the detection of a high-energy neutrino by IceCube. Initial observations started $\sim$4 hours after the circulation of the neutrino alert~\cite{GCN21916}. A preliminary on-site analysis did not reveal any significant \g-ray emission~\cite{2017ATel10787....1D} for this data set. A second set of observations was acquired during the nights of 27 September, 2017 and 28 September, 2017 following the announcement of \Fermi-LAT that TXS~0506+056 was in an active state and positionally coincident with the direction of the neutrino event~\cite{2017Atel_Fermi}. In total 3.25 hours of high-quality observations including the central large telescope (CT5) were obtained at zenith angles ranging from 31\,deg to 46\,deg. 

The 3.25 hours of CT5 data were analyzed in mono mode using the Model Analysis~\cite{ModelAnalysis} with {\it loose} cuts to achieve a low energy threshold. No \g-ray emission at a significant level was detected and upper limits on the VHE \g-ray flux have been calculated. The best fit spectral index of $-3.9$ as measured from the MAGIC data was used as the spectral assumption. Limits at 95\% C.L. were derived using the Rolke method~\cite{Lundberg:2009iu} and assuming a systematic uncertainty of 30\%. Negative excess fluctuations of the measured counts in the signal region were taken into account by replacing them with the measured background counts, scaled with the signal region exposure time.

The limits and fluxes were calculated for each night of the data set individually. They are shown in Figure~\ref{fig:CombinedLC} and summarized in Table~\ref{tab:hess_nightly_flux_limits} above
an energy threshold of 175 GeV. The table and figure also include flux upper limits from two archival observation campaigns from September 2015 and December 2015 to January 2016.
Additionally, differential flux upper limits were calculated for the whole 3.25~hour dataset. They are depicted in Figure~\ref{fig:txs0506_sed} and summarized in Table~\ref{tab:hess_diff_limits}. All results have been cross-checked with an independent calibration and analysis chain~\cite{ImPACT}, which showed consistent results.

\begin{table}[htp]
\begin{center}
\caption{{\bf Flux upper limits from H.E.S.S. for TXS~0506+056.} Archival and nightly \g-ray flux upper limits at 95\% confidence level for TXS~0506+056 derived from the H.E.S.S. observations assuming an $E^{-3.9}$ energy spectrum.
\label{tab:hess_nightly_flux_limits}}
\begin{tabular}{ccc}
\hline
MJD & Observation time  & Flux $>$ 175 GeV  \\
$[{\rm days}]$ & $[{\rm h}]$  & [$\mathrm{ph\,cm^{-2}\,s^{-1}}$]
\\
\hline
57286 -- 57288 & 5.4 & $<7.2 \times 10^{-12}$ \\
57358 -- 57390 & 4.4 & $<1.1 \times 10^{-11}$ \\
58019.07 & 1.35  & $<1.0 \times 10^{-11}$ \\
58024.08 & 0.48  & $<1.8 \times 10^{-11}$ \\
58025.08 & 1.65 &  $<1.8 \times 10^{-11}$ \\

\hline
\end{tabular}
\end{center}
\end{table}%

\begin{table}[htp]
\begin{center}
\caption{{\bf H.E.S.S. differential \g-ray flux upper limits for TXS~0506+056.} Flux upper limits ($f_\gamma$) at $95\%$ C.L. obtained for the full TXS~0506+056 H.E.S.S. data set and assuming an $E^{-3.9}$ energy spectrum. $E_{\rm min}~$ and $E_{\rm max}$
define the energy range over which the differential flux upper limit is derived.
\label{tab:hess_diff_limits}}
\begin{tabular}{ccc}
\hline
$E_{\rm min}~$  & $E_{\rm max}$ & $f_\gamma$ \\
 $[{\rm TeV}]$	& $[{\rm TeV}]$ & [$\mathrm{cm^{-2}\,s^{-1}\,TeV^{-1}}$] \\ \hline
 0.16 &  0.28 & $ < 6.6 \times 10^{-11} $ \\
 0.28 &  0.48 & $ < 2.1 \times 10^{-11} $ \\
 0.48 &  0.85 & $ < 4.5 \times 10^{-12} $ \\
 0.85 &  1.50 & $ < 1.8 \times 10^{-12} $ \\
 1.50 &  2.63 & $ < 5.9 \times 10^{-13} $ \\
 2.63 &  4.62 & $ < 3.3 \times 10^{-13} $ \\
 \hline
\end{tabular}
\end{center}
\end{table}%

\paragraph*{VERITAS}
\addcontentsline{toc}{subsection}{\protect\numberline{}VERITAS}%

The Very Energetic Radiation Imaging Telescope Array System (VERITAS)~\cite{VTS}, was used to perform follow-up observations of IceCube-170922A. Observations started on 23 September 2017 at 09:06 UTC, 12.2 hours after the IceCube detection, accumulating an exposure of one hour under partial cloud coverage in normal observation mode. Additional VERITAS observations were collected following the \Fermi-LAT report of the detection of a hard GeV \g-ray flare from the blazar TXS~0506+056 located within the neutrino error region. Five additional hours were collected during the period between 28 September 2017 at 08:57 UTC and 30 September 2017 at 11:04 UTC (5.5 to 7.6 days after the neutrino detection), resulting in a total exposure of 5.5 hours for the entire data set after quality cuts.

An analysis of the data optimized for soft-spectrum sources shows no evidence of \g-ray emission at the blazar location or anywhere else in the 3.5\,deg VERITAS field of view. The integral \g-ray flux upper limit derived from the VERITAS observations at the TXS~0506+056 position is $1.2 \times 10^{-11}$ cm$^{-2}$ s$^{-1}$ at 95\% C.L. above an energy threshold of 175 GeV assuming the power-law photon spectral index of -3.9 from the MAGIC data.

All limits were calculated using the method described in \cite{2001NIMPA.458..745R} with the requirement of a minimum of 10 events present in the off-source region to reduce the uncertainty in the estimation of the background rate. The systematic uncertainty in the energy scale of VERITAS is 15\% to 20\%~\cite{2015ICRC...34..771P}.
Differential \g-ray flux upper limits are listed in Table~\ref{vts_uls} at 95\% C.L. for observations obtained within two weeks of the neutrino alert. These observations, with the addition of historical observations of the blazar TXS~0506+056 performed by VERITAS prior to the detection of IceCube-170922A, were used to calculate light-curve integral flux upper limits above a threshold of 175~GeV which are listed in Table~\ref{vts_lc}. For observation periods where the VERITAS energy threshold was higher than 175~GeV, the upper limits were scaled to this value by conservatively assuming a photon spectral index of $-4.3$ based on a 1$\sigma$ deviation of the index from the MAGIC data ($-3.9\pm$0.4).

VERITAS observations of IceCube-170922A were performed as part of the VERITAS neutrino follow-up program~\cite{2016chep.confE..85S}. Prompt follow-up observations performed by VERITAS under this program in response to alerts prior to the IceCube-170922A are included in Table~\ref{tab:alerts}.

\bigskip

\begin{table}[ht]
\small
\centering
\caption{{\bf Flux upper limits from VERITAS for TXS~0506+056.}
Nightly \g-ray flux upper limits at 95\% confidence level from VERITAS above an energy threshold of 175 GeV, assuming an $E^{-3.9}$ energy spectrum.}
\begin{tabular}{ccc}
\hline
MJD & Time window (half width) & Flux $>$ 175 GeV \\
$[{\rm days}]$ & $[{\rm days}]$ & [ph cm$^{-2}$ s$^{-1}$] \\
\hline
 57685.4392  &  $\pm$0.0104 &  $< 6.8 \times 10^{-12}$ \\
 57686.4500  &  $\pm$0.0200 &  $< 5.7 \times 10^{-12}$ \\
 57786.1544  &  $\pm$0.0142 &  $< 1.1 \times 10^{-11}$ \\
 58019.3971  &  $\pm$0.0124 &  $< 2.1 \times 10^{-10}$ \\
 58024.4380  &  $\pm$0.0653 &  $< 1.4 \times 10^{-11}$ \\
 58025.3932  &  $\pm$0.0219 &  $< 5.2 \times 10^{-11}$ \\
 58026.4399  &  $\pm$0.0211 &  $< 1.1 \times 10^{-11}$ \\
\hline
\end{tabular}
\label{vts_lc}
\end{table}

\begin{table}[ht]
\small
\centering
\caption{{\bf VERITAS differential \g-ray flux upper limits for TXS~0506+056.}
Differential \g-ray flux upper limits ($f_\gamma$) derived from VERITAS observations of the TXS~0506+056 blazar position. $E_{\rm min}~$ and $E_{\rm max}$
define the energy range over which the differential flux upper limit is derived and are based on observations obtained within 2 weeks of the neutrino alert.}
\begin{tabular}{ccc}
\hline
$E_{\mathrm{min}}$ & $E_{\mathrm{max}}$ & $f_\gamma$ \\
$[{\rm TeV}]$	& $[{\rm TeV}]$ & [cm$^{-2}$ s$^{-1}$ TeV$^{-1}$] \\
\hline
  0.141  &  0.316 &  $< 5.4 \times 10^{-11}$ \\
  0.316  &  0.708 &  $< 6.4 \times 10^{-12}$ \\
  0.708  &  1.585 &  $< 5.3 \times 10^{-13}$ \\
  1.585  &  3.548 &  $< 8.0 \times 10^{-14}$ \\
\hline
\end{tabular}
\label{vts_uls}
\end{table}

\begin{table}[ht]
\small
\centering
\caption{{\bf Neutrino follow-up observations performed by VERITAS.}}
\begin{tabular}{ccccc}
\hline
IceCube & UTC Date & Obs delay & Exposure  & VERITAS \\
Alert ID & $  $ & $[\rm {hr}]$  & $[\rm {hr}]$ & publication\\
\hline
 IceCube-160427A  &  27 April 2016 & 0.05 & 3.15 & \cite{GCN19377}  \\
 IceCube-161103A  &  3 November 2016 & 0.06 & 1.5 & -  \\
 IceCube-170321A  &  21 March 2017 & 19.3 & 0.5 & -  \\
 IceCube-170922A &  22 September 2017 & 12.2 & 5.5 & \cite{2017ATel10833....1M} \\
\hline
\end{tabular}
\label{tab:alerts}
\end{table}

\paragraph*{HAWC}
\addcontentsline{toc}{subsection}{\protect\numberline{}HAWC}%
The High-Altitude Water Cherenkov (HAWC) \g-ray observatory \cite{CrabHAWC} has a very wide field of view ($\sim$ 2~sr) and operates with $>$95\% uptime, enabling it to survey 2/3 of the sky above $\sim$1~TeV. The IceCube-170922A location was not in the field of view of HAWC at the time of the event. Three time periods were searched for VHE \g-rays with no evidence for a source in any of the studies. For all these searches, the 95\% C.L. upper limits on the flux above 1~TeV assume a spectral index of $-3.9$. The 1~TeV threshold for flux limits was chosen so that the limit depends very weakly on the spectral index.

First, a time integrated search at the location of TXS~0506+056 was made using archival data from 26 November 2014 to 27 August 2017. The upper limit on the flux is 1.6$\times 10^{-13}$~cm$^{-2}$~s$^{-1}$. Second, the transits of TXS~0506+056, in HAWC, right before and right after the time stamp of IceCube-170922A (22 September 2017, from 08:37:15 to 14:29:45 UTC and 23 September 2017, from 08:33:19 to 14:25:49 UTC) were used. The upper limit on the flux is $3.6\times 10^{-12}$~cm$^{-2}$~s$^{-1}$. Finally, data from 9 September 2017 09:28:22 to 19 September 2017 14:41:33 UTC and from 21 September 2017 08:41:11 to 6 October 2017 13:34:43 UTC was used (the time gap was due to a power outage after Mexico's earthquake on 19 September 2017), roughly coinciding with the \Fermi-LAT reported flare~\cite{2017Atel_Fermi}, results in an upper limit of the flux of $2.1\times 10^{-12}$~cm$^{-2}$~s$^{-1}$. Quasi-differential upper limits on $E^2 dN/dE$ using HAWC data are presented in Figure~\ref{fig:txs0506_sed} using the method described in \cite{2017A&A...607A.115I}.

\subsection*{Radio, optical and X-ray observations}
\addcontentsline{toc}{section}{\protect\numberline{}Radio, optical and X-ray observations}%

\paragraph*{VLA}
\addcontentsline{toc}{subsection}{\protect\numberline{}VLA}%
The Karl G. Jansky Very Large Array (VLA)~\cite{2011ApJ...739L...1P} was used to obtain radio frequency observations 
of the blazar TXS~0506$+$056, following its identification as the potential astrophysical origin of IceCube-170922A~\cite{GCN21916}. The VLA observations were taken over six epochs between 5 October 2017 and 21 November 2017, in {\it S} ($2-4\,{\rm GHz}$), {\it C} ($4-8\,{\rm  GHz}$) and {\it X} ($8-12\,{\rm GHz}$) bands. The array was
split into 3 sub-arrays, with 8 antennas observing at {\it C} band, 9 antennas at {\it X} band, and 10 antennas observing at {\it S} band, to simultaneously sample the source flux density across the three receiver sets. The antennas for each sub-array were selected so that all sub-arrays had similar beam patterns. A total of 10.8 minutes on target were acquired in each band, per epoch, cycling continuously between the calibrator (1 min) and target (5.4 min).
All observations were made with the 8-bit samplers, using 2 base-bands with 8 spectral windows of 64 2 MHz channels each, giving a total bandwidth of 1.024 GHz per base-band. The flagging, calibration, and imaging were carried out within the Common Astronomy Software Application package
(\textsc{casa}, v5.1.1;\cite{mcmullin2007}) using standard procedures.

For all sub-arrays, 3C~138~(QSO~J0521$+$166) was applied as the
flux calibrator, and QSO~J0502$+$0609 as the phase calibrator. When imaging, a natural weighting scheme was used to maximize
sensitivity. A phase-only self-calibration was performed on the data (with 10 second solution intervals) to correct for phase de-correlation of the unresolved emission.  No
self-calibration was performed when the preliminary flux densities of the first 4 epochs were reported in \cite{2017ATel10861....1T}.
As expected, phase de-correlation was strongest at higher frequencies, and on 6 October 2017, that was measured to
have significantly stronger tropospheric contribution to the interferometric phase.

TXS~0506+056 was detected significantly in all bands/epochs. Observation times and flux densities for TXS~0506+056 are shown in Table~\ref{table:vla_obs}, where a point source was fitted in the image plane (with the \textsc{casa} imfit task) to obtain each of these measurements. The reported uncertainties on the flux density do not include the $\sim5\%$ systematic uncertainty on the absolute flux scale calibration. This uncertainty should be included when comparing these flux density measurements to those with other facilities. 
Flux densities and the measured $\nu^{\sim0.2}$ spectra were relatively constant from 5 - 12 October 2017 (epochs 1–4). The source brightened slightly above 4 GHz on 24 October 2017. The final data (on 21 November 2017) indicate a ($\sim$20\%) brighter source at frequencies below 6~GHz, and spectral steepening at higher frequencies ($\nu^{\sim0.2}$ at lower frequencies and $\nu^{\sim-0.1}$ at higher frequencies).
This peak may be due to synchrotron self-absorption. An injection of energy ({\it e.g.}, jet ejecta) that moves downstream and reaches the radio photosphere for $\nu \sim 10$ GHz by 21 November 2017, could provide a spectral turnover.

\newcommand{\phn}{\phantom{0}}%
\begin{table*}[ht]
\centering
\caption{{\bf VLA Radio Frequency Flux Densities of TXS~0506+056.} A $\sim 5\%$ systematic uncertainty in the absolute flux scale should be included when comparing these to flux densities measured with other facilities.}
\begin{tabular}{ ccccc }
\hline\hline
{Epoch}&{MJD}&{Sub-array}&{Frequency}&{Flux Density}\\
{}&{}&{Receiver Band}&{(GHz)}&{mJy}\\\hline
1&$58031.6429\pm0.0104$&S&\phn2.50&$519.7\pm1.3$\\
&&S&\phn3.50&$540.4\pm0.9$\\
&&C&\phn5.25&$565.3\pm1.0$\\
&&C&\phn7.45&$624.9\pm1.1$\\
&&X&\phn9.00&$663.2\pm1.5$\\
&&X&11.00&$695.9\pm1.4$\\
2&$58032.3724\pm0.0104$&S&\phn2.50&$522.8\pm0.8$\\
&&S&\phn3.50&$543.6\pm0.5$\\
&&C&\phn5.25&$569.8\pm1.0$\\
&&C&\phn7.45&$640.2\pm1.3$\\
&&X&\phn9.00&$662.8\pm3.5$\\
&&X&11.00&$725.7\pm6.1$\\
3&$58035.5662\pm0.0104$&S&2.50&$507.5\pm0.9$\\
&&S&\phn3.50&$529.4\pm0.6$\\
&&C&\phn5.25&$563.3\pm1.1$\\
&&C&\phn7.45&$625.9\pm1.5$\\
&&X&\phn9.00&$650.1\pm1.2$\\
&&X&11.00&$670.9\pm1.1$\\
4&$58038.3585\pm0.0104$&S&2.50&$520.4\pm0.8$\\
&&S&\phn3.50&$535.0\pm0.6$\\
&&C&\phn5.25&$571.3\pm1.0$\\
&&C&\phn7.45&$631.9\pm1.0$\\
&&X&\phn9.00&$661.9\pm1.1$\\
&&X&11.00&$722.8\pm1.1$\\
5&$58050.3048\pm0.0104$&S&\phn2.50&$511.4\pm1.6$\\
&&S&\phn3.50&$549.7\pm0.8$\\
&&C&\phn5.25&$607.2\pm1.4$\\
&&C&\phn7.45&$699.4\pm1.6$\\
&&X&\phn9.00&$723.8\pm1.7$\\
&&X&11.00&$753.0\pm1.6$\\
6&$58078.5534\pm0.0104$&S&\phn2.50&$606.0\pm1.6$\\
&&S&\phn3.50&$658.4\pm1.3$\\
&&C&\phn5.25&$696.2\pm1.2$\\
&&C&\phn7.45&$669.3\pm1.5$\\
&&X&\phn9.00&$667.0\pm1.4$\\
&&X&11.00&$646.2\pm1.0$\\\hline
\end{tabular}\\
\label{table:vla_obs}
\end{table*}

\paragraph*{ASAS-SN}
\addcontentsline{toc}{subsection}{\protect\numberline{}ASAS-SN}%

The All-Sky Automated Survey for Supernovae (ASAS-SN)
consists of two units --- one at Haleakala, Hawaii and one at Cerro Tololo in Chile --- each comprising four robotic 14\,cm telescopes. ASAS-SN has been monitoring the visible sky to $\sim 17$~mag in the {\it V} band on a $2$--$3$~day cadence~\cite{2014ApJ...788...48S}.

An initially extracted light curve spanning from August 2013 to October 2017 from the ASAS-SN Sky Patrol~\cite{2017PASP..129j4502K} interface indicated that TXS~0506$+$056 had brightened by $\Delta V \sim 0.5$~mag over the 50 days prior to the neutrino event~\cite{2017ATel10794....1F}.
While the source shows significant variability, the recent data indicate this is the brightest this object has been in several years. This source was observed at higher cadence compared to the regular survey for a few days after the neutrino detection thanks to automated ASAS-SN target-of-opportunity observations triggered by the public IceCube alerts.

The ASAS-SN {\it V} band light curve is shown in Figure~\ref{fig:CombinedLC} and is extracted using aperture photometry on the difference images and combining the multiple images obtained at each epoch. Proximity to the Sun prevented observations from April to July 2017. The source had a relatively steady flux of $V \sim 14.5$~mag in the previous observing season (August 2016 to April 2017) and brightened from a minimum of $V \simeq 15.0$~mag in the season before that (July 2015 to March 2016). ASAS-SN images have $\sim 15$'' FWHM PSF. There is a modest dilution of the variability amplitude through the contributions of a nearby source to the photometry aperture.

\paragraph*{Kanata/HONIR} 
\addcontentsline{toc}{subsection}{\protect\numberline{}Kanata/HONIR}%
TXS~0506+056 was monitored in the imaging mode with the Hiroshima Optical and
Near-InfraRed camera (HONIR)~\cite{2014SPIE.9147E..4OA} installed at the Cassegrain focus
of the 1.5-m Kanata telescope at the Higashi-Hiroshima Observatory, starting $\sim 20$ hours after the IceCube-170922A alert.
Polarimetric observation with HONIR was also performed since 30 September 2017, revealing that TXS 0506+056 was highly polarized ($\sim 7$\% in {\it R} band). The magnitude was measured by relative photometry with respect to the nearby reference stars listed in
the AAVSO Photometric All-Sky Survey~\cite{2009AAS...21440702H}. For polarimetry of
TXS~0506+056, strongly-polarized standard stars (BD +64 106 and
BD +59 389; \cite{1992AJ....104.1563S}), and several unpolarized standard stars, were
observed for calibration. The
instrumental polarization for HONIR was confirmed to be negligibly small
($\lesssim 0.1$\%), and no correction was applied.
The foreground Galactic extinction ($E_{B-V}=0.096$,\cite{2011ApJ...737..103S}) indicates that the interstellar polarization is $p_{R}\leq 0.9$\% \cite{1975ApJ...196..261S}, suggesting the observed
large polarization is predominantly intrinsic to
the blazar. 
Possible blazar candidates were selected from flat spectrum radio sources of $\alpha > -0.5$ (where $\alpha$ is a spectral index defined as $F_{\nu} \propto \nu^{\alpha}$) from 0.15~GHz TGSS~\cite{2017A&A...598A..78I} and 1.4~GHz NVSS~\cite{1998AJ....115.1693C} catalogs. Four flat-radio-spectrum objects (including TXS 0506+056) found within the IceCube-170922A error region were examined. Visual inspection of differenced images taken on 23 September 2017 and 24 September 2017 showed a clear fading of blazar TXS 0506+056~\cite{2017ATel10844....1Y}.

\paragraph*{Kiso/KWFC} 
\addcontentsline{toc}{subsection}{\protect\numberline{}Kiso/KWFC}%
Monitoring observations of TXS~0506+056 in optical $g$, $r$, and $i$ bands commenced after the neutrino alert with the Kiso Wide Field Camera~\cite{2012SPIE.8446E..6LS} attached to the 1.05~m Kiso Schmidt telescope. The point spread function sizes are 4-6 arcsec FWHM, well separated from a nearby bright star in the KWFC images. The data were reduced in the same manner as that for a previous supernova survey~\cite{2014PASJ...66..114M}. Compared with previous observations~\cite{2016arXiv161205560C}, the flux of the object increased by approximately 1.0 mag and remained in a bright phase until at least November 2017.

\begin{figure*}
\centering
\includegraphics[width=0.7\textwidth]{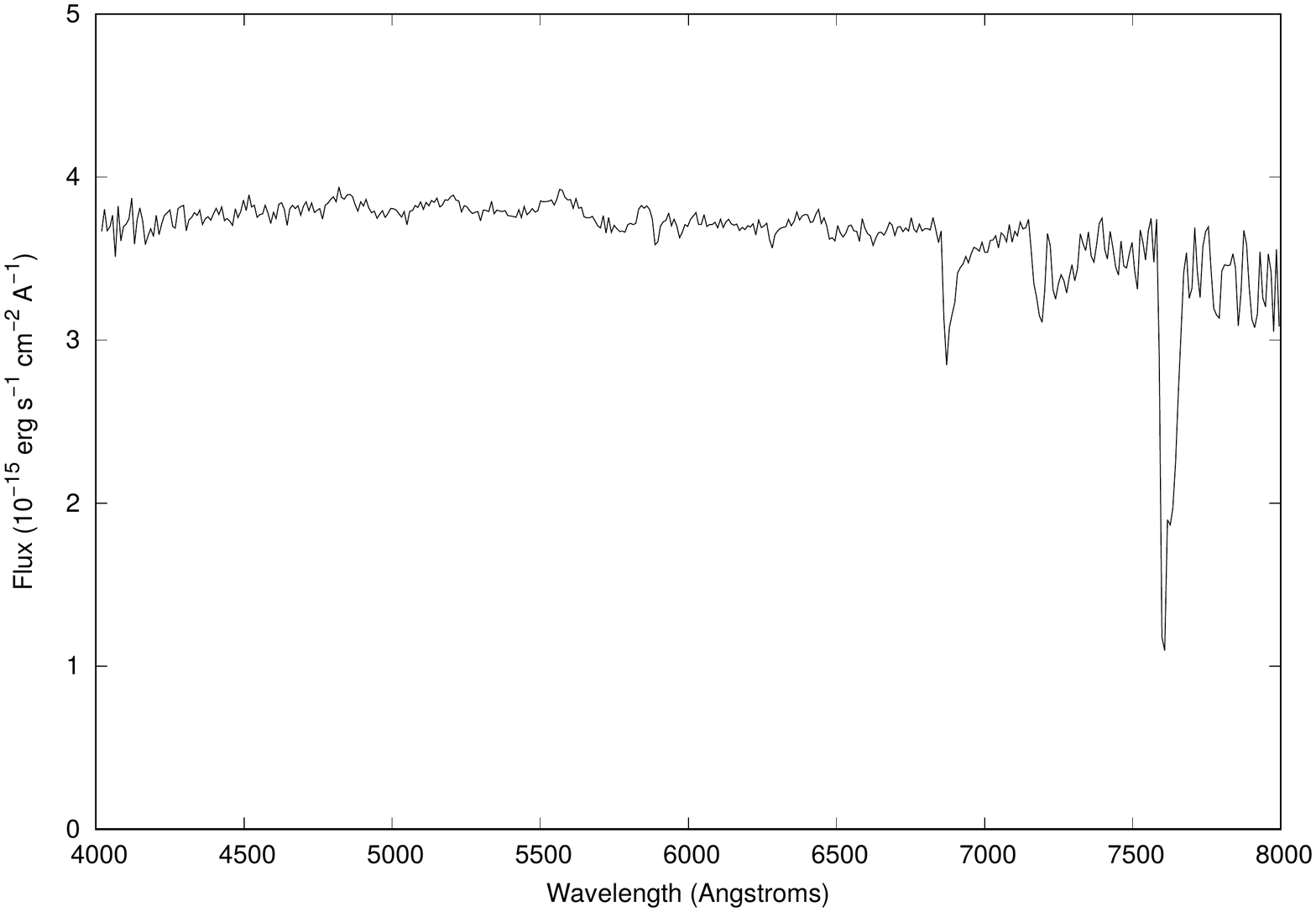}
 \caption{{\bf  Liverpool Telescope measured spectrum of TXS~0506+056.}  Spectrum of TXS 0506+056 obtained with the low resolution (R$\approx$350) spectrograph SPRAT on the Liverpool Telescope on 29 September 2017. The spectrum shows a smooth continuum typical of a BL Lac object, with absorption features that are telluric or attributable to Galactic interstellar Na~\textsc{I} absorption.}
 \label{fig_liverpoolspec}
\end{figure*}

\paragraph*{Liverpool Telescope} 
\addcontentsline{toc}{subsection}{\protect\numberline{}Liverpool Telescope}%
Low resolution (resolving power $\sim350$, $4000-8000$ {\AA}) optical spectra of TXS~0506+056 were obtained with the SPRAT spectrograph of the 2.0\,m Liverpool Telescope~\cite{2004SPIE.5489..679S} on 29 September 2017 02:31 UTC and 30 September 2017 02:15 UTC. The spectra, shown in Figure~\ref{fig_liverpoolspec}, showed no sign of variation and are typical of a BL Lac object, showing a smooth continuum. The only feature seen is attributable to Galactic interstellar Na~\textsc{I} absorption. No redshift measurement is possible from the optical spectra. Both spectra have a flat spectral energy distribution with $F_\lambda \sim 4\times10^{-15}$ ergs s$^{-1}$cm$^{-2}${\AA}$^{-1}$, which is bluer than the (similarly featureless) spectrum presented in~\cite{2003AJ....125..572H}, indicating a likely ``bluer-when-brighter" behavior.

\begin{figure*}
\centering
\includegraphics[width=0.9\textwidth]{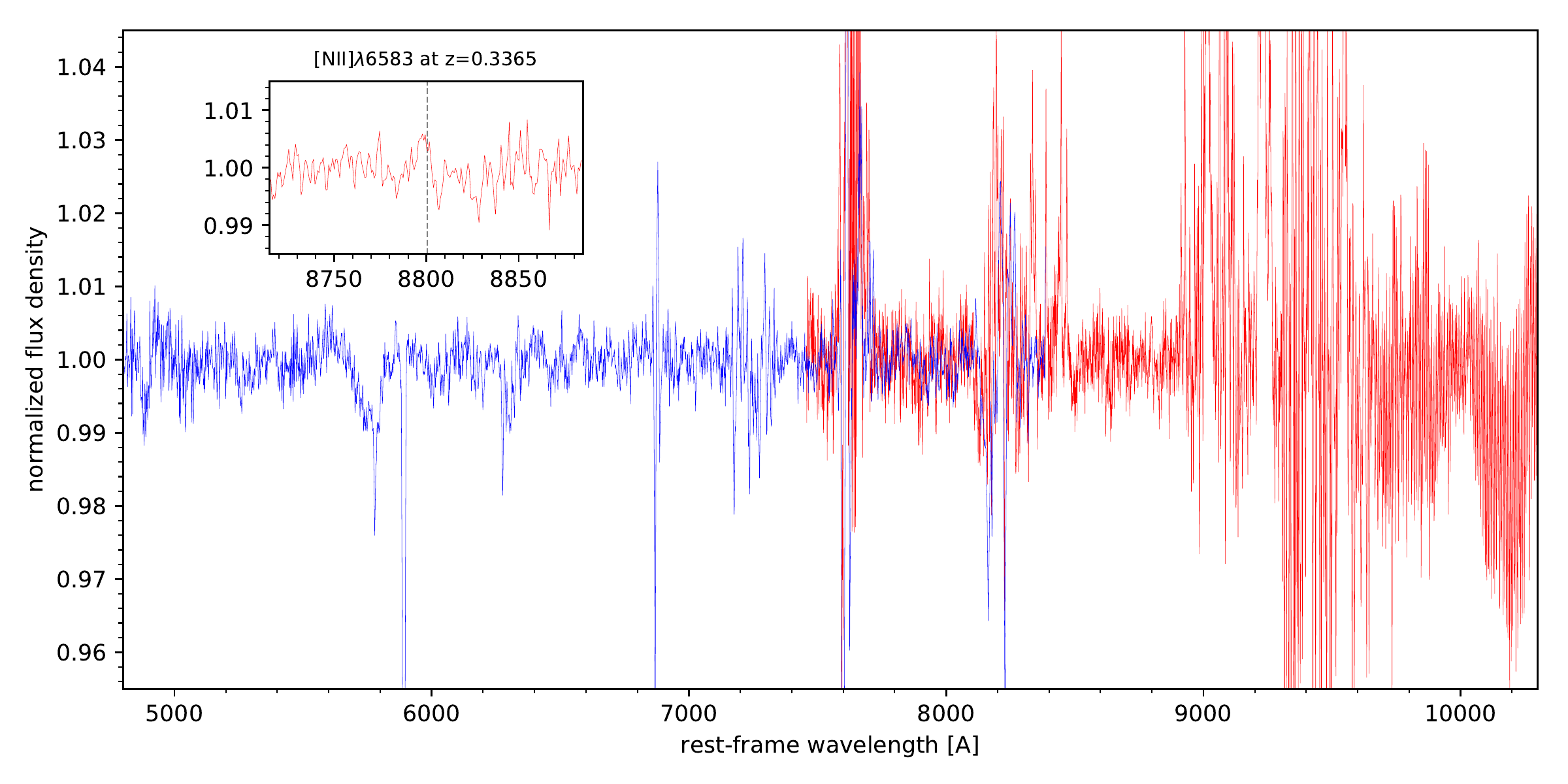}
\caption{{\bf Subaru/FOCAS spectra of TXS~0506+056.} Normalized spectra taken with FOCAS on the 8.2-m Subaru telescope on 30 September 2017 (blue) and
1 October 2017 (red), in two different settings of the grism and order-sort filters. Note that some atmospheric absorption effects
remain. The [NII] line detected by Paiano et al.~\cite{Paiano:2018qeq} is marginally detected as shown in the inset figure. }
 \label{fig_subaruspec}
\end{figure*}

\paragraph*{Subaru/FOCAS} 
\addcontentsline{toc}{subsection}{\protect\numberline{}Subaru/FOCAS}%
Low resolution (resolving power $\sim400$ in 4700-8200~\AA~ and $\sim1200$ in 7500-10500~\AA) optical spectra were obtained with the Faint Object Camera and Spectrograph~\cite{2002PASJ...54..819K} on the 8.2-m Subaru telescope on 30 September 2017 and 1 October 2017, respectively, and are shown in Figure~\ref{fig_subaruspec}. The data was reduced with IRAF software in the standard manner~\cite{irafref}.
The signal-to-noise ratios are roughly 350 per pixel. The spectra are almost featureless smooth continua over the entire wavelength range except for a weak emission line is marginally detected around 8,800A, corresponding [NII] detected in \cite{Paiano:2018qeq}.

\newcommand{\icnu}{\mbox{IceCube-170922A}}
\newcommand{\qsotxs}{\mbox{TXS~0506+056}}
\newcommand{\fermi}{\textit{Fermi}}
\newcommand{\egamma}{\mbox{$\varepsilon_\gamma$}}

\newcommand{\dg}{deg}
\newcommand{\ergcms}{\mbox{erg cm$^{-2}$ s$^{-1}$}}
\def\arcmin{\hbox{$^\prime$}}
\newcommand{\ctksec}{\mbox{ct ks$^{-1}$}}
\newcommand{\etal}{\mbox{et al.}}
\newcommand{\xray}{\mbox{X-ray}}
\newcommand{\percmsq}{\mbox{cm$^{-2}$}}

\paragraph*{ \swift$~$ and \nustar}
\addcontentsline{toc}{subsection}{\protect\numberline{}\swift$~$ and \nustar}%

\swift\ carried out rapid-response follow-up observations of \icnu\ as a mosaic of 19~pointings beginning 3.25~hours after the neutrino detection, lasting 22.5~hours, and accumulating approximately 800~s exposure per pointing.
The tiled \swift\ XRT observations together cover a roughly circular region
centered on RA, Dec (J2000) = (77.2866\,deg, +5.7537\,deg), with radius
of approximately 0.8\,deg and sky area 2.1\,deg$^2$.
XRT data was analyzed automatically as data was received at the University of Leicester, via the reduction routines described in~\cite{Evans:2008wp,Evans:2013daa}.
Nine sources were detected in the covered region down to a typical achieved depth of $3.8\times 10^{-13}$\,\ergcms\ (0.3\,keV -- 10.0\,keV).
All of the detected sources were identified as counterparts to known and cataloged stars, \xray\ sources, or radio sources~\cite{GCN21930}.
Source~2 from these observations, located 0.077\,deg from the center of the neutrino localization, was identified as the likely \xray\ counterpart to \qsotxs.

Following the \Fermi-LAT report that \qsotxs\ was in an enhanced GeV-flaring state, a \swift\ monitoring campaign was initiated~\cite{2017ATel10792....1K} and a single \nustar\ observation~\cite{2017ATel10845....1F} was requested.
\swift\ monitoring observations began on 27 September 2017 with 12 epochs (and 24.7~ks total exposure time) completed by 23 October 2017 (Table~\ref{table:xrt-monitoring}).
\nustar\ observations over 02:23 to 17:48 UTC on 29 September 2017 yielded 23.9~ks (24.5~ks) exposure in the A (B) units, respectively, after processing with \nustar\ standard software tools~\cite{1995ASPC...77..367B} (\texttt{SAAMODE=strict}). With count rates of 21.3\,\ctksec\ (20.8\,\ctksec) in the A (B) units, \qsotxs\ is well detected in these data.

For joint analysis purposes, the \swift\ XRT data from the 27 September 2017 and 30 September 2017 monitoring observations were selected; processing of these data occurred with the online tools of the UK \swift\ Science Data Centre~\cite{Evans:2008wp}, yielding a 6.9~ks exposure, with the source exhibiting a count rate of 88\,\ctksec.

\begin{figure*}
\begin{center}
\includegraphics[scale=0.5]{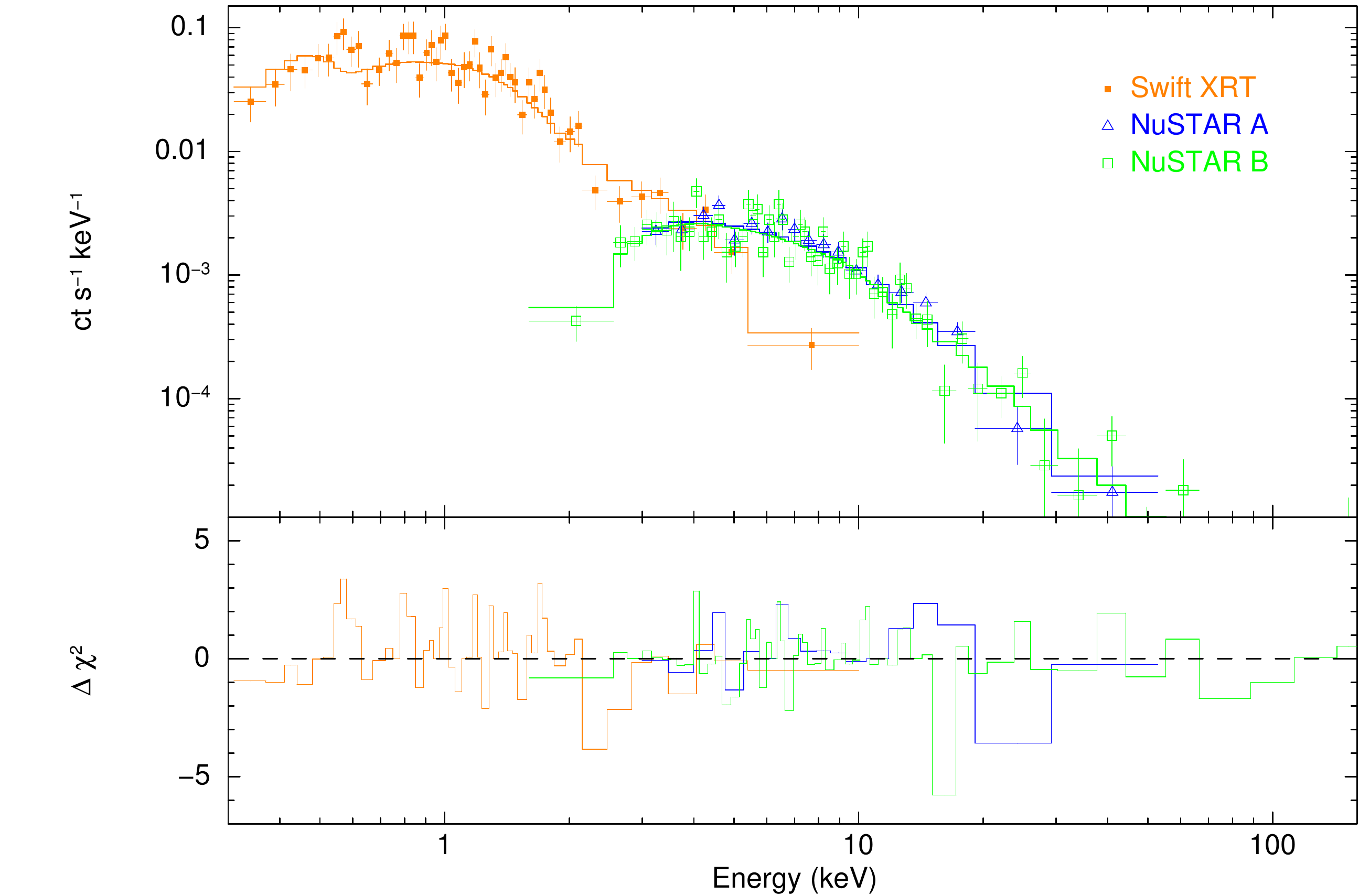}
\caption{{\bf \swift-XRT and \nustar\ observations of \qsotxs.}
The spectrum of TXS 0506+056 from a joint fit to Swift-XRT data with the \nustar\ spectrum of \qsotxs. \swift-XRT data are shown with orange markers and the two \nustar\ units are shown with blue and green markers.}
\label{fig:xrt-nustar}
\end{center}
\end{figure*}

The resulting \qsotxs\ spectrum over 0.3\,keV -- 100\,keV is adequately fit with a double power-law spectral model with the galactic hydrogen column density($N_{\rm H}$) fixed at the expected Galactic value, $N_{\rm H}=1.11\times 10^{21}$\,\percmsq\ (Fig. ~\ref{fig:xrt-nustar}).

The soft power-law component has the photon index $\Gamma_s=-2.78\pm0.30$ and yields a flux of $(1.78\pm0.41)\times10^{-12}$\,\ergcms\ over 0.3\,keV -- 10\,keV, while the hard power-law component has the photon index $\Gamma_h=-1.43\pm0.25$ and yields a flux of $(4.7\pm2.4)\times10^{-12}$\,\ergcms\ over 3\,keV -- 100\,keV.
The hard power-law component that dominates over the \nustar\ bandpass extrapolates to $\nu F_\nu = 5.8\times 10^{-11}$\,\ergcms\ at $E\gamma=20$\,MeV, consistent with the flux observed over the \Fermi-LAT bandpass (0.1\,GeV -- 100\,GeV) at that epoch.

To characterize the \xray\ flux and spectral variability of \qsotxs, we performed a power-law fit to each individual \swift\ XRT observation (Table~\ref{table:xrt-monitoring}), as well as to the summed spectrum from all listed epochs.
The 14 October 2017 observation is excluded from the spectral analysis due to low exposure time. The summed spectrum (24.7~ks total exposure) is adequately fitted with a single power-law spectral model having $N_{\rm H}=1.11\times 10^{21}$\,\percmsq, resulting in a photon index $\Gamma=-2.46 \pm 0.06$ and mean flux of $3.06\times 10^{-12}$\,\ergcms (0.3\,keV -- 10.0\,keV). This source has been observed on multiple previous occasions with the \swift\ XRT~\cite{Evans:2013daa}. In past observations, \qsotxs\ exhibits a typical count rate of 40\,\ctksec, with one observation at approximately 90 \,\ctksec. The source was therefore in an active \xray\ flaring state by comparison to historical \xray\ measurements (cf. Figure ~\ref{fig:CombinedLC}).

Individual monitoring observations show evidence of spectral variability; photon indices and \xray\ flux measurements for each epoch are provided in Table~\ref{table:xrt-monitoring}, and the variations in photon index are shown in Figure ~\ref{fig:CombinedLC}. The $\chi^{2}$ statistic for photon index variations (compared to a fixed $\Gamma=-2.46$ from the summed spectrum) is $33.91$ for 11 degrees of freedom ($p=3.7\times 10^{-4}$).

\begin{sidewaystable}[ht]
\begin{center}
\caption{{\bf \swift\ XRT Monitoring Campaigns of \qsotxs.} $R_{\rm X}$ and $F_{{\rm X},-12}$ indicate count rate and energy flux (0.3-10 keV), in units of \ctksec\ and $10^{-12}$\,\ergcms, respectively. Uncertainties are 90\%-confidence regions. The * indicates that UT end was during the following day. The energy flux reported here is not corrected for Galactic absorption.}
\def\arraystretch{1.5}
\begin{tabular}{ccccccc}
\hline
Epoch & UT start & UT end & Exposure [ks] & $R_{{\rm X}}$ [\ctksec] & Photon Index & $F_{{\rm X},-12}$ [$10^{-12}$\,\ergcms] \\
\hline
2017-09-23 & 00:09:16 & 22:24:03 & 0.8 &	$65.8 \pm10.1$  & $1.83^{+0.43}_{-0.42}$ & $2.33^{+1.07}_{-0.72}$\\
2017-09-27 & 18:51:08 & 22:04:26 & 4.9 &	$121.2 \pm 5.3$ & $2.43 \pm 0.12$ & $3.51^{+0.32}_{-0.29}$\\
2017-09-30 & 04:27:57 & 06:24:00 & 2.0 &	$66.2 \pm 7.8$ & $2.30 \pm 0.33$ & $1.55^{+0.43}_{-0.33}$\\
2017-10-02	& 15:04:30 & 15:41:32 & 2.0 & $117.3 \pm 8.2$ & $2.73 \pm 0.20$ & $2.92^{+0.40}_{-0.35}$\\
2017-10-03 & 13:38:57 & 18:41:00 & 1.1 & $182.4 \pm 14.1$	& $2.46^{+0.22}_{-0.21}$ & $4.96^{+0.80}_{-0.70}$\\
2017-10-04	& 16:42:57 & 18:37:00 & 1.2 & $186 \pm 16.1$	& $2.82^{+0.26}_{-0.25}$ & $4.41^{+0.74}_{-0.65}$\\
2017-10-05	& 16:32:57 & 21:32:00 & 2.3 & $255.1 \pm 11.7$	& $2.64 \pm 0.13$ & $6.47^{+0.57}_{-0.53}$\\
2017-10-06	& 13:16:57 & 05:48:00* & 2.1 & $90.1 \pm 7.1$ & $2.36 \pm 0.21$ & $2.56^{+0.44}_{-0.37}$\\
2017-10-08	& 09:51:57 & 11:34:00 & 1.9 & $108.1 \pm 8.1$	& $2.53 \pm 0.21$ & $2.87^{+0.44}_{-0.38}$\\
2017-10-16 & 14:21:57 & 22:30:00 & 2.2 & $61.9 \pm 5.8$ & $2.0 \pm 0.27$ & $2.20^{+0.59}_{-0.44}$\\
2017-10-18 & 01:23:57 & 04:48:00 & 1.8 & $52.6 \pm 6.0$ & $2.10 \pm 0.31$ & $1.70^{+0.49}_{-0.37}$\\
2017-10-20 & 13:47:57	& 20:23:00* & 2.3 & $49.6 \pm 5.1$ & $2.11 \pm 0.28$ & $1.61^{+0.41}_{-0.32}$\\
\hline
\end{tabular}
\label{table:xrt-monitoring}
\end{center}
\end{sidewaystable}

\newcommand{\ecs}[1]{#1~\mbox{erg cm$^{-2}$ s$^{-1}$}}
\paragraph*{\it INTEGRAL}
\addcontentsline{toc}{subsection}{\protect\numberline{}{\it INTEGRAL}}%
The {\it INTEGRAL} observatory \cite{Winkler:2003nn} has surveyed the sky in hard
X-rays and soft \g-rays at energies above 20~keV since October 2002.
At the time of the IceCube-170922A detection, {\it INTEGRAL} was performing a slew
between two pointings, and the sensitivity to emission from the
IceCube neutrino direction depends on time. Combining data from the Spectrometer on {\it INTEGRAL} - Anti-Coincidence Shield (SPI-ACS) and Imager on Board the {\it INTEGRAL} Satellite - Veto (IBIS/Veto) instruments, we set a limit on the 8-second peak flux at any time within $\pm$30~minutes from the time of the alert at a level of \ecs{10$^{-7}$}, assuming a power-law spectrum with a slope of -2~\cite{GCN21917}.

The location of IceCube-170922A was serendipitously in the field of view of {\it INTEGRAL} from 30 September 2017, 05:36:04 UTC (MJD 58026.23) to 24 October 2017, 16:20:25 UTC (MJD 58050.68). Due to the large off-axis angle, the resulting effective exposure was only 32 ks. In the combined mosaicked images of {\it INTEGRAL} Soft Gamma-Ray Imager data we did not detect the source, and set an upper limit (3$\sigma$ C.L.) on the average flux from the position of TXS~0506+056 of
\ecs{7.1$\times$10$^{-11}$} in the 20\,keV --\,80\,keV energy range and \ecs{9.8$\times$10$^{-11}$}
in the 80\,keV --\,250\,keV energy range.

\subsection*{Neutrino-blazar coincidence analysis}
\addcontentsline{toc}{section}{\protect\numberline{}Neutrino-blazar coincidence analysis}%
In order to calculate the chance probability of a coincidence between a neutrino alert, such as IceCube-170922A, and a flaring blazar, several hypothesis tests have been performed covering a range of assumptions on the spatial and temporal signal
distribution and neutrino emission scenarios. For each hypothesis we create a test statistic (TS) that we use in a likelihood ratio test to compare the signal hypothesis to the null hypothesis.
In each case our null hypothesis assumes no correlation between a cataloged \g-ray source and high-energy neutrino events (including atmospheric neutrinos and misidentified muons, and the astrophysical neutrinos).
The signal hypothesis assumes that neutrino events originate from cataloged \Fermi-LAT blazars, given a particular model for the correlation between the neutrino and \g-ray emission.

As a common framework for all the analyses, we start with an unbinned likelihood function defined in a similar way to previous IceCube point source analyses~\cite{Aartsen:2016oji,Aartsen:2016lir}:

\begin{equation}
\mathcal{L} = \prod_{i}^N \left( \frac{n_s}{N} \mathcal{S} + (1-\frac{n_s}{N}) \mathcal{B} \right),
\end{equation}

\noindent with signal and background probability density functions (PDFs) denoted as $\mathcal{S}$ and $\mathcal{B}$, respectively. $N$ is the total number of events and $n_s$ the number of signal events.
Additionally, there is one constrained nuisance parameter included in the likelihood for the \g-ray energy flux (or flux ratio) normalization of each source using the method from~\cite{Ackermann:2015zua}, treating the flux error as a normal distribution. Including the nuisance factor does not have a significant influence on the results. In the simple case considered here, only a single event enters the analysis, i.e. $N=1$.

We define a test statistic of the form

\begin{equation}
TS = 2 \log \frac{\mathcal{L}(n_s=1)}{\mathcal{L}(n_s=0)}  = 2 \log \frac{\mathcal{S}}{\mathcal{B}}.
\label{eq:TS_def}
\end{equation}

\noindent In this definition the likelihood ratio test reduces to a test between two fixed alternate hypotheses and TS can take negative values for background-like events. The signal PDF consists of three independent parts, a spatial factor, a flux weight factor, and a factor for the detector acceptance:

\begin{equation}
\mathcal{S}(\vec{x},t) =  \sum_{s} \frac{1}{2\pi\sigma^2}e^{-|\vec{x}_s-\vec{x}|^2/(2\sigma^2)} \, w_s(t) \, w_{\rm acc}(
\theta_s),
\end{equation}
where the sum runs over all 2257 extragalactic \Fermi-LAT sources, $s$.
Their light curves were constructed as described above for the analysis of the \Fermi-LAT light curves for TXS~0506+056, where 28-day wide bins are used to characterize the \g-ray activity at time $t$.  The term $w_{\rm acc}$ is the IceCube acceptance as a function of zenith angle (normalized over all zenith angles, $\theta_s$), assuming a neutrino signal spectral index $\gamma=-2.13$. This factor accounts for the zenith-dependent sensitivity of the IceCube detector. The function $w_s(t)$ derived from the \Fermi-LAT light curve, describes the model-dependent relation between the \g-ray emission and the expected neutrino flux from source $s$ as a function of time.

The leading factor inside the summation is the spatial weight accounting for the distance of a source at position, $\vec{x}_s$, to the reconstructed neutrino direction, $\vec{x}$, in terms of the reconstruction uncertainty $\sigma$ of the neutrino direction, which is found on a per-event basis~\cite{Aartsen:2016oji}. The uncertainty of the \g-ray source position is negligible compared to the neutrino angular uncertainty. Sources at large angular distances from the neutrino are assigned a negligible weight by the spatial factor, which models the IceCube point-spread function (PSF).
The ``signalness'' of a neutrino event, as mentioned in the main text, is a quantity constructed by the realtime system from the energy and zenith angle estimates, to rapidly allow an assessment of whether an event is a worthy target of opportunity. It does not enter into the likelihood.

The background PDF is described by the zenith acceptance, $\mathcal{P}_{BG}(\sin\theta)$, which is a probability density function describing the zenith distribution of the alert events that are due to background.

\begin{equation}
\mathcal{B}(\vec{x}) = \frac{\mathcal{P}_{BG}(\sin\theta)}{2\pi},
\end{equation}

where $\theta$ is the zenith angle of the reconstructed neutrino direction $\vec{x}$. To construct a background TS distribution we randomly draw neutrino events from an IceCube all-sky Monte Carlo sample containing muon-neutrinos and misidentified muons from air-showers and astrophysical neutrinos with energies according to the spectral shape presented in~\cite{Aartsen:2016xlq}.

The final p-value is then determined by calculating the fraction of background TS values larger than the measured one for IceCube-170922A.
Note that the overall normalization of $\mathcal{S}$ and $\mathcal{B}$ does not influence the final p-value, but only shifts the TS distribution.

As the production mechanisms of neutrinos and \g-rays
in astrophysical environments are poorly understood, three models connecting the \g-ray and the neutrino flux are considered for $w_s(t)$. All models are based on the assumption that at least part of the \g-ray emission is of hadronic origin. In all cases the extragalactic sources from the \Fermi-LAT catalog are used.

\begin{enumerate}
\item [Model 1:] The neutrino energy flux is proportional to the \g-ray energy flux of the source in the time bin where the neutrino arrives~\cite{Aartsen:2016lir}.
This is motivated by the fact that a similar amount of energy is expected to be channeled into the neutrino and \g-ray emission, if pion decays from pp or p\g~ interactions dominate at high energies. Alternatively, it can be relevant even if emission from electrons dominate, as long as protons and electrons are accelerated at a fixed power ratio.

In this case the weight is equal to the \g-ray energy flux defined as:
\begin{equation}
w_s(t) = \phi_E(t)
= \int^{\textrm{100~GeV}}_{\textrm{1~GeV}} E_\gamma \frac{d\phi_\gamma (t)}{dE_\gamma}dE\gamma,
\end{equation}
where $\phi_\gamma (t)$ is the photon flux from the \g-ray light curves, at time $t$.
The resulting pre-trial p-value is $2.1 \cdot 10^{-5}$, corresponding to a Gaussian equivalent one-sided probability of $4.1\sigma$.

\item [Model 2:]
The neutrino production and detection probability depend only on the relative flux change of the \g-ray source emission around the neutrino event time, $t$.
This prevents missing a correlation with \g-dim sources that may be much brighter in neutrinos than \g-rays, at the cost of some sensitivity to bright sources.

Here,
\begin{equation}
w_s(t) = \phi_{\gamma}(t)/
\left< \phi_{\gamma} \right>,
\end{equation}
where $\left< \phi_{\gamma} \right>$ is the time averaged \g-ray flux from the source. 
The resulting pre-trial p-value is $2.5 \cdot 10^{-5}$ ($4.1\sigma$).
\item [Model 3:]
The neutrino energy flux is proportional to the \g-ray energy flux predicted in the very-high-energy (VHE) \g-ray regime (100\,GeV -- 1\,TeV). This approach is triggered by the detection of VHE \g-ray emission by MAGIC.

If a similar amount of energy is channeled into neutrinos and \g-rays in the sources (as in Model 1) the energy flux is expected to be correlated with the neutrino energy flux. The VHE \g-ray emission is closer in energy to the observed neutrino and might therefore be a better indicator for high-energy particle acceleration.

As no unbiased survey of the sky exists at energies above $100$\,GeV, the 2257 extragalactic \Fermi-LAT sources were considered. The VHE spectral functional form was obtained through extrapolations of the spectrum measured by \Fermi-LAT in the energy range from 1\,GeV -- 100\,GeV over the entire 9.5-year \Fermi-LAT exposure.  The VHE spectral normalization
was scaled to match each monthly bin of the \Fermi-LAT light-curve.
Since any additional softening of the spectrum in the VHE energy band due to limitations in the acceleration capabilities of the source, limitations in the radiative efficiency, absorption within the source or in the extragalactic background light (EBL) would yield a lower flux, these extrapolations represent a conservative assumption.

The pre-trial p-value in this case is $4.9\cdot 10^{-5}$ ($3.9\sigma$). Including absorption by the extragalactic background light for the extrapolations does not change the p-value significantly.
Extrapolations to VHE energies are potentially uncertain for weaker, hard-spectrum sources with a spectral shape not well constrained in the high-energy band. For the sources within the Fermi 3FHL catalog, the
results obtained with the best fit from 9.5 years of Fermi-LAT data
({\it e.g.} power-law or log-parabola or power-law with exponential cutoff) were compared to extrapolations based on the power-law fit. Minimal impact was found on the weights in the energy band considered.
\end{enumerate}

A test was applied to assess the impact of the flux weight on the chance coincidence probability, and to quantify the probability of a simple spatial coincidence between a neutrino alert and a cataloged source. To achieve that, the flux weight was set to one for all cataloged sources ($w_s(t)= 1$). This choice implies that the intensities of the neutrino and \g-ray emission are not correlated for LAT catalog sources. In this case, the pre-trial p-value is reduced to $0.0017$ (2.9$\sigma$).
Another set of tests was applied to check the impact of the spatial factor in the likelihood description above, given that IceCube-170922A was found very close to the source (at a distance of 0.1 deg, much smaller than the 90\% angular error typical for IceCube through-going track events). The Gaussian PSF factor is replaced by 1 in this test for neutrino events within 0.5 deg of the source and by 0 otherwise. If the full PSF information is not used, the significance values drop by $0.4 \sigma - 0.5 \sigma$ for the three models described above.

Prior to 22 September 2017, IceCube had publicly issued 9 alerts. In addition 41 archival events (before April 2016) were inspected, which would have triggered alerts if the realtime system had been operational. %
Since no \Fermi-LAT source comparable in energy flux to TXS~0506+056 was found within the 90\% error region of any of the potential previous alerts, the global p-value, corrected for all trials, can be obtained from the pre-trial local p-value $p_{\rm global}  = 1 - (1 - p_{\rm local})^N$, where $N=51$ is the number of trials. For $p_{\rm local} \ll 1$ this simplifies to $p_{\rm global} \approx p_{\rm local}N$.
For Model 1 and 2, the trial factor correction yields a global p-value of $3.0\sigma$. Five of the 10 IceCube alerts were followed-up by VHE observations. With the exception of IceCube-170922A, no alert has been observed by an IACT more than 3.2 hours. Considering 5 alerts only, the global p-value for Model 3 based on the formula above becomes $3.5\sigma$. For 10 (51) alerts the corresponding p-values are $3.3\sigma$ ($2.8\sigma$). Since IACTs do not follow up all IceCube alerts, it is not clear if these values are relevant in this case.

\paragraph*{Correlation analysis sensitivity}
\addcontentsline{toc}{subsection}{\protect\numberline{}Correlation analysis sensitivity}%
To demonstrate that the test described above is sensitive to our signal hypothesis,  an ensemble of simulated IceCube observations was generated assuming a proportionality between the instantaneous \g-ray and neutrino emission, corresponding to Model 1 above.

In these simulations, the signal normalization is set such that the expected number of IceCube real-time alerts originating from cataloged extragalactic \Fermi-LAT sources is equal to 1.
Here we have assumed that the purity of the IceCube real-time stream is $\sim 40\%$ and that blazars produce $\sim5\%$ of the diffuse neutrino flux, which is consistent with the current upper limits in~\cite{Aartsen:2016lir}. With a total of 51 alerts, we expect roughly one coincidence. In each realization of these signal simulations, one IceCube event is injected on a cataloged extragalactic \Fermi-LAT source. The probability to select each individual source is set proportional to its energy flux.

For each injected IceCube event, we calculate the TS as described above using the observed \g-ray light curves.

\begin{figure}[ht]
  \centering
    \begin{subfigure}[b]{0.7\textwidth}
      \caption{}
      \includegraphics[width=\textwidth]{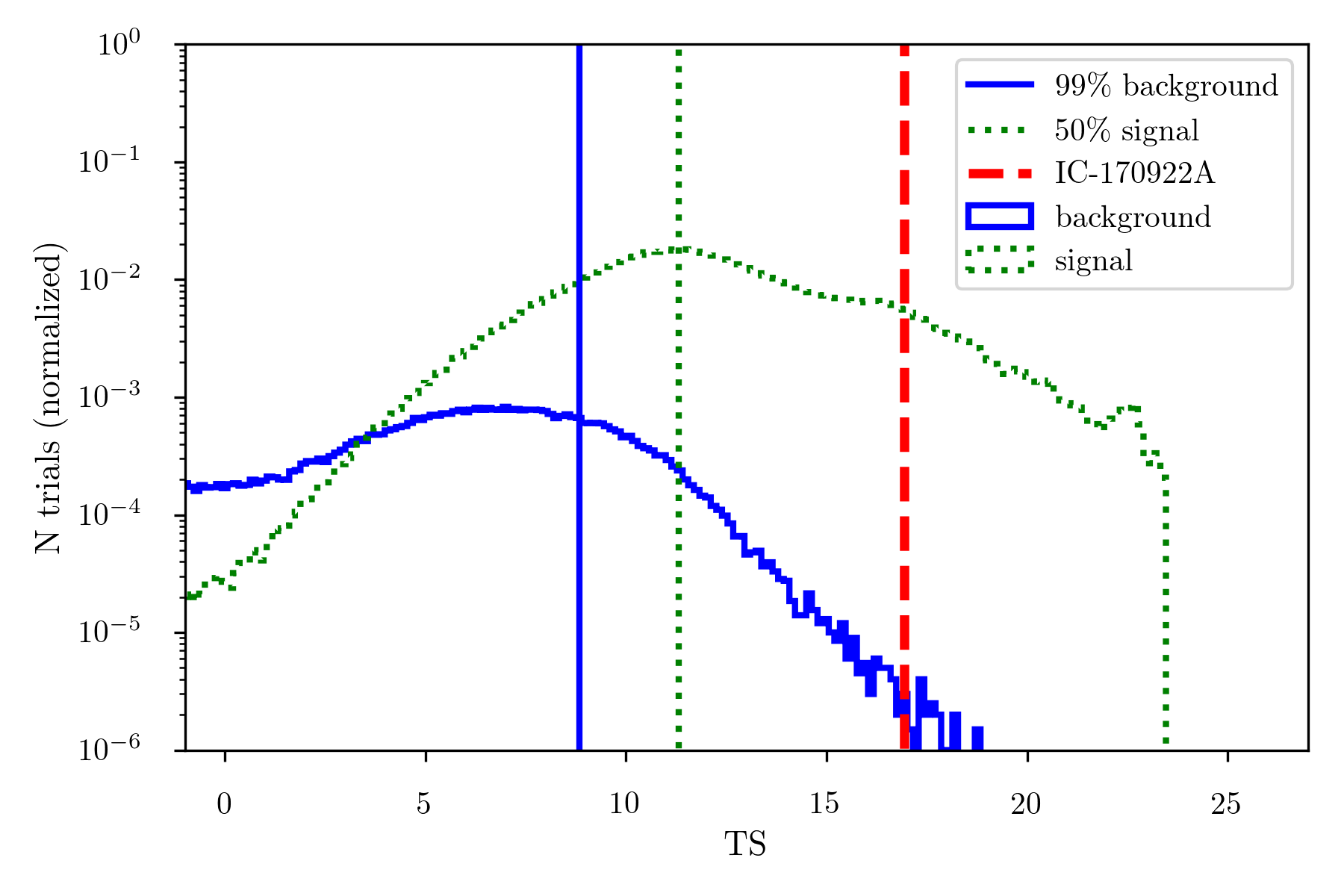}
    \end{subfigure}
    \begin{subfigure}[b]{0.7\textwidth}
      \caption{}
      \includegraphics[width=\textwidth]{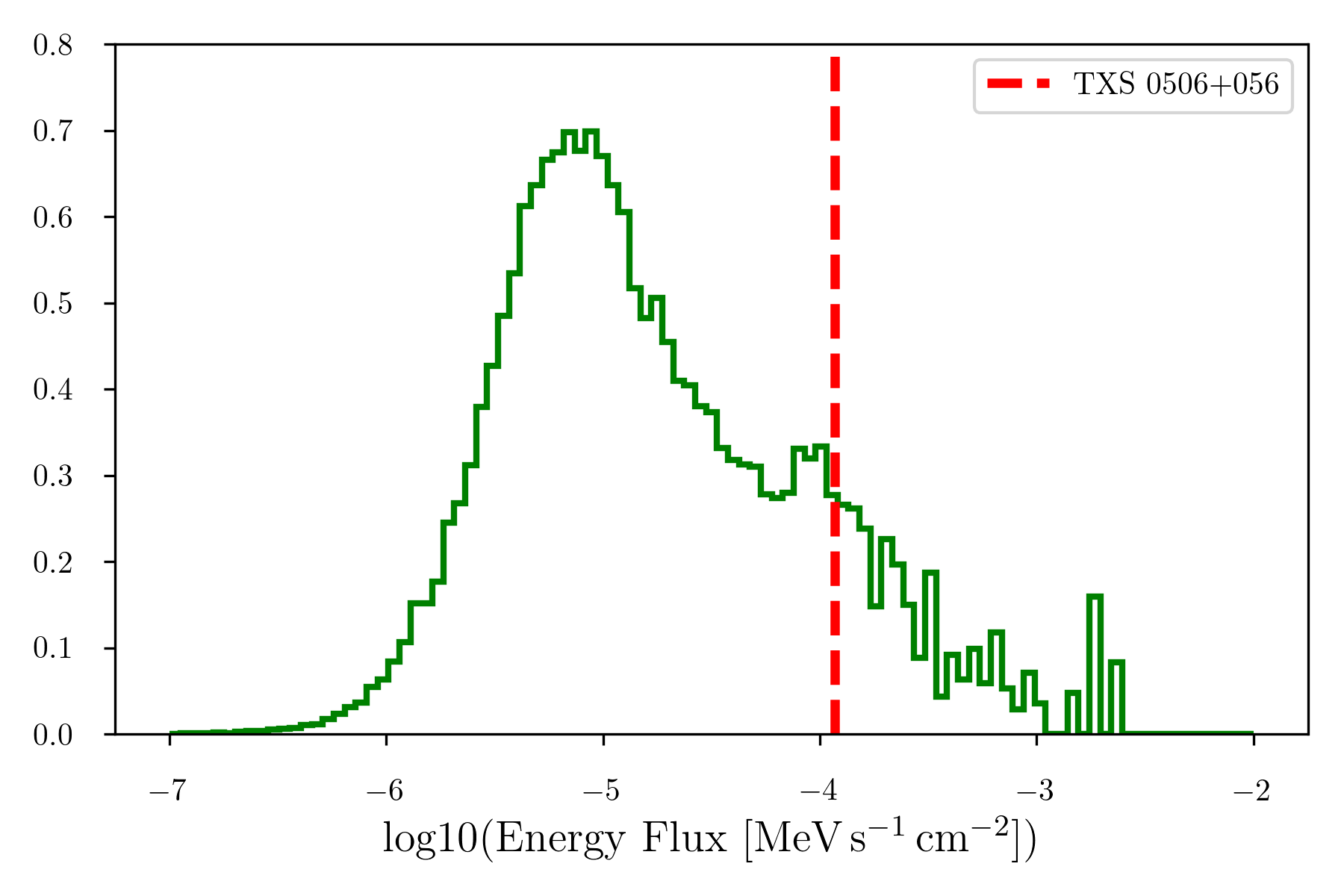}
    \end{subfigure}
  \caption{{\bf \g-ray energy flux and neutrino correlation study sensitivity.}
Panel~A: $TS$ distribution for background trials (blue) and signal trials (green dashed) assuming a linear correlation of \g-ray energy flux and neutrino flux. The solid blue line indicates the TS value below which 99\% of the background trials lie. The green dotted line shows the median TS of the signal trial distribution. The red dashed line shows the measured TS value for IceCube-170922A. The x-axis is suppressed in order to show only the relevant tail of the background distribution. Panel~B: distribution of \g-ray energy flux (for \g-ray energies $>$1 GeV) for found neutrino \g-ray correlations assuming that all sources produce neutrinos proportionately to their energy fluxes in the range 1\,GeV -- 100\,GeV.}
  \label{fig:TS}
\end{figure}

The resulting signal and background TS distributions are compared in Figure~\ref{fig:TS}.
Defining sensitivity as the fraction of the signal realizations correctly identified as such, the figure shows that for a reasonable cut at e.g. $TS=9$, a sensitivity $>50$\% is obtained for a p-value less than 1\%. The pre-trial p-value shown for IceCube-170922A is $2.1 \cdot 10^{-5}$, corresponding to Model 1.
This test has been applied for the assumption of one signal event and can therefore be directly compared with the pre-trial p-values of the energy flux weighting scenario described above.

Furthermore the energy flux distribution of \g-ray sources with simulated neutrino coincidences was inspected assuming that all sources produce neutrinos proportional to their \g-ray energy flux. It is found that $14\%$ of all sources have an equal to or larger \g-ray energy flux than TXS~0506+056 during the time of IceCube-170922A (see Fig.~\ref{fig:TS}). This shows that if the neutrino flux is in fact highly correlated with the \g-ray flux it is not surprising that we detect a neutrino in coincidence with this particular source and flaring incident.
This test neglects dim \g-ray sources below the detection threshold of the \Fermi-LAT, which would also contribute to the neutrino flux.

\paragraph*{Previous high-energy IceCube events}
\addcontentsline{toc}{subsection}{\protect\numberline{}Previous high-energy IceCube events}%
Prior to IceCube-170922A, the IceCube real-time system sent 9 public high-energy neutrino alerts. IceCube data recorded prior to the start of the real-time system in April 2016, starting from 2010, has been inspected for events that would have passed the selection criteria of the real-time stream. An additional 41 events were identified. The 90\% error contours of all 51 events were searched for \g-ray sources in spatial coincidence. The angular resolution for those events varies strongly with the topology and energy of the event. Only events with an angular uncertainty of less than 5\,deg$^2$ are considered, excluding 4 events from the pre-alert time period.  Events with larger uncertainty would get a small spatial weight assigned in the likelihood analysis and would not yield a significant p-value. One neutrino (9 December 2014) is found with a 90\% location uncertainty region of 1.76\,deg$^2$ in spatial coincidence with the \g-ray source 3FGL J1040.4+0615. The best-fitting neutrino position is 0.27 deg from the position of the \g-ray source.
In the monthly time bin around the neutrino arrival time, the source was detected with an energy flux of $1.3 \times 10^{-11}$~erg cm$^{-2}$ s$^{-1}$ between 1~GeV  and 100~GeV, more than an order of magnitude lower than the energy flux at TXS~0506+056 during the time of IceCube-170922A of $1.9 \times 10^{-10}$~erg cm$^{-2}$ s$^{-1}$, and about a factor of 2 below the brightest emission period ($2.7 \times 10^{-11}$~erg cm$^{-2}$ s$^{-1}$) observed for this particular source. Therefore, this event would have produced substantially lower test statistic values in the statistical tests for chance coincidence described above, where a correlation between the gamma-ray and neutrino emission is assumed.

\end{document}